\documentclass[10pt]{article}

\usepackage{graphicx} 
\usepackage{enumitem}
\usepackage[a4paper, total={6.5in, 9in}]{geometry}
\usepackage[ruled,vlined]{algorithm2e}
\usepackage{amsmath}
\usepackage{amssymb}
\usepackage{amsthm}
\usepackage{bbm}
\usepackage{booktabs}
\usepackage{multirow}
\usepackage{natbib}
\usepackage{bm}
\usepackage{float}
\usepackage{xr}

\usepackage{hyperref}

\newcommand{\ind}{\perp\!\!\!\!\perp}

\let\vec\boldsymbol

\newtheorem{theorem}{Theorem}
\newtheorem{lemma}{Lemma}

\title{Bayesian Model Averaging in Causal Instrumental Variable Models}
\author{Gregor Steiner\thanks{gregor.steiner@warwick.ac.uk} \and Mark Steel\thanks{m.steel@warwick.ac.uk}}
\date{Department of Statistics, University of Warwick \\[12pt] \today}

\begin{document}

\maketitle

\begin{abstract}
Instrumental variables are a popular tool to infer causal effects under unobserved confounding, but choosing suitable instruments is challenging in practice. We propose gIVBMA, a Bayesian model averaging procedure that addresses this challenge by averaging across different sets of instrumental variables and covariates in a structural equation model. This allows for data-driven selection of valid and relevant instruments and provides additional robustness against invalid instruments. Our approach extends previous work through a scale-invariant prior structure and accommodates non-Gaussian outcomes and treatments, offering greater flexibility than existing methods. The computational strategy uses conditional Bayes factors to update models separately for the outcome and treatments. We prove that this model selection procedure is consistent. In simulation experiments, gIVBMA outperforms current state-of-the-art methods. We demonstrate its usefulness in two empirical applications: the effects of malaria and institutions on income per capita and the returns to schooling. A software implementation of gIVBMA is available in Julia.
\end{abstract}

\noindent%

\section{Introduction}

Instrumental variables (IV) offer a way to infer causal effects in the presence of unobserved confounding. To be suitable, an instrumental variable must meet two criteria: It must not be affected by the unobserved confounder and must not affect the outcome directly (valid), and it must be associated with the regressor of interest (relevant). In practice, finding variables that fulfill these assumptions is challenging. Furthermore, when multiple instrumental variables are available, the results can be highly sensitive to the specific instruments chosen. In many applications, there is a large degree of uncertainty over which instruments to use in addition to the usual uncertainty over covariate inclusion in a regression model. To address this uncertainty, we propose a Bayesian model averaging (BMA) procedure that averages across different sets of instrumental variables and covariates in a multivariate sampling model with correlated residuals. In our context,  endogenous variables are treatments that are affected by unobserved confounding.

An important contribution of this paper is a data-driven approach for selecting valid and relevant instruments. The user specifies a matrix of potential instruments and covariates, and the BMA framework identifies valid and relevant instruments as those variables included in the treatment model but excluded from the outcome model. This approach provides robustness against the use of invalid or irrelevant instruments. We establish a consistency result showing that the posterior
asymptotically concentrates on the correct model.

We build on a large existing literature. Key ideas and methodology for Bayesian variable selection in multivariate regression models can be found in \cite{Brown_etal_98}, while \cite{rossi_bayesian_2005} propose a simple Bayesian approach to instrumental variables based on conjugate normal and inverse Wishart priors and a Gibbs sampler iterating between the outcome, treatment, and covariance parameters. \citet{koop_bayesian_2012} were among the first to propose BMA in IV models, offering a general approach for simultaneous equations. However, their complex reversible jump MCMC scheme (which can mix quite slowly in practice) makes their method less accessible for practitioners. \citet{karl_instrumental_2012} build on the Bayesian IV framework of \citet{rossi_bayesian_2005}, using conditional Bayes factors to iteratively update the outcome and treatment models within a Gibbs sampler. Subsequent work \citep{lenkoski_sovereign_2020,kourtellos2020measuring,lee_incorporating_2022} extends their approach to allow for non-Gaussian and multiple endogenous variables. \citet{lenkoski_two-stage_2014} introduce a hybrid Bayesian version of the classical Two-Stage Least Squares (TSLS) estimator. 

Our proposed methodology follows \cite{karl_instrumental_2012}, and we take close inspiration from their computational strategy: we use a Gibbs sampler to update the outcome, treatment, and covariance parameters separately and use conditional Bayes factors in the model updates. However, we also provide a careful analysis of prior structures, and we use suitably adapted $g$-priors on the regression coefficients, which make the analysis scale-invariant, and a flexible Cholesky-based prior on the structural covariance \citep{lopes_bayesian_2014}. Our approach uses independent priors on the outcome and treatment model space without enforcing identification through zero prior probability on non-identified models. We accommodate non-Gaussian outcomes and endogenous variables by adapting the latent Gaussian framework with univariate link functions (ULLGM) proposed in \cite{steel_model_2024}. This greatly extends the applicability of our method while not substantially adding to the computational cost. We prove that our model selection procedure is consistent in the sense that the conditional Bayes factors used in the model updates tend to infinity in favor of the true model as the sample size grows. Finally, we provide software in the accompanying \texttt{gIVBMA.jl} package written in the Julia language. While \cite{karl_instrumental_2012} previously released an R package for their method, it has since been withdrawn from CRAN. We refer to the method of \cite{karl_instrumental_2012} as IVBMA and call our proposed procedure generalised IVBMA (gIVBMA).


Related ideas and methods can also be found in the context of Bayesian inference on binary treatment effects under endogenous treatment selection. Methodology for these models with Bayesian variable selection was developed by \cite{jacobi_etal_16} using a single latent factor, and by \cite{wagner_etal_23} from a bi-factor model perspective. 

An important strand of the classical literature focuses on identification and estimation when some instrumental variables are invalid \citep[e.g.][]{ditraglia_using_2016, kang_instrumental_2016, windmeijer_use_2019, windmeijer_confidence_2021}.  \cite{kang_instrumental_2016} show that classical identification is still possible as long as a plurality rule holds, {i.e.~}the valid instruments outnumber the invalid ones. Our approach performs well in settings with potentially invalid instruments, but is not bound by the plurality rule.

Our proposed method also performs well in cases with many individually weak instruments. Traditional classical estimators tend to exhibit substantial bias and can even become inconsistent in such settings. To address these issues, common approaches include first-stage regularisation\footnote{Inspired by the TSLS procedure, ``first-stage'' refers to the estimation of the treatment model, while ``second-stage'' relates to inference in the outcome model.} \citep[e.g.][]{okui_instrumental_2011, belloni_sparse_2012, carrasco_regularization_2012}, using jackknife-fitted values in the first stage \citep[e.g.][]{angrist_jackknife_1999, hansen_instrumental_2014}, selecting instruments based on minimizing a mean-square error criterion \citep[e.g.][]{donald_choosing_2001}, or model-averaging to obtain model-averaged first-stage predictions of the endogenous variable \citep[e.g.][]{kuersteiner_constructing_2010}.

Through extensive experiments on real and simulated data, we evaluate gIVBMA against IVBMA, BMA, and a number of classical methods tailored for this problem. The gIVBMA methodology works well in settings with potentially invalid instruments and many individually weak instruments. Another advantage of gIVBMA (and IVBMA) over most classical methods is that we obtain the posterior distribution of the structural covariance matrix which informs us on the degree of endogeneity. This also provides these methods with the ability to borrow strength from the model for the endogenous variables.

Section {\ref{sec:sampling}} introduces the sampling model and discusses model uncertainty and identification. Sections {\ref{sec:prior}} and \ref{sec:Posterior_Inference} respectively describe the prior specification and posterior inference, while Section \ref{sec:consistency} provides the model consistency results. Simulated and real data are analysed in Sections {\ref{sec:simulation}} and \ref{sec:empirical}. Finally, Section {\ref{sec:conclusion}} concludes. 


\section{The sampling model}\label{sec:sampling}

\subsection{Model Specification}

We consider a structural model for $n$ observations of a single outcome and $l\ge 1$ potentially endogenous variables:
\begin{align} \label{eq:StructMod}
\begin{aligned}
    \vec{y} &= \alpha \vec{\iota} + \vec{X} \vec{\tau} + \vec{Z} \vec{\beta} + \vec{\epsilon} \\
    \vec{X} &=  \vec{\iota} \vec{\Gamma} + \vec{Z} \vec{\Delta} + \vec{H},
\end{aligned}
\end{align}
where $\vec{y}$ is an $n \times 1$ outcome vector, $\vec{X} = \vec{[x_1 : \ldots : x_l]}$ is an $n \times l$ matrix of causes or treatments that are potentially endogenous, $\vec{Z}$ is an $n \times p$ matrix of instruments and exogenous covariates (i.e.~$\vec{Z}\ind \vec{\epsilon}$), $\vec{\epsilon}$ is an $n \times 1$ vector of outcome residuals, and $\vec{H = [\eta_1 : \ldots : \eta_l]}$ is an $n \times l$ matrix of treatment residuals. The outcome model is parameterized by an intercept $\alpha$, an $l \times 1$ vector of ``effects'' $\vec{\tau}$, and a $p \times 1$ vector of instrument (or covariate) coefficients $\vec{\beta}$. The treatment model is parameterized by a $1 \times l$ (row) vector $\vec{\Gamma}$ and a $p \times l$ matrix of slope coefficients $\vec{\Delta}$. The main parameter of interest is the vector of treatment effects $\vec{\tau}$.

We allow all instruments to enter the outcome model, which is essential for our model selection procedure to identify valid and relevant instruments in a data-driven manner. For valid instruments, $\vec{\beta}$ will have most of its posterior mass on (or very close to) zero, but we do not enforce this a priori. This point will become clearer in the next section, where we introduce model uncertainty.

This setup can be motivated by the potential outcomes framework \citep[here we follow the exposition in][]{imbens_instrumental_2014}. Suppose that $y_i({x})$ is the potential outcome for the $i$-th outcome observation with conditional expectation 
$    \mathbb{E} \left[ y_i({x}) \mid \vec{Z_i} \right] = \alpha + {x} {\tau} + \vec{Z_i \beta}
$
such that the average treatment effect (ATE) of increasing $x$ by one unit is exactly $\tau$. If we have multiple endogenous variables, i.e.~$\vec{x}$ is a vector, any component of $\vec{\tau}$ is the ATE of only increasing the corresponding component of $\vec{x}$ by 1 while the other components remain constant. Define the outcome residual $\epsilon_i = y_i(x) - \mathbb{E} \left[ y_i(x) \mid \vec{Z_i} \right]$, which is by definition uncorrelated with the exogenous variables $\vec{Z_i}$. Given observations $(y_i, \vec{X}_i, \vec{Z}_i)$, this potential outcomes model implies the observed data model
$    y_i = y_i(\vec{X}_i) = \alpha + \vec{X}_i \vec{\tau} + \vec{Z}_i \vec{\beta} + \epsilon_i,
$
which is our outcome equation for a single observation. Thus, $\vec{\tau}$ has a causal interpretation. Subsection \ref{sec:identification} contains more discussion on the identification of $\vec{\tau}$.

We assume the residuals follow a matrix normal distribution (defined in supplementary Section \ref{sec:notation})
$   \vec{[\epsilon : H]} \sim MN(\vec{0}, \vec{I_n}, \vec{\Sigma})$,
where the structural covariance matrix $\vec{\Sigma}$ can be partitioned into
\begin{align*}
    \vec{\Sigma} = \begin{pmatrix}
        \sigma_{yy} & \vec{\Sigma_{yx}} \\ \vec{\Sigma_{yx}}^\intercal & \vec{\Sigma_{xx}}
    \end{pmatrix} =
   \begin{pmatrix}
        \text{var}(\epsilon_i)& \text{cov}(\epsilon_i,\vec{H_i}) \\ \text{cov}(\vec{H_i},\epsilon_i)  & \text{var}(\vec{H_i})
    \end{pmatrix}.
\end{align*}
If $\vec{\Sigma_{yx}} \neq 0$, 
the outcome and the treatment residuals are correlated, signaling the presence of unobserved confounding or endogeneity. We assume the errors to be homoskedastic and serially uncorrelated across observations 
but this could be generalised relatively easily. 

The conditional distribution of $\vec{y} \mid \vec{X}$ is given by (see supplementary Section \ref{sec:deriv_cond})
\begin{align*}
    \vec{y} \mid \vec{X} \sim N(\alpha \vec{\iota} + \vec{X} \vec{\tau} + \vec{Z} \vec{\beta} + \vec{H} \vec{\Sigma_{xx}^{-1}} \vec{\Sigma_{yx}}^\intercal, \; \sigma_{y|x} \vec{I_n}),
\end{align*}
where $\sigma_{y|x} = \sigma_{yy} - \vec{\Sigma_{yx} \Sigma_{xx}^{-1} \Sigma_{yx}}^\intercal$ and $\vec{H} = \vec{X} - (\vec{\iota} \vec{\Gamma} + \vec{Z} \vec{\Delta})$. The marginal distribution of the treatment matrix,
$
    \vec{X} \sim  MN( \vec{\iota \Gamma + Z \Delta }, \vec{I_n}, \vec{\Sigma_{xx}})
$
completes the joint distribution of $\vec{y}$ and $\vec{X}$.

\subsection{Model uncertainty}\label{sec:modeluncertainty}

This paper aims to incorporate model uncertainty into the framework outlined above (or its extension to non-Gaussian distributions as described in Subsection \ref{sec:nonGaussian}). In particular, a model refers to the exclusion of a specific set of covariates or instruments, or, equivalently, exact zero restrictions on the corresponding regression coefficients. We consider such uncertainty in both the treatment model and the outcome model.

Let $L \in \mathcal{L}$ denote a possible outcome model and let $M \in \mathcal{M}$ denote a possible treatment model, where $\mathcal{L}$ and $\mathcal{M}$ are the sets of all models considered. Then, the likelihood conditional on the models is given by
\begin{align*}
    \vec{y} \mid \vec{X}, L, M &\sim N(\alpha \vec{\iota} + \vec{X} \vec{\tau} + \vec{Z}_L \vec{\beta}_L + \vec{H \Sigma_{xx}^{-1} \Sigma_{yx}}^\intercal, \; \sigma_{y|x} \vec{I_n}) \\
    \vec{X} \mid M &\sim MN(\vec{\iota \Gamma} + \vec{Z}_M \vec{\Delta}_M, \vec{I_n}, \vec{\Sigma_{xx}}),
\end{align*}
where the subscripts indicate the appropriate subset of the columns of the design matrices and equivalent zero restrictions on the coefficients. Note that the outcome also depends on the treatment model $M$ through the residual matrix $\vec{H}$. Both the number of potential outcome  and treatment models are $2^p$. We do not introduce any cross-restrictions, so the number of combinations of outcome and treatment models is $2^{2p}$.

Unlike \cite{karl_instrumental_2012}, we choose to always include the endogenous variables in the outcome model. This is a conscious choice and not necessary, as our prior specification and computational implementation would also easily allow for excluding the endogenous variables. We believe, however, that this is more natural as our primary focus is on estimating the effects of the endogenous variables while accounting for other covariates. This implies that we obtain a non-zero effect estimate in every model we consider. Compared to \cite{lee_incorporating_2022}, our treatment model $M$ puts the same zero restrictions on all columns of the coefficient matrix $\vec{\Delta}$. This results in slightly less flexibility (for $l>1$) but leads to more interpretable results and substantially reduces the computational cost.

We also allow the user to fix certain variables to be instruments, meaning they are always excluded from the outcome model. In this variant, the user specifies two matrices: the pre-specified instruments $\vec{Z}$ and the covariates $\vec{W}$. The columns of $\vec{Z}$ are restricted to the treatment model, while the columns of $\vec{W}$ may enter either model. All of the results below continue to hold for this variant with slight modifications. Further details are provided in supplementary Section  \ref{sec:fixed_Z}.

\subsection{Identification} \label{sec:identification}

The regression coefficient of $\vec{X}$ in the conditional model $\vec{y} \mid \vec{X}$ is generally not equal to $\vec{\tau}$. To see this, write the mean of the conditional distribution as
\begin{align*}
    \mathbb{E}\left[ \vec{y} \mid \vec{X}, L, M \right]
    &= \alpha \vec{\iota} + \vec{X} \left(\vec{\tau + \Sigma_{xx}^{-1} \Sigma_{yx}}^\intercal \right)+ \vec{Z}_L \vec{\beta}_L -  \left( \vec{\iota \Gamma} + \vec{Z}_M \vec{\Delta}_M \right)\vec{\Sigma_{xx}}^{-1} \vec{\Sigma_{yx}}^\intercal.
\end{align*}
Naively regressing $\vec{y}$ on $\vec{X}$ targets $\left(\vec{\tau + \Sigma_{xx}^{-1} \Sigma_{yx}}^\intercal \right)$ instead of $\vec{\tau}$ itself. Whenever $\vec{\Sigma_{yx}} \neq 0$, this approach leads to biased results. This illustrates why it is necessary to consider the joint distribution of $\vec{y}$ and $\vec{X}$ instead of only the conditional distribution of $\vec{y} \mid {\vec{X}}$. Inference on $\vec\tau$ is still possible if we have suitable instruments (i.e.~columns of $\vec{Z}_M$ that are not in $\vec{Z}_L$). Then, one can use the inference on  $\vec{\Delta}$ from the treatment model to infer the covariance ``ratio'' $\vec{\Sigma_{xx}^{-1} \Sigma_{yx}}^\intercal$ and subsequently identify $\vec{\tau}$.

Our approach does not do this explicitly, but instead, in the outcome model, we condition on $\vec{H \Sigma_{xx}^{-1} \Sigma_{yx}}^\intercal$, which is known conditionally on the treatment and covariance parameters. The treatment residual $\vec{H}$ contains all the variation in $\vec{X}$ that the instruments and covariates do not explain. Therefore,  including $\vec{H}$ acts as a control for the unobserved confounding. This is similar to classical control function approaches \citep[see e.g.][]{wooldridge_control_2015}, but the coefficient of the control function is fixed (conditional on $\vec{\Sigma}$) and implied by our modeling assumptions. The treatment effect  $\vec{\tau}$ is identified in a ``classical'' sense if at least $l$ instruments satisfy the following:
\begin{enumerate}
    \item \textbf{Relevance:} The $j$-th instrument $\vec{z}_j$ is relevant if it is associated with at least one of the endogenous variables, that is, $\vec{\Delta}_j$ is not the zero-vector.
    \item \textbf{Validity/Exogeneity:} The instruments are valid (or exogenous) if they are conditionally independent of the error in the outcome model, $\vec{Z}_M \ind \vec{\epsilon} \mid \vec{Z}_L$.
\end{enumerate}
The exogeneity assumption above combines the unconfoundedness and exclusion restriction assumptions usually featured in a potential outcomes setup. The monotonicity assumption is trivially satisfied in our setup as the treatment equation is linear in the instruments \citep[for more details see e.g.][]{imbens_instrumental_2014}. 

For a given combination of $L$ and $M$, the model is 
just-identified if the number of valid and relevant instruments implied by $L$ and $M$ is exactly equal to $l$. The model is 
under-identified if there are fewer instruments than $l$ and over-identified if there are more.

It is important to emphasize that identification is not a binary characteristic as it tends to be in classical inference for two reasons. First, if the model is under-identified, then the likelihood does not add information on some components of $\vec{\Sigma_{xx}^{-1} \Sigma_{yx}}^\intercal$. However, with a proper prior, we can still obtain a proper posterior distribution, albeit one that may be quite diffuse in some components. This carries through to the inference on $\vec{\tau}$. The second point is specific to our BMA approach: Each combination of outcome and treatment models can have different instruments with varying degrees of instrument strength. While some of these models will be well-identified, others might be very uninformative. Consequently, the marginal posterior of $\vec{\tau}$ can be very diffuse if the majority of the posterior weight is concentrated on these uninformative models. This merely reflects that we are not learning much about $\vec{\tau}$ from the available data, but it does not prevent us from conducting inference.

\subsection{Non-Gaussian models}\label{sec:nonGaussian}

We can relax the Gaussianity assumption using the Univariate Link Latent Gaussian Models (ULLGM) framework proposed in \cite{steel_model_2024}. The idea is to assign a latent Gaussian representation to any column of $\vec{[y : X]}$ that is not Gaussian and then perform posterior inference conditional on the latent Gaussian. Unlike \cite{lee_incorporating_2022}, we do not need to rely on approximations.

More precisely, assume $\vec{y}$ and $\vec{X}$ are non-Gaussian and let the $n \times 1$ vector $\vec{q}$ and the $n \times l$ matrix $\vec{Q}$ be their latent Gaussian representations, respectively. Then, we model  $(y_i, \vec{X_i}), \; i= 1,\ldots, n$, independently as
\begin{align*}
    y_i \mid q_i, r_y \sim F^{(y)}_{h_y\left(q_i \right), r_y}, \quad \vec{X_i} \mid \vec{Q_i}, \vec{r_x} \sim \prod_{j=1}^l F^{(x_j)}_{h_{x_j}\left(Q_{ij} \right), r_{x_j}}    
\end{align*}
where
\begin{align*}
    q_i \mid \vec{X_i, Q_i} &\sim N \left(\alpha + \vec{X_i \tau + Z_i \beta} + \left(\vec{Q_i} - \vec{\Gamma}^\intercal - \vec{Z_i} \vec{\Delta} \right) \vec{\Sigma_{xx}^{-1} \Sigma_{yx}}^\intercal, \sigma_{y|x} \right) \\
    \vec{Q} &\sim MN(\vec{\iota \Gamma + Z \Delta}, \vec{I_n}, \vec{\Sigma_{xx}}),
\end{align*}
$F^{(y)}$ and $F^{(x_j)}$ are the distributions of $\vec{y}$ and $\vec{x_j}$, and $h_y$ and $h_{x_j}$ are invertible univariate link functions that map the latent Gaussian to the appropriate parameter of $F^{(y)}$ and $F^{(x_j)}, j=1\dots,l$. If required, $r_y$ and $r_{x_j}$ group any additional parameters of $F^{(y)}$ and $F^{(x_j)}$, which are assumed to be the same for all observations. We also assume the endogenous variables are independent conditional on their latent Gaussian representation, that is, all the dependence is in the Gaussian part. Many distributions can be expressed as members of the ULLGM family. Two examples that we use in our simulations and applications are \citep[for more examples see][]{steel_model_2024}:
\begin{itemize}
    \item \textbf{Poisson-Log-Normal:} $F = \text{Poisson}(\lambda_i), \lambda_i = h(q_i) = \exp(q_i)$
    \item \textbf{Beta-Logistic:} $F = \text{Beta}\left( \mu_i, r \right), \mu_i = h(q_i) = \exp(q_i) / (1 + \exp(q_i))$ , where $\mu_i$ represents the mean and $r$ a dispersion parameter \citep[using the alternative parameterization of the Beta distribution introduced in][]{ferrari_beta_2004}.
\end{itemize}
The interpretation of the Gaussian parameters varies with the distributions. For instance, in a Poisson-Log-Normal distribution for the outcome, the outcome regression parameters have a log-linear interpretation. The causal interpretation of $\vec{\tau}$ is then that of an expected log-ratio of expected potential outcomes (assuming $l=1$ for simplicity),
\begin{align*}
   \vec{\tau} = \mathbb{E}_q\left[ q_i(x+1) - q_i(x) \right] = \mathbb{E}_q\left[ \log \left( \frac{\mathbb{E}_{y\mid q}\left[ y_i(x+1) \right]}{\mathbb{E}_{y \mid q}\left[ y_i(x) \right]} \right) \right]=
   \mathbb{E}_q\left[ \log \left( \frac{\lambda_i(x+1)}{\lambda_i(x)} \right) \right],
\end{align*}
where $q_i(x) = \log \mathbb{E}_{y \mid q}\left[ y_i(x) \right]=\log \lambda_i(x)$ is the latent Gaussian potential outcome and $y_i(x)$ is the potential outcome on the observed level for a given value of $x$. Note that the causal interpretation of $\vec{\tau}$ is not affected by the choice of $F^{(x_j)}_{h_{x_j}(\cdot)}, j=1,\dots,l$, as $\vec{X}_i$ (and not $\vec{Q}_i$) multiplies $\vec{\tau}$ in the mean for $q_i$. If we choose to parameterize a location parameter of $\vec{X}_i$ by the latent Gaussian process $\vec{Q}_i$, then the invertibility of $h_{x_j}, j=1,\dots,l$, gives us a stochastic version of the monotonicity condition. 

Model uncertainty is now introduced in the equations for the latent variables, similar to the discussion in Subsection \ref{sec:modeluncertainty}. 

\section{Prior Specification}\label{sec:prior}

\subsection{The prior on the regression coefficients}\label{sec:priorCoeff}

As a prior on the regression coefficients, we adopt a version of the widely used $g$-prior \citep{zellner_assessing_1986}. However, we incorporate the intercepts into the $g$-prior as we cannot guarantee posterior propriety with an improper prior. In the outcome model, the prior distribution under model $L$ on the coefficient vector $\vec{\rho} = (\alpha, \vec{\tau}^\intercal, \vec{\beta}^\intercal)^\intercal$ then has the following continuous prior on $\vec{\rho}_L = (\alpha, \vec{\tau}^\intercal, \vec{\beta}_L^\intercal)^\intercal$, while the other elements of $\vec{\rho}$ are exactly zero:
\begin{align}\label{eq:priorL}
     \vec{\rho}_L \mid L, \vec{\Sigma} \sim N \left(\vec{0}, g_L \sigma_{y|x} (\vec{U}_L^\intercal \vec{U}_L)^{-1}\right),
\end{align}
where $\vec{U}_L = [\vec{\iota : X : Z}_L]$. In the treatment model, we use a matrix version of the $g$-prior on the coefficient matrix $\vec{\Lambda} = [\vec{\Gamma}^\intercal : \vec{\Delta}^\intercal]^\intercal$ given by a matrix normal prior on $\vec{\Lambda}_M = [\vec{\Gamma}^\intercal : \vec{\Delta}_M^\intercal]^\intercal$
\begin{align}\label{eq:priorM}
    \vec{\Lambda}_M \mid M, \vec{\Sigma} \sim MN(\vec{0}, g_M (\vec{V}_M^\intercal \vec{V}_M)^{-1}, \vec{\Sigma_{xx}}),
\end{align}
where $\vec{V}_M = [\vec{\iota : Z}_M]$ and the other elements of $\vec{\Lambda}$ are zero. 
We will assume that for all models that we consider, the matrices $\vec{U}_L$ and $\vec{V}_M$ are of full column rank. This will ensure that the inverses in (\ref{eq:priorL}) and (\ref{eq:priorM}) exist for all models. Since the $g$-priors include the intercept, we recommend centering all the Gaussian components of $[\vec{y} : \vec{X}]$ to ensure that the zero mean prior on the intercept is reasonable. As stated in Subsection \ref{sec:modeluncertainty}, we impose the same zero restrictions on all treatment model equations if $l>1$. This means that the matrix normal prior in (\ref{eq:priorM}) can be adopted, but in case we would want to allow potentially different covariates in each treatment model equation, we would require more flexible priors such as {\it e.g.}~the Recursive Extended Natural Conjugate prior of \cite{RichardSteel}. 
In case we would want to impose asymmetric shrinkage across treatment equations, we could adopt the Asymmetric Conjugate prior introduced by  \cite{Chan22} in the context of VARs.   
Of course, the size of the model space and the complexity of the analysis would increase substantially and, for our purposes in this paper, we are typically not really interested in the coefficients in the treatment model. \cite{Hahn_etal_18} propose a factor shrinkage prior for the instrument coefficients when $l=1$, based on an approximation to the  horseshoe prior (see Subsection \ref{sec:competing}). 

The priors in (\ref{eq:priorL}) and (\ref{eq:priorM}) are conditionally conjugate for our sampling model, so we obtain closed-form expressions for conditional posteriors and marginal likelihoods (see Subsection \ref{sec:CML}). In addition, we only have to elicit two scalar hyperparameters, $g_L$ and $g_M$. The choice of $g_L,g_M>0$ controls the prior variance and the complexity penalty of the Bayes factors.

We can either fix $g_L$ and $g_M$ or put hyperpriors on them. We will focus on two possible choices: an adapted version of the benchmark or BRIC prior \citep{fernandez_benchmark_2001} and the hyper-$g/n$ prior \citep{liang_mixtures_2008}. In the adapted benchmark prior, we fix
$
    g_L = \max \left\{n, (p+l+1)^2 \right\}$ 
and $g_M = \max \left\{n, (p+1)^2 \right\}$. 
In the context of the standard normal linear regression model, the benchmark prior results in model selection consistency \citep{fernandez_benchmark_2001}, which will be examined in our setting in Subsection \ref{sec:consistency}.

The hyper-$g/n$ prior is characterised by its pdf,
$$
    p(g) = \frac{a-2}{2n} \left( 1 + \frac{g}{n} \right)^{-a/2},
$$
where $a>2$.
This prior (unlike the regular hyper-$g$ prior) leads to consistent model selection in normal linear regression but does not yield analytic expressions for the marginal likelihood. 
Typical choices are $a=3$ and $a=4$, which behave similarly. 

On any additional parameters $r_y$ and $r_{x_j}$ we adopt proper priors to ensure posterior propriety. 

\subsection{The prior on the covariance matrix}

As in \cite{karl_instrumental_2012}, we first put an inverse Wishart prior (as defined in supplementary Section \ref{sec:notation}) on the structural covariance matrix, 
$   \vec{\Sigma} \sim IW(\nu, \vec{I_{l+1}})$.
Centering the prior over the (diagonal) identity matrix reflects the fact that we want inference on the degree of endogeneity to be primarily driven by the data. We recommend standardising the outcome and endogenous variables (or at least selecting appropriate units of measurement) to ensure the prior scale is not in strong conflict with the likelihood. The degrees of freedom parameter $\nu$ controls the amount of information in the prior and must satisfy $\nu > l$ so that the prior is proper. The choice of $\nu$ can be quite influential as {\it{i.a.}}~it controls how tight the prior on all off-diagonal elements of $\vec{\Sigma}$ is around zero. It can substantially impact the inference on the marginal variances and the degree of endogeneity. To avoid being too dogmatic, we assume a hyperprior on $\nu$.
An Exponential prior that is shifted by at least $l$, such that $\nu > l$, works well in our experience. For more numerical stability, it may be desirable to shift it slightly further. We use an Exponential with scale $1$ shifted by $l + 1$ as our default choice throughout the remainder of the paper. This prior is quite uninformative on $\nu$ itself while implying a prior on the covariance ratio $\vec{\Sigma_{xx}^{-1} \Sigma_{yx}}^\intercal$ that is (almost) identical to fixing $\nu = 3$ \citep[as done by][]{karl_instrumental_2012}. Figure \ref{fig:implied_prior_sigma} in the supplementary Subsection \ref{sec:prior_nu} illustrates the implied priors on the covariance ratio and the outcome variance for shifted Exponential priors with different scales.

As a more flexible alternative, we also consider the Cholesky-based prior proposed in \cite{lopes_bayesian_2014}. Write $\vec{\Sigma}$ as
\begin{align} \label{eq:sigma_cholesky}
    \vec{\Sigma} = \begin{pmatrix}
    1 & \vec{a_{yx}} \\ 0 & \vec{I_l}
\end{pmatrix} \begin{pmatrix}
    \sigma_{y \mid x} & 0 \\ 0 & \vec{\Sigma_{xx}}
\end{pmatrix} \begin{pmatrix}
    1 & 0 \\ \vec{ a_{yx}}^\intercal & \vec{I_l}
\end{pmatrix},
\end{align}
where $\vec{a_{yx}} = \vec{\Sigma_{yx}} \vec{ \Sigma_{xx}}^{-1}$ denotes the scaled residual covariance. We specify the prior on $\vec{\Sigma}$ in three parts by assigning priors to the conditional outcome variance $\sigma_{y \mid x}$, the scaled covariance $\vec{a_{yx}}$, and the treatment variance $\vec{\Sigma_{xx}}$. This allows us to separately quantify the prior uncertainty about these individual covariance components, a potential  advantage over the inverse Wishart prior used in \cite{karl_instrumental_2012}.

A convenient choice is a Gaussian prior on the scaled covariance, $\vec{a_{yx}}^\intercal \sim N(0, \vec{\Omega}_a)$, an inverse Gamma prior (as in Section \ref{sec:notation}) on the outcome variance, $\sigma_{y \mid x} \sim IG(c, d)$, and an inverse Wishart prior on the treatment covariance matrix, $\vec{\Sigma_{xx}} \sim IW(\xi, \vec{I_l})$. These priors lead to conditional closed-form updates. As default choices for the variance hyperparameters, we use $c = \nu/2, d = 1/2, \xi = \nu-1$, where $\nu$ denotes the degrees-of-freedom parameter of the full inverse Wishart prior. Again, it can be useful to put a hyperprior on $\nu$. For the (scaled) residual covariance, we set $\vec{\Omega_a} = \omega_a \vec{I_l}$, where higher $\omega_a$ indicates a weaker prior belief in exogeneity. The key difference with the inverse Wishart  prior is that we can now freely choose $\omega_a$, whereas it is restricted to $\omega_a = \sigma_{y \mid x}$ in the prior implied by the full  inverse Wishart on $\vec\Sigma$.

\subsection{The prior on the model space}
\label{sec:priormodels}

We use a model prior based on the independent inclusion of variables in both the outcome and treatment equations. That is, we have inclusion probabilities, say $w_L$ and $w_M$, and the prior probability of model $L_j$ is 
$    p(L_j) = w_L^{p_j} (1 - w_L)^{p - p_j}$,
where $p$ is the number of potential instruments and covariates 
and $p_j$ is the number included in model $L_j$. The choice of $w_L$ can be very influential, so we recommend a hyperprior on $w_L$ to make the procedure more adaptive. Using a Beta$(a,b_L)$ hyperprior results in the Beta-binomial prior
\begin{align*}
    p(L_j) = \frac{\Gamma(a+b_L)}{\Gamma(a) \Gamma(b_L)} \frac{\Gamma(a+p_j) \Gamma(b_L+p-p_j)}{\Gamma(a+b_L + p)}.
\end{align*}
Following \cite{ley_effect_2009}, we choose $a = 1$ and $b_L = (p-m_L)/m_L$, 
where the user specifies the prior mean outcome model size $m_L$. If not stated otherwise, we use $m_L = p/2$ as our default choice. For the treatment model we use a similar procedure, leading to uniform hyperpriors on the prior inclusion probabilities, as in \cite{ScottBerger10}.

Note that this prior puts positive probability on all potential models. In particular, this means that we do not exclude underidentified models \citep[unlike][]{karl_instrumental_2012}. As discussed above, in an underidentified model, the likelihood provides no information about certain components of the covariance. With a proper prior, however, we can still obtain a well-defined posterior, though some components may remain quite diffuse.

It also means that we allow for variables to enter the outcome model without entering the treatment model, which is generally discouraged in classical IV analysis \citep[see for example][Section 4.6.1]{angrist2009mostly}. The reason is that the first-stage residuals are not necessarily uncorrelated with the covariates excluded from the treatment equation. This correlation leads to inconsistent coefficient estimates that can spill over into the treatment effect of interest. In our setting, however, this issue is less problematic for the following reason: If variables are excluded from the treatment equation, their correlation with the treatment is typically weak, meaning that any resulting second-stage bias has only a small effect on the treatment estimate. In addition, such a variable can still contribute to reducing residual variance in the outcome model and therefore increase precision.

A related question is what prior is induced on the number of valid and relevant instruments.  
This is described in supplementary Subsection \ref{sec:priorValidInstr} and compared with the prior implied by IVBMA (with $l=1$). 

\section{Posterior inference} \label{sec:Posterior_Inference}

\subsection{Computational strategy}
\label{sec:comput}

Obtaining a tractable joint posterior distribution of all parameters and models across both the outcome and treatment models is not possible. Instead, we tackle the problem conditionally and take inspiration from the computational strategy in \cite{karl_instrumental_2012}. We use a Gibbs sampler to iteratively sample the outcome parameters, treatment parameters, and the covariance matrix. Within the first and second steps, we update the models via a Metropolis-Hastings (MH) step in model space, update $g$ if it is not fixed, and then, given the model, we draw the parameters from their conditional posteriors. 
We describe each of these steps for updating the outcome model in more detail below, and the treatment model update is 
analogous. Algorithm \ref{algo} summarises the Gibbs sampler. 

\begingroup
\begin{algorithm}[h] \label{algo}
\caption{The gIVBMA Gibbs Sampling Algorithm}
\KwIn{Data $\mathcal{D} = (\vec{y}, \vec{X}, \vec{Z}, \vec{W})$, number of posterior samples $S$, prior mean model sizes $\left(m_L, m_M \right)$, initial values $\alpha^{(0)}, \vec{\tau}^{(0)}, \vec{\beta}^{(0)}, \vec{\Gamma}^{(0)}, \vec{\Delta}^{(0)}, \vec{\Sigma}^{(0)}, L^{(0)}, M^{(0)}$, and optionally $\vec{q}^{(0)}, \vec{Q}^{(0)}, r_y^{(0)}, r_x^{(0)}, g_L^{(0)}, g_M^{(0)}, \nu^{(0)}$.}

\For{$s = 1, \ldots, S$}{
(a) \textbf{Latent Gaussian Update}: 
\begin{enumerate}[noitemsep, nolistsep]
    \item If any column of $[\vec{y} : \vec{X}]$ is not Gaussian, draw a latent Gaussian representation $\vec{q}^{(s)}$ or $\vec{Q}^{(s)}$ via an MH step using a Barker proposal.
    \item If required, draw additional parameters $r_y^{(s)}$ or $r_x^{(s)}$ via an MH step.
\end{enumerate}
(b) \textbf{Outcome Model Update}:
\begin{enumerate}[noitemsep, nolistsep]
    \item Sample the inclusion indicators $L^{(s)}$ via an MH step in model space.
    \item If $g_L$ is random, draw $g_L^{(s)}$ via an MH step.
    \item Sample the outcome model parameters $\left(\alpha^{(s)}, \vec{\tau}^{(s)}, \vec{\beta}^{(s)} \right)$ from their full conditional distribution in (\ref{eq:cond_post_rho}).
\end{enumerate}
(c) \textbf{Treatment Model Update}:
\begin{enumerate}[noitemsep, nolistsep]
    \item Sample the inclusion indicators $M^{(s)}$ via an MH step in model space.
    \item If $g_M$ is random, draw $g_M^{(s)}$ via an MH step.
    \item Sample the treatment model parameters $\left( \vec{\Gamma}^{(s)}, \vec{\Delta}^{(s)} \right)$ from their full conditional distribution in (\ref{eq:cond_post_lambda}).
\end{enumerate}
(d) \textbf{Covariance Matrix Update}:
\begin{enumerate}[noitemsep, nolistsep]
    \item If $\nu$ is random, draw $\nu^{(s)}$ via an MH step.
    \item Sample the covariance matrix $\vec{\Sigma}^{(s)}$ from its full conditional distribution.
\end{enumerate}
}

\KwOut{A posterior sample for s = 1,\ldots, S: $\left(\alpha^{(s)}, \vec{\tau}^{(s)}, \vec{\beta}^{(s)}, \vec{\Gamma}^{(s)}, \vec{\Delta}^{(s)}, \vec{\Sigma}^{(s)}, L^{(s)}, M^{(s)}, \vec{q}^{(s)}, \vec{Q}^{(s)}, r_y^{(s)}, r_x^{(s)}, g_L^{(s)}, g_M^{(s)}, \nu^{(s)} \right)$.}
\end{algorithm}
\endgroup

Moving in model space is implemented through an MH step where the probability of accepting the proposed model $L'$ given the current model $L$ is
\begin{align*}
    a(L', L) = \min\left(\frac{p(\vec{y} \mid \vec{X}, \vec{\Lambda}, \vec{\Sigma}, L', M)}{p(\vec{y} \mid \vec{X}, \vec{\Lambda}, \vec{\Sigma}, L, M)} \frac{p(L')}{p(L)} \frac{h(L \mid L')}{h(L' \mid L)}, \;1 \right),
\end{align*}
where $h$ is a proposal kernel. Throughout, models are proposed by randomly permuting the inclusion index of one covariate. This proposal is symmetric in the sense that $h(L \mid L')/h(L' \mid L) = 1$, so the acceptance probability reduces to the (conditional) Bayes factor times the prior ratio (capped at 1). This proposal can lead to slow mixing in very high dimensions, where other proposals might be more attractive, but we leave this to future work.

If $g_L$ is random, we update it using an MH step with a lognormal proposal. 
The proposal scale is tuned adaptively, targeting an acceptance rate of $0.234$. 

Finally, we  update the parameter vector $\vec{\rho} = (\alpha, \vec{\tau}^\intercal, \vec{\beta}^\intercal)^\intercal$ conditional on $L$ and $g_L$. The conditionally conjugate priors lead to a known distribution for the conditional posterior, as described in (\ref{eq:cond_post_rho}) below in Subsection \ref{sec:CML}.


\subsection{Conditional Bayes factors and posteriors}\label{sec:CML}

We can obtain closed-form conditional posteriors and marginal likelihoods based on the prior specification described above. A detailed derivation is provided in supplementary Section \ref{sec:posteriors}. In the outcome model, we have the conditional posterior 
\begin{align} \label{eq:cond_post_rho}
    \vec{\rho}_L \mid \vec{\Lambda, \Sigma}, L, M, \vec{y}, \vec{X} &\sim N\left( \frac{g_L}{g_L + 1} (\vec{U}_L^\intercal \vec{U}_L)^{-1} \vec{U}_L^\intercal \vec{\Tilde{y}}, \; \sigma_{y|x} \frac{g_L}{g_L + 1} (\vec{U}_L^\intercal \vec{U}_L)^{-1} \right),
\end{align}
where $\vec{\Tilde{y}} = \vec{ y - H \Sigma_{xx}^{-1} \Sigma_{yx}}^\intercal$ is the endogeneity corrected outcome. The conditional Bayes factor (CBF) of model $L_i$ versus model $L_j$ is given by
\begin{align*}
    \text{CBF}(L_i, L_j) = \frac{p(\vec{y}  \mid \vec{X}, \vec{\Lambda}, \vec{\Sigma}, L_i, M)}{p(\vec{y}  \mid \vec{X}, \vec{\Lambda}, \vec{\Sigma}, L_j, M)} = (g_L + 1)^{(d_{U_j} - d_{U_i})/2} \exp\left( -\frac{1}{2 \sigma_{y|x}} \frac{g_L}{g_L + 1} \vec{\Tilde{y}}^\intercal (\vec{P_{U_j}} - \vec{P_{U_i}}) \vec{\Tilde{y}} \right),
\end{align*}
where $\vec{U_i}, d_{U_i}$ and $\vec{U_j}, d_{U_j}$ are the design matrices and their number of columns in models $L_i$ and $L_j$ respectively.

In the treatment model, we obtain  the (conditional) posterior 
\begin{align} \label{eq:cond_post_lambda}
\begin{aligned}
    \vec{\Lambda}_M \mid \vec{\rho, \Sigma}, L, M, \vec{y}, \vec{X} &\sim \\ MN\left( (\vec{V}_M^\intercal \vec{V}_M)^{-1} \vec{V}_M^\intercal \vec{\Tilde{X}} \left( \left(\vec{I_l} + g_M^{-1} \vec{B_\Sigma^{-1}} \right)^{-1} \right)^\intercal  \right. &, \left. \; (\vec{V}_M^\intercal \vec{V}_M)^{-1}, \; \left(\vec{B_\Sigma} + g_M^{-1} \vec{I_l} \right)^{-1} \vec{\Sigma_{xx}} \right),
\end{aligned}
\end{align}
where $\vec{B_\Sigma} = \vec{I_l} + \frac{1}{\sigma_{y|x}} \vec{\Sigma_{yx}}^\intercal \vec{\Sigma_{yx} \Sigma_{xx}^{-1}}$ and $\vec{\Tilde{X}} = \vec{X} - \frac{1}{\sigma_{y|x}} \vec{\epsilon \Sigma_{yx} }\left(\vec{B_\Sigma^{-1}}\right)^\intercal$.
The CBF of model $M_i$ versus model $M_j$ is
\begin{align}\label{eq:CBF_M}
    \text{CBF}(M_i, M_j) = | g_M \vec{B_\Sigma} + \vec{I_l}|^{(d_{V_j} - d_{V_i}) / 2} \exp \left( -\frac{1}{2} \text{tr} \left( \vec{A_\Sigma \Tilde{X}}^\intercal (\vec{P_{V_j}} - \vec{P_{V_i}}) \vec{\Tilde{X}} \right) \right),
\end{align}
where $\vec{A_\Sigma} = \left( \left( \vec{I_l} + g_M^{-1} \vec{B_\Sigma}^{-1} \right)^{-1} \right)^\intercal \vec{\Sigma_{xx}}^{-1} \vec{B_\Sigma}$ and $\vec{V}_i, d_{V_i}$ and $\vec{V}_j, d_{V_j}$ are the design matrices and their number of columns in models $M_i$ and $M_j$ respectively.

The treatment model uses the joint distribution of $\vec{y}$ and $\vec{X}$ for its marginal likelihood, while the outcome model uses the conditional distribution $\vec{y} \mid \vec{X}$. This distinction arises because the marginal distribution of $\vec{X}$ is independent of outcome parameters, but the outcome model $\vec{y} \mid \vec{X}$ depends on the treatment parameters. 

Using the IW prior the conditional posterior for the covariance matrix is 
\begin{align} \label{eq:cond_post_sigma}
    \vec{\Sigma} \mid \vec{\rho, \Lambda}, L, M, \vec{y, X} \sim IW(\nu + n, \vec{I_{l+1}} + \vec{[\epsilon : H]}^\intercal \vec{[\epsilon : H]}).
\end{align}
Conditional on the data and the outcome and treatment parameters, the residuals $\vec{\epsilon}$ and $\vec{H}$ are known, so the posterior scale matrix can be computed within the Gibbs steps. 

For the Cholesky-based prior, we can use the conditional posteriors for the three covariance components, 
\begin{align}
    \vec{a_{yx}}^\intercal \mid \sigma_{y \mid x}, \vec{\rho, \Lambda}, L, M, \vec{y, X} &\sim N\left( \left( \vec{H}^\intercal \vec{H} + \frac{\sigma_{y \mid x}}{\omega_a} \vec{I_l} \right)^{-1} \vec{H}^\intercal \vec{\epsilon}, \sigma_{y \mid x} \left( \vec{H}^\intercal \vec{H} + \frac{\sigma_{y \mid x}}{\omega_a} \vec{I_l} \right)^{-1} \right), \\
    \sigma_{y \mid x} \mid \vec{a_{yx}}, \vec{\rho, \Lambda}, L, M, \vec{y, X} &\sim IG \left( \frac{n + \nu}{2}, \frac{1}{2} \left( \left(\vec{\epsilon} - \vec{H} \vec{a_{yx}}^\intercal \right)^\intercal \left(\vec{\epsilon} - \vec{H} \vec{a_{yx}}^\intercal \right) + 1 \right) \right), \\
    \vec{\Sigma_{xx}} \mid \vec{\rho, \Lambda}, L, M, \vec{y, X} &\sim IW\left(n + \nu - 1, \vec{H}^\intercal \vec{H} + \vec{I_l} \right).
\end{align}
A draw of the covariance matrix $\Sigma$ is then implied by equation (\ref{eq:sigma_cholesky}).

If we put a hyperprior on $\nu$, we add an extra step to the Gibbs sampler updating $\nu$. Given $\vec{\Sigma}$, $\nu$  is independent of everything else, so the full conditional is proportional to the prior on $\vec{\Sigma}$ given $\nu$ times the prior on $\nu$.
We use an adaptive MH step, targeting an acceptance rate of $0.234$.

\subsection{The non-Gaussian case}

Here, we discuss the computational strategy to deal with the ULLGM models introduced in Subsection \ref{sec:nonGaussian}. We add an extra step to the Gibbs sampler that draws the latent Gaussian (and potentially additional parameters $r_y$ or $r_x$) and then perform posterior inference on the Gaussian parameters conditional on the latent Gaussian representation. \cite{steel_model_2024} propose to use an MH step with an adaptive Barker proposal \citep{livingstone_barker_2022} to sample the latent Gaussians. This is a good compromise between using gradient information to increase the mixing speed and maintaining robustness.

Define the residuals based on the latent Gaussian $\epsilon_i = q_i -  \left(\alpha + \vec{X_i \tau} + \vec{Z_i \beta} \right)$ and $\vec{H_i} = \vec{Q_i} - \vec{\Gamma}^\intercal - \vec{Z_i \Delta}$. For the outcome, the gradient used in the Barker proposal is 
\begin{align*}
    \frac{\partial \log p\left(q_i \mid \vec{\rho, \Lambda, \Sigma, y, X} \right)}{\partial q_i} = \frac{\partial \log p \left(y_i \mid q_i, r_y \right)}{\partial q_i} - \frac{\epsilon_i - \vec{H_i \Sigma_{xx}^{-1} \Sigma_{yx}}^\intercal }{\sigma_{y|x}},
\end{align*}
where the first term depends on the specific distribution of $y_i$, and the second term comes from the Gaussian ``prior''. For the $j$-th endogenous variable, the gradient is
\begin{align*}
    \frac{\partial \log p\left(Q_{ij} \mid \vec{\rho, \Lambda, \Sigma}, q_i, \vec{X} \right)}{\partial Q_{ij}} = \frac{\partial \log p \left(X_{ij} \mid Q_{ij}, r_{x_j} \right)}{\partial Q_{ij}} + \frac{\left[ \vec{\Sigma_{xx}^{-1} \Sigma_{yx}}^\intercal \right]_j}{\sigma_{y|x}} \left( \epsilon_i - \vec{H_i \Sigma_{xx}^{-1} \Sigma_{yx}}^\intercal \right) - [\vec{\Sigma_{xx}^{-1} Q_i}]_j.
\end{align*}
The first term depends on the distribution of the $j$-th endogenous variable, the second arises from the outcome distribution, and the last is the contribution of the Gaussian ``prior''.

If additional parameters $r_y$ or $r_{x_j}$ are required, we add an extra step to update them separately after the respective component of the latent Gaussian. This is done through an adaptive MH step, targeting an acceptance rate of $0.234$.

After updating the latent Gaussian representation and any additional parameters, we can proceed with the same steps as in the Gaussian case. The formulas from the latter remain valid using the latent Gaussian representation in place of any non-Gaussian variables.

\section{Model selection consistency of gIVBMA} \label{sec:consistency}

It is important to know that if the sample size $n$ increases, we end up putting more and more posterior mass on the ``correct'' model. We are assuming here that the model that actually generated the data lies within the model space considered.\footnote{This is what is often referred to as an $M$-closed setting. In most situations, model selection consistency naturally extends to the $M$-open framework by selecting the model that is closest to the true model according to a suitable metric \citep{Mukhopadhyay}. We leave a detailed treatment of this for future work. 
}

Let us assume that the data are generated from models $L_i$ and $M_i$ for, respectively, the outcome and treatment equations. Then we say that our gIVBMA procedure is model selection consistent if both CBF($L_i,L_j$) and CBF($M_i,M_j$) tend to $\infty$ with $n$ for any $L_j\ne L_i$ and $M_j\ne M_i$.  We show that the following results hold: 
\begin{theorem}\label{Th:consistency}
The procedure gIVBMA, detailed in Sections \ref{sec:sampling}
and \ref{sec:prior}, 
is model selection consistent in the Gaussian case if and only if the following conditions are satisfied:
\begin{itemize}
\item $\lim_{n\to \infty} g=\infty$ for fixed $g\in\{g_L,g_M\}$
\item if we assume a hyperprior $p(g_L)$, we have that  $$\lim_{n\to \infty}\int_{\Re_+} (1+g_L)^{1/2} p(g_L) dg_L=\infty$$
\item under a hyperprior $p(g_M)$, we have for $c\in\{l,2l,\dots,pl\}$
$$\lim_{n\to \infty}\int_{\Re_+} g_M^{c/2} p(g_M) dg_M=\infty$$
\end{itemize}


If (some of) the components in $(\vec{y},\vec{X})$ are assigned a non-Gaussian sampling distribution as in Section \ref{sec:nonGaussian}, with $(\vec{q},\vec{Q})$ the latent Gaussian counterparts, then model selection consistency is assured if, in addition to the conditions on $g$ above, we have independence between any additional parameters $r_y,\vec{r}_x$ and $(\vec{q},\vec{Q},L_i, M_j)$. 
\end{theorem}
\begin{proof}
    See supplementary Section \ref{sec:consistencyProof}.
\end{proof}

For both the fixed (BRIC) and random $g$ (hyper-$g/n$) options used throughout this paper, we can thus show that model selection consistency holds. This result provides additional justification for data-driven instrument selection. If the sample size is large enough, posterior mass concentrates on the ``correct'' models and, therefore, instruments and covariates will be correctly separated.

\section{Simulation experiments} \label{sec:simulation}

We evaluate the performance of gIVBMA in three different scenarios and compare it to naive BMA, IVBMA, and several classical methods. Throughout, we consider the median absolute error (MAE) and median bias of the point estimates of $\vec{\tau}$, (credible or confidence) interval coverage of $\vec{\tau}$, and the log predictive score (LPS) on a separate holdout dataset. For all Bayesian methods, we use the posterior mean as our point estimate for the MAE and bias calculations. While MAE, bias, and coverage relate specifically to the quality of the estimation of 
$\vec{\tau}$, LPS is a measure of predictive adequacy, focused on probabilistic predictions of the outcomes. In each scenario, we simulate $100$ independent datasets of size $n$, and for the LPS computation, we also generate holdout datasets of size $n/5$. Supplementary Section \ref{sec:performance} provides more details on the different measures used, while Section \ref{sec:competing} describes the different estimators and their implementation.

The predictive comparison is perhaps not entirely fair to some of the classical methods (in particular those that are two-stage in nature) as they reduce the bias in $\vec{\tau}$ but lack a correction term like gIVBMA. Consequently, when making predictions, these methods use an estimate of $\vec{\tau}$ instead of $\vec{\tau + \Sigma_{xx}^{-1} \Sigma_{yx}}^\intercal$, which would be the appropriate coefficient for the conditional model $\vec{y} \mid \vec{X}$. This can lead to poor predictions when $\vec{\Sigma_{xx}^{-1} \Sigma_{yx}}^\intercal$ becomes large. However, in practice, these methods might not be employed when prediction is the primary objective. Methods that do not correct for endogeneity (e.g., naive BMA) target the biased coefficient $\vec{\tau + \Sigma_{xx}^{-1} \Sigma_{yx}}^\intercal$, but provide good predictions as this is the right coefficient for predicting $\vec{y}$ conditional on $\vec{X}$. This discrepancy reflects the fact that predicting observed and counterfactual outcomes are different problems under endogeneity. Our probabilistic method allows us to provide endogeneity-corrected inference on $\vec{\tau}$ while maintaining good predictive power in the observational model.

\subsection{Invalid instruments} \label{sec:sim-invalid}

To investigate the performance in the presence of invalid instruments, we consider a simulation setup similar to the ones in \cite{kang_instrumental_2016} and \cite{windmeijer_use_2019}. We consider two different sample sizes $n \in \{50, 500\}$ and $p = 10$ potential instruments simulated from independent standard Normals. Their coefficients in the outcome model are set to $\vec{\beta} = (1, \ldots, 1, 0, \ldots, 0)$ where the first $s$ coefficients are one (and, therefore, the first $s$ instruments are invalid). We vary $s \in \{3, 6\}$ such that the plurality rule only holds in the first scenario. The instruments' treatment coefficient is set to $\vec{\delta} = c \vec{\iota}$, where $c>0$ is chosen such that the first-stage $R^2_f$ is approximately $0.2$. The treatment effect is set to $\tau = 0.1$. Then the data is generated from (\ref{eq:StructMod}), while $\sigma_{yy}=\sigma_{xx}=1$ and $\sigma_{yx}=1/2$.

To compare the performance of our different prior specifications, we evaluate several variants of our method using BRIC or hyper-$g/n$ priors on the regression coefficients, and either the full inverse-Wishart or the Cholesky-based covariance prior, the latter with different hyperparameters $\omega_a \in \{0.1, 1, 10\}$. We benchmark our approach against IVBMA \citep{karl_instrumental_2012}, TSLS with all instruments, oracle TSLS, and the sisVIVE estimator \citep{kang_instrumental_2016}. We also use a Bayesian IV method that includes all instruments in both models with horseshoe priors \citep{carvalho2009handling} on their regression coefficients. See supplementary Section~\ref{sec:competing} for more details.

Table~\ref{tab:Kang_Sim_Instruments} reports the instrument selection performance of the different gIVBMA variants. The true number of valid and relevant instruments implied by the data-generating process is \(N_Z = p - s\). For \(n = 50\), all variants distribute their posterior mass between \(N_Z = 0\) (exceeding the prior probability) and \(0 < N_Z < p - s\), placing only negligible mass on the true value (or above). When \(s = 3\), the mass is roughly evenly split, whereas for \(s = 6\), the models assign more weight to non-identified specifications. 
In contrast, for \(n = 500\), all variants allocate only negligible posterior mass to non-identified models. Overall, the hyper-\(g/n\) specifications yield consistently better performance, while 
smaller values of $\omega_a$ seem to do better for $n=500$. 

\begin{table}
\small
\centering
\begin{tabular}{l*{8}{r}}
\toprule
 & \multicolumn{8}{c}{$n = 50$} \\
 & \multicolumn{4}{c}{$s = 3$} & \multicolumn{4}{c}{$s = 6$} \\
\cmidrule(lr){2-5}\cmidrule(lr){6-9}
 & $0$ & $(0, p-s) $ & $p-s$ & $ (p-s, p] $ & $0$ & $(0, p-s) $ & $p-s$ & $ (p-s, p] $  \\
 \midrule
 Prior probabilities
& 0.275 & 0.634 & 0.039 & 0.053
& 0.275 & 0.43 & 0.085 & 0.211 \\
\midrule
gIVBMA (BRIC, IW) & 0.573 & 0.427 & 0.0 & 0.0 & 0.807 & 0.183 & 0.005 & 0.004 \\
gIVBMA (h-$g/n$, IW) & 0.488 & 0.508 & 0.003 & 0.0 & 0.732 & 0.235 & 0.017 & 0.016 \\
gIVBMA (BRIC, $\omega_a = 0.1$) & 0.603 & 0.397 & 0.0 & 0.0 & 0.815 & 0.184 & 0.001 & 0.0 \\
gIVBMA (h-$g/n$, $\omega_a = 0.1$) & 0.488 & 0.507 & 0.005 & 0.0 & 0.706 & 0.285 & 0.008 & 0.0 \\
gIVBMA (BRIC, $\omega_a = 1$) & 0.595 & 0.405 & 0.0 & 0.0 & 0.779 & 0.195 & 0.007 & 0.019 \\
gIVBMA (h-$g/n$, $\omega_a = 1$) & 0.51 & 0.487 & 0.003 & 0.0 & 0.705 & 0.224 & 0.017 & 0.054 \\
gIVBMA (BRIC, $\omega_a = 10$) & 0.58 & 0.42 & 0.0 & 0.0 & 0.735 & 0.236 & 0.022 & 0.008 \\
gIVBMA (h-$g/n$, $\omega_a = 10$) & 0.494 & 0.503 & 0.003 & 0.0 & 0.651 & 0.19 & 0.036 & 0.123 \\
\midrule
 & \multicolumn{8}{c}{$n = 500$} \\
 & \multicolumn{4}{c}{$s = 3$} & \multicolumn{4}{c}{$s = 6$} \\
\cmidrule(lr){2-5}\cmidrule(lr){6-9}
 & $0$ & $(0, p-s) $ & $p-s$ & $ (p-s, p] $ & $0$ & $(0, p-s) $ & $p-s$ & $ (p-s, p] $  \\
\midrule
gIVBMA (BRIC, IW) & 0.053 & 0.495 & 0.453 & 0.0 & 0.159 & 0.537 & 0.304 & 0.0 \\
gIVBMA (h-$g/n$, IW) & 0.05 & 0.417 & 0.533 & 0.0 & 0.158 & 0.468 & 0.355 & 0.019 \\
gIVBMA (BRIC, $\omega_a = 0.1$) & 0.001 & 0.529 & 0.471 & 0.0 & 0.038 & 0.606 & 0.355 & 0.0 \\
gIVBMA (h-$g/n$, $\omega_a = 0.1$) & 0.001 & 0.419 & 0.58 & 0.0 & 0.034 & 0.542 & 0.424 & 0.0 \\
gIVBMA (BRIC, $\omega_a = 1$) & 0.05 & 0.498 & 0.452 & 0.0 & 0.205 & 0.507 & 0.288 & 0.0 \\
gIVBMA (h-$g/n$, $\omega_a = 1$) & 0.065 & 0.397 & 0.538 & 0.0 & 0.168 & 0.446 & 0.352 & 0.034 \\
gIVBMA (BRIC, $\omega_a = 10$) & 0.083 & 0.474 & 0.443 & 0.0 & 0.278 & 0.463 & 0.257 & 0.003 \\
gIVBMA (h-$g/n$, $\omega_a = 10$) & 0.098 & 0.393 & 0.509 & 0.0 & 0.227 & 0.401 & 0.271 & 0.101 \\
\bottomrule
\end{tabular}
\caption{\textbf{Invalid instruments:} Mean posterior probabilities of the number of valid and relevant instruments $N_Z$ for different variants of gIVBMA across $100$ simulated datasets with $n$ observations and $s$ invalid instruments. The true number of valid and relevant instruments is $N_Z = p-s$.}
\label{tab:Kang_Sim_Instruments}
\end{table}

Table \ref{tab:Kang_Sim} shows that all gIVBMA variants predict well and lead to small bias and MAE. For small $n$, the BRIC specifications  mostly have a higher MAE and median bias than the hyper-$g/n$ variants. The IVBMA method also predicts well and has low MAE and median bias, but overcovers for small $n$ and significantly undercovers for large $n$. TSLS results in a very large positive bias, very small coverage, and poor prediction. The oracle version of TSLS does better but is slightly worse than the gIVBMA methods for small $n$. All the methods above appear largely unaffected by the plurality rule. The latter does affect sisVIVE, however, which performs very poorly when $s=6$, as expected. The Bayesian horseshoe (BayesHS) method incurs substantial bias and does not cover well at all.

\begin{table}
\centering
\begin{tabular}{l*{8}{r}}
\toprule
 & \multicolumn{8}{c}{$n = 50$} \\
 & \multicolumn{4}{c}{$s = 3$} & \multicolumn{4}{c}{$s = 6$} \\
\cmidrule(lr){2-5}\cmidrule(lr){6-9}
 & \textbf{MAE} & \textbf{Bias} & \textbf{Cov.} & \textbf{LPS} & \textbf{MAE} & \textbf{Bias} & \textbf{Cov.} & \textbf{LPS} \\
\midrule
BMA (hyper-g/n) & 0.47 & 0.47 & 0.1 & 1.39 & 0.47 & 0.47 & 0.08 & \textbf{1.41} \\
gIVBMA (BRIC, IW) & 0.3 & 0.29 & \textbf{0.93} & 1.39 & 0.24 & 0.22 & 0.99 & \textbf{1.41} \\
gIVBMA (hyper-$g/n$, IW) & 0.14 & 0.06 & 0.9 & 1.4 & 0.17 & \textbf{0.01} & 0.99 & \textbf{1.41} \\
gIVBMA (BRIC, $\omega_a = 0.1$) & 0.42 & 0.42 & 0.85 & 1.38 & 0.42 & 0.42 & \textbf{0.95} & \textbf{1.41} \\
gIVBMA (hyper-$g/n$, $\omega_a = 0.1$) & 0.33 & 0.33 & 0.92 & 1.39 & 0.33 & 0.33 & 0.99 & \textbf{1.41} \\
gIVBMA (BRIC, $\omega_a = 1$) & 0.24 & 0.23 & 0.9 & 1.39 & 0.26 & 0.2 & 0.97 & \textbf{1.41} \\
gIVBMA (hyper-$g/n$, $\omega_a = 1$) & 0.18 & \textbf{0.01} & 0.91 & 1.4 & 0.24 & 0.09 & \textbf{0.95} & \textbf{1.41} \\
gIVBMA (BRIC, $\omega_a = 10$) & 0.47 & 0.02 & 0.83 & 1.39 & 0.59 & 0.03 & 0.92 & \textbf{1.41} \\
gIVBMA (hyper-$g/n$, $\omega_a = 10$) & 0.33 & 0.21 & 0.86 & 1.4 & 0.62 & 0.28 & 0.81 & \textbf{1.41} \\
BayesHS & 0.37 & 0.37 & 0.38 & 1.78 & 0.49 & 0.49 & 0.37 & 1.83 \\
TSLS & 1.25 & 1.25 & 0.18 & 2.17 & 2.43 & 2.43 & 0.09 & 2.41 \\
O-TSLS & 0.21 & 0.2 & 0.79 & 1.43 & 0.3 & 0.27 & 0.89 & 1.54 \\
IVBMA & \textbf{0.08} & 0.04 & 0.99 & \textbf{1.37} & \textbf{0.13} & 0.12 & 1.0 & 1.46 \\
sisVIVE & 0.6 & 0.6 & - & 1.64 & 2.2 & 2.2 & - & 2.37 \\
\midrule
 & \multicolumn{8}{c}{$n = 500$} \\
 & \multicolumn{4}{c}{$s = 3$} & \multicolumn{4}{c}{$s = 6$} \\
\cmidrule(lr){2-5}\cmidrule(lr){6-9}
 & \textbf{MAE} & \textbf{Bias} & \textbf{Cov.} & \textbf{LPS} & \textbf{MAE} & \textbf{Bias} & \textbf{Cov.} & \textbf{LPS} \\
\midrule
BMA (hyper-g/n) & 0.46 & 0.46 & 0.0 & 1.3 & 0.48 & 0.48 & 0.0 & \textbf{1.28} \\
gIVBMA (BRIC, IW) & 0.1 & 0.06 & 0.98 & 1.29 & 0.21 & 0.2 & 0.97 & \textbf{1.28} \\
gIVBMA (hyper-$g/n$, IW) & 0.08 & 0.06 & 0.94 & 1.29 & 0.21 & 0.18 & \textbf{0.94} & \textbf{1.28} \\
gIVBMA (BRIC, $\omega_a = 0.1$) & 0.1 & 0.09 & 0.93 & 1.29 & 0.16 & 0.16 & 0.89 & \textbf{1.28} \\
gIVBMA (hyper-$g/n$, $\omega_a = 0.1$) & 0.09 & 0.08 & 0.92 & 1.29 & 0.17 & 0.17 & 0.89 & \textbf{1.28} \\
gIVBMA (BRIC, $\omega_a = 1$) & 0.08 & 0.06 & 0.93 & 1.29 & 0.2 & 0.18 & 0.96 & \textbf{1.28} \\
gIVBMA (hyper-$g/n$, $\omega_a = 1$) & 0.1 & 0.07 & \textbf{0.95} & 1.29 & 0.19 & 0.18 & 0.93 & \textbf{1.28} \\
gIVBMA (BRIC, $\omega_a = 10$) & 0.1 & 0.07 & 0.97 & 1.29 & 0.32 & 0.31 & 0.89 & \textbf{1.28} \\
gIVBMA (hyper-$g/n$, $\omega_a = 10$) & 0.11 & 0.08 & 0.93 & 1.29 & 0.52 & 0.52 & 0.7 & \textbf{1.28} \\
BayesHS & 0.35 & 0.35 & 0.04 & 1.34 & 0.51 & 0.51 & 0.02 & 1.3 \\
TSLS & 1.82 & 1.82 & 0.0 & 2.2 & 3.63 & 3.63 & 0.0 & 2.7 \\
O-TSLS & \textbf{0.07} & \textbf{0.04} & 0.92 & 1.42 & \textbf{0.08} & \textbf{0.05} & 0.96 & 1.4 \\
IVBMA & 0.09 & 0.08 & 0.78 & \textbf{1.28} & 0.09 & 0.06 & 0.74 & 1.3 \\
sisVIVE & 0.21 & 0.21 & - & 1.61 & 4.36 & 4.36 & - & 3.07 \\
\bottomrule
\end{tabular}
\caption{\textbf{Invalid instruments:} MAE, median bias, coverage, and mean LPS on 100 simulated datasets with $s$ invalid instruments. The best values in each column are printed in bold. The sisVIVE estimator does not provide any uncertainty quantification, so we do not report any coverage results.}
\label{tab:Kang_Sim}
\end{table}

In the subsequent simulations, we shall focus on the hyper-$g/n$ prior on the regression coefficients, using either the full inverse-Wishart covariance prior or the Cholesky-based prior with $\omega_a = 0.1$, as these performed best in this experiment. 
The Cholesky prior with $ \omega_a = 1$ also performed well, but behaved similarly to the full inverse-Wishart prior. This is not surprising, since $\sigma_{y \mid x}$ is roughly of the order 1 in the IW prior (see Supplementary Figure \ref{fig:implied_prior_sigma}).

\subsection{Multiple endogenous variables with correlated instruments} \label{sec:mult_end_sim}

We consider an example with two endogenous variables, a Gaussian and a Beta, and correlated instruments. We generate 15 valid instruments (five of which are also relevant) with a correlation structure similar to \cite{fernandez_benchmark_2001}. The data-generating process is given in detail in supplementary Section \ref{sec:sim_mult}. We vary the sample size $n \in \{50, 500\}$. 

We compare gIVBMA against naive BMA, IVBMA, OLS, TSLS (using all instruments and an oracle version), and MATSLS. The latter refers to the model-averaged TSLS estimator  with unrestricted weights by \cite{kuersteiner_constructing_2010} and is included as it is the only feasible  classical method that is competitive with oracle TSLS in all scenarios 
of Subsection \ref{sec:SimManyWeak}. 
IVBMA cannot directly model the Beta-distributed endogenous variable and instead relies on a Gaussian approximation.

Supplementary Table \ref{tab:SimResMultEndo} presents the full estimation error, coverage, and LPS results. Both retained gIVBMA variants perform best in terms of MAE and median bias, for both sample sizes. BMA is  severely biased as expected, but coverage for the second endogenous variable is adequate (in contrast with the coverage for $\vec{X}_1$). This is a consequence of the non-linear transformation from the latent Gaussian to the Beta variable, diluting the endogeneity. TSLS, oracle TSLS, and MATSLS all perform similarly well. IVBMA results in a substantial bias for both sample sizes, yet it predicts and covers relatively well.  

\begin{figure}[h!]
    \centering
    \includegraphics[width=0.8\linewidth]{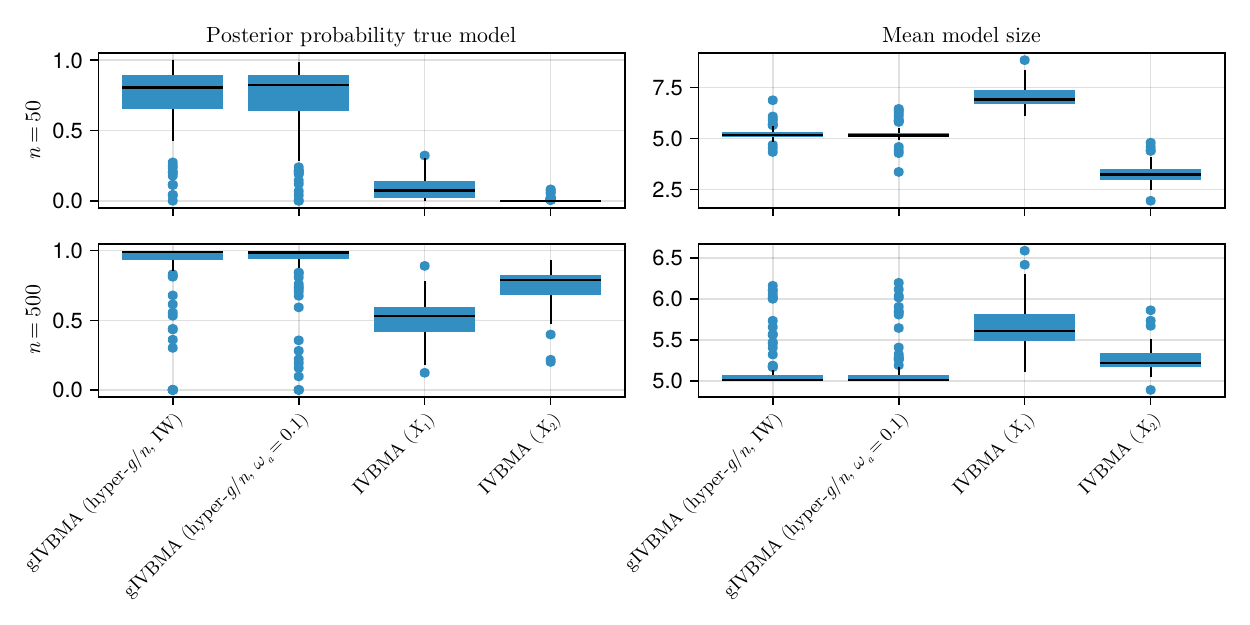}
    \caption{ {\bf Multiple endogenous variables with correlated instruments}: Posterior probabilities of the true treatment model and mean model sizes for $100$ simulated datasets of size $n = 50$ (top) and $n=500$ (bottom). The true treatment model size is 5. 
    IVBMA uses separate treatment models for the two endogenous variables $X_1$ and $X_2$.}
    \label{fig:MultEndSim_model_selection}
\end{figure}

Figure \ref{fig:MultEndSim_model_selection} presents the distributions of the posterior probability of the true treatment model and the mean treatment model size. Unlike gIVBMA, IVBMA allows for  separate treatment models for the two endogenous variables (see Subsection \ref{sec:modeluncertainty}), and we report the results separately for $X_1$ and $X_2$. Both gIVBMA variants put substantial posterior mass on the true model, and their mean model size is accordingly very close to that of the true model (which is $5$). 
For $n=50$, IVBMA tends to select too many variables for $X_1$ and too few for $X_2$. The performance of IVBMA is better in the $n=500$ setting, but it still cannot compete with gIVBMA. Supplementary Table \ref{tab:SimMultEnd_PIPs} presents median posterior inclusion probabilities (PIP) for all instruments. The selection performance of gIVBMA is clearly superior for both sample sizes.

\subsection{Many weak instruments} \label{sec:SimManyWeak}

We also consider a setting with many weak instruments, similar to the simulation study in \cite{kuersteiner_constructing_2010}, and a variant with a Poisson endogenous variable.  In this case, the instruments are assumed
to be valid a priori and cannot be included in the outcome model, using the gIVBMA version described in Section \ref{sec:fixed_Z}. In the Gaussian setting, the gIVBMA variants deliver good coverage and predictions, but are worse than IVBMA in terms of MAE. By contrast, in the Poisson setting, gIVBMA is clearly superior in MAE, and IVBMA has very poor coverage. The horseshoe prior yields lower MAE than gIVBMA in most scenarios, but tends to overcover and predicts poorly. In addition, gIVBMA (like all the other methods with exact zero restrictions) formally deals with identification and has an easier interpretation. 
MATSLS is competitive for large $n$ but predicts substantially worse than gIVBMA throughout. Full details are provided in supplementary Section \ref{sec:many_weak}.

\section{Empirical Examples}\label{sec:empirical}

\subsection{Geography or institutions?}
\label{sec:growth}

\cite{carstensen_primacy_2006} show that geographic factors such as disease ecology, particularly malaria prevalence, have a substantial negative impact on income. \cite{ditraglia_using_2016} reanalyzes these data to illustrate the usefulness of their proposed Focused Moment Selection Criterion. They consider the regression model
$    \log \text{gdpc}_i = \beta_1 + \beta_2  \text{rule}_i + \beta_3  \text{malfal}_i + \epsilon_i$, 
where $\text{gdpc}$ is real GDP per capita in 1995 prices, 
$\text{rule}$ is an average governance indicator measuring the quality of institutions, and $\text{malfal}$ is the fraction of the population at risk of malaria transmission in 1994. The quality of institutions and the prevalence of malaria are likely endogenous. The dataset (for 44 countries) contains various potential instruments such as historical settler mortality (lnmort), malaria transmission stability (maleco), winter frost levels (frost), the maximum temperature during peak humidity (humid), latitude (distance from equator), the proportion of Western Europeans and English speakers (eurfrac and engfrac), proximity to coast (coast), and predicted trade share (trade). Settler mortality (lnmort) and malaria ecology (maleco) are very plausible to be exogenous, 
but there is uncertainty about the exogeneity of all other potential instruments. 

We will consider both institutional quality and malaria prevalence as joint endogenous variables (i.e. $l = 2$) and compare the PIP of all instruments in the outcome and treatment model to assess their validity. The rule variable is approximately Gaussian. The malfal variable is a proportion and only takes values in $[0, 1]$, so we model it using a Beta distribution. We put an Exponential prior with rate $1$ on the additional dispersion parameter of the Beta distribution and draw it in an MH step with a Gaussian proposal. One challenge is that several countries in our sample have recorded malfal values of exactly 0 or 1. This is a problem for our algorithm as the logistic function (e.g., used in computing the gradient for the proposal) is not finite at 0 or 1. We deal with this incompatibility of our observations with the sampling model by using set observations \citep{fernandez_multivariate_1999}. We treat zero observations (and analogously ones) as belonging to a set $(0, 0.0005)$ and add an extra step to draw the value from its sampling distribution truncated to that set. This matches these observations to the dominating measure (Lebesgue) of the sampling model. 

\begin{table}[h]
\centering
\begin{tabular}{lcccccc}
\toprule
& \multicolumn{2}{c}{\textbf{gIVBMA (IW)}} & \multicolumn{2}{c}{\textbf{gIVBMA ($\omega_a = 0.1$)}} & \multicolumn{2}{c}{\textbf{BMA (hyper-g/n)}}\\
\cmidrule(lr){2-3} \cmidrule(lr){4-5} \cmidrule(lr){6-7}
& Mean & 95\% CI & Mean & 95\% CI & Mean & 95\% CI \\
\midrule
 rule & 0.85 &  [0.6, 1.11]  & 0.86 & [0.61, 1.11] & 0.79 & [0.45, 1.12] \\
 malfal & -1.05 &  [-1.5, -0.6]  & -1.03 & [-1.46, -0.59] & -1.06 & [-1.67, -0.43] \\
 $\sigma_{12}$ & -0.03 &  [-0.09, 0.03]  & -0.03 & [-0.09, 0.03] & - & -\\
 $\sigma_{13}$ & 0.02 &  [-0.11, 0.17]  & 0.01 & [-0.12, 0.13] & - & -\\
\midrule
& PIP L & PIP M & PIP L & PIP M & PIP L & PIP M \\
\midrule
maleco & 0.0 & 1.0 & 0.0 & 0.99 &  0.02 & - \\
lnmort & 0.01 & 1.0 & 0.0 & 1.0 &  0.02 & - \\
frost & 0.01 & 0.14 & 0.0 & 0.16 &  0.02 & - \\
humid & 0.01 & 0.99 & 0.0 & 0.99 &  0.04 & - \\
latitude & 0.0 & 0.24 & 0.0 & 0.21 &  0.02 & - \\
eurfrac & 0.01 & 0.6 & 0.0 & 0.56 &  0.03 & - \\
engfrac & 0.0 & 0.19 & 0.01 & 0.17 &  0.02 & - \\
coast & 0.0 & 0.99 & 0.01 & 1.0 &  0.03 & - \\
trade & 0.0 & 0.38 & 0.0 & 0.42 &  0.03 & - \\
\bottomrule
\end{tabular}
\caption{\textbf{Geography or institutions?} Treatment effect estimates (posterior mean and 95\% credible interval) and posterior inclusion probabilities (PIP) in outcome (L) and treatment (M) models for rule and malfal as endogenous variables. The algorithm was run for 10,000 iterations (the first 2,000 of which were discarded as burn-in).}
\label{tab:CG_results}
\end{table}

Table \ref{tab:CG_results} shows the results. All outcome PIPs are very low, while four of the instruments are included in the treatment model with high probability, indicating strong support for overidentification.  
In particular, maleco and lnmort are valid and relevant instruments, as expected, while the humidity and coast variables are valid and highly relevant as well. The hyper-$g/n$ prior tends to yield a larger $g$ in the outcome model and a smaller $g$ in the treatment model than the fixed $g$ corresponding to BRIC, leading to lower and higher PIP, respectively. Our treatment effect estimates are in line with those in \cite{ditraglia_using_2016}. Their point estimates for rule and malfal are between $0.81$ and $0.97$ and between $-1.16$ and $-0.9$, respectively, depending on which instrument set they use. The posterior distributions of both components of the covariance vector, 
$\sigma_{12}$ and $\sigma_{13}$,  concentrate near zero. Thus, the gIVBMA results are similar to naive BMA, though gIVBMA credible intervals are slightly narrower, reflecting the borrowing of strength from the treatment equation.

To assess the predictive performance on this dataset, we compute the LPS using leave-one-out cross-validation. For each observation, the model is trained excluding that observation, and the LPS is then calculated for the excluded data point. We compare the performance of gIVBMA, IVBMA,  
BMA, and TSLS. Although sisVIVE would be relevant given potential concerns about instrument validity (at least a priori), we exclude it as this approach 
does not accommodate multiple endogenous variables. Methods designed for many weak instruments were also omitted since their predictive performance was consistently inferior to TSLS in our simulations. Table \ref{tab:LOOCV_LPS} presents the mean LPS values. The gIVBMA variants perform best, closely followed by naive BMA and TSLS, while IVBMA predicts considerably worse.

\begin{table}[h]
\centering
\begin{tabular}{lc}
\toprule
Method & Mean LPS \\
\midrule
gIVBMA (IW) & 0.656 \\
gIVBMA ($\omega_a = 0.1$) & \textbf{0.654} \\
IVBMA & 0.696 \\
BMA (hyper-g/n) & 0.658 \\
TSLS & 0.666 \\
\bottomrule
\end{tabular}
\caption{{\bf Geography or institutions?} The mean LPS calculated across all iterations of leave-one-out cross-validation on the \cite{carstensen_primacy_2006} dataset, where each iteration uses a single observation as the holdout set and the rest as the training set.}
\label{tab:LOOCV_LPS}
\end{table}

\subsection{Returns to schooling}

A prominent problem in microeconomics is estimating the causal relationship between educational attainment and earnings. Education levels are not randomly assigned but are likely influenced by unobservable characteristics that simultaneously affect earning potential. Therefore, the observed correlation between education and wages may not accurately reflect the true economic returns to schooling. Many econometricians have tried to isolate the causal effect by finding instrumental variables that provide plausibly exogenous variation in educational attainment, such as quarter of birth \citep{angrist_does_1991} or geographic variation in college proximity \citep{card1995collegeproximity}.

We revisit this problem using the dataset from \cite{card1995collegeproximity}, a subset of the National Longitudinal Survey of Young Men, containing all men who were interviewed in 1976 and provided valid wage and education responses. Our outcome is the logarithm of hourly wages 
with years of schooling as the single endogenous explanatory variable of interest. 
Our set of potential exogenous covariates and instruments includes experience, experience squared, college proximity (distance to a 2 or 4-year college),  
variables on family background, marital status, race, and regional indicators. The dataset also includes information on parents' educational attainment, which have a substantial proportion of missing values. As in \cite{card1995collegeproximity}, we impute those missing values using their mean and include indicators for missingness. After imputing missing values, the data consists of $3,003$ observations and $23$ potential instruments and exogenous controls (see supplementary Table \ref{tab:card_def}). We use fully data-driven instrument selection such that all variables can be used as instruments or covariates.

\begin{figure}[h!]
    \centering
    \includegraphics[width=0.8\linewidth]{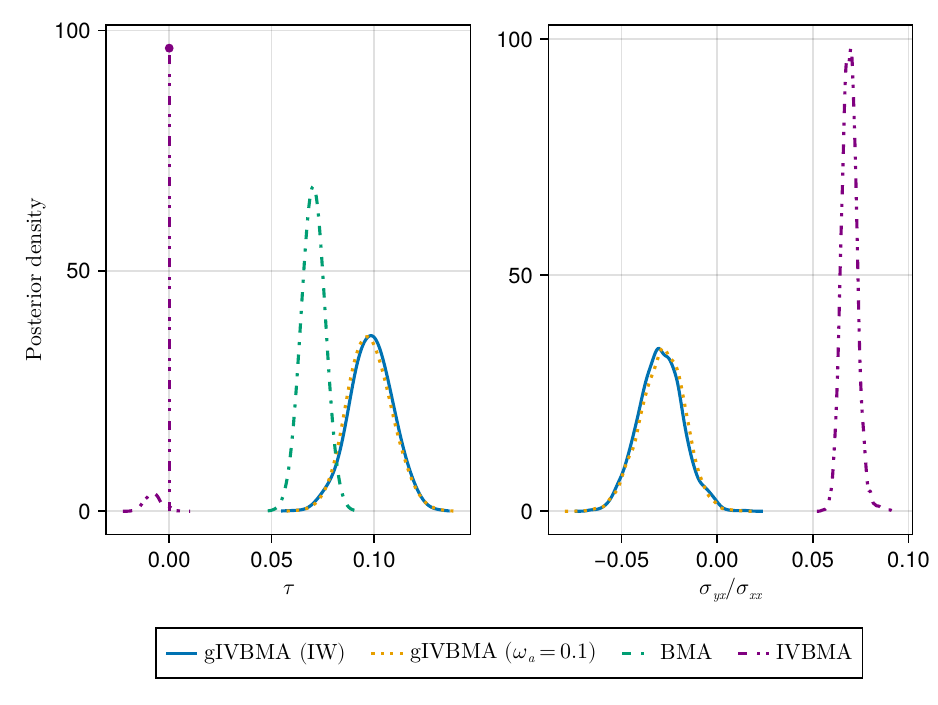}
    \caption{{\bf {Returns to schooling}}: Posterior distributions of the treatment effect of education and the covariance ratio $\sigma_{yx} / \sigma_{xx}$ based on the imputed dataset ($n = 3,003$). The algorithm was run for $5,000$ iterations, the first $500$ of which were discarded as burn-in. The vertical line for IVBMA indicates a point mass at zero of $0.962$.
    }
\label{fig:posterior_schooling}
\end{figure}

Figure \ref{fig:posterior_schooling} presents our posterior results.\footnote{Fitting each specification requires less than an hour on a standard 12th Gen Intel i7 CPU. Thus, computational costs are not a limiting factor for this approach, making gIVBMA practical for datasets with several thousand observations.} Consistent with \cite{card1995collegeproximity}, the gIVBMA posterior for the treatment effect $\tau$ indicates higher returns to schooling than BMA models that lack an endogeneity correction. For comparison, the TSLS estimate when using only proximity to a four-year college as an instrument and all other variables (except proximity to a two-year college) as exogenous controls is $\hat{\tau} = 0.142$ ($95\%$ CI: $[0.028, 0.256]$). Thus, our posterior results support slightly lower returns to schooling with higher estimation precision, a discrepancy likely driven in part by differences in instrument selection. While our algorithm frequently selects proximity to a four-year college as an instrument, it always includes parental education as well. In contrast, IVBMA yields implausible results by concentrating most of its posterior mass exactly at zero. This might occur because IVBMA fails to select experience in the outcome model and uses it as an instrument instead (see supplementary Table \ref{tab:card_pip_1}). IVBMA tends to have a higher number of instruments as it selects fewer variables in the outcome model (see supplementary Figure~\ref{fig:card_instruments}). For all methods, the posterior probability of non-identification (no instruments) is zero.

As a robustness check, we re-estimated the model using only observations with complete parental education data ($n=2,215$; see supplementary Section \ref{sec:details_schooling}). The results remain qualitatively similar to the main specification. The primary difference is that college proximity is selected as an instrument less frequently, while parental education remains a constant choice. Notably, the IVBMA results appear more plausible in this subsample, as the model shifts toward using parental education as its primary instrument (see supplementary Table \ref{tab:card_pip_2}).

Supplementary Table \ref{tab:schooling_5_fold_LPS} shows that the predictive  performance of gIVBMA, IVBMA and naive BMA is relatively similar on both versions of the dataset, while TSLS does worse.

\section{Conclusion}\label{sec:conclusion}

We have introduced the gIVBMA method, which performs posterior inference in structural equation models with at least one endogenous regressor, averaging over all possible sets of potential instruments and covariates. We allow for the model to freely choose its instruments based on the data, which provides additional robustness against invalid instruments. Our computational strategy relies on updating the outcome model, the treatment model, and the covariance matrix separately using conditional Bayes factors in the model updates. Exploiting the ULLGM framework \citep{steel_model_2024}, we can accommodate non-Gaussian outcomes and endogenous variables. We provide simple and unrestrictive necessary and sufficient conditions for model selection consistency of gIVBMA. In both simulation and real data experiments, gIVBMA outperforms the earlier method proposed by \cite{karl_instrumental_2012} and many classical estimators considered in the literature.

Future work could extend the framework to more complex covariance structures, 
possibly rendering gIVBMA available for settings such as time series or network data. 

\section*{Supplementary Material}

All derivations and proofs, along with further details on the variant with fixed instruments, prior specification, simulation experiments, and empirical examples, are provided in the Supplementary Material. The code and data used are available at \texttt{https://github.com/gregorsteiner/gIVBMA-Code}.

\bibliographystyle{apalike}
\bibliography{references}

\appendix

\renewcommand{\thesection}{S.\arabic{section}}

\renewcommand{\thefigure}{S.\arabic{figure}}
\renewcommand{\thetable}{S.\arabic{table}}
\setcounter{figure}{0}
\setcounter{table}{0}

\section{Notation}
\label{sec:notation}

Let $\vec{\iota} = (1, \ldots, 1)^\intercal$ denote a vector of ones of length $n$, while $\vec{I_n}$ is the $n \times n$ identity matrix. For any vector $\vec{v}$, let $v_i$ denote the $i$-th component of $\vec{v}$. For any matrix $\vec{V}$, let $\vec{V_i}$ be the $i$-th row of $\vec{V}$ and $V_{ij}$ the element in row $i$ and column $j$. For an $n \times k$ matrix $\vec{R}$ of full column rank, let $\vec{P_R} = \vec{R} (\vec{R}^\intercal \vec{R})^{-1} \vec{R}^\intercal$ be the $n \times n$ projection matrix and $\vec{M_R} = \vec{I_n} - \vec{P_R}$ be the projection onto the orthogonal space. For matrices $\vec{S}$ and $\vec{T}$ with the same number of rows, $\vec{R = [S : T]}$ is the matrix that results from horizontally concatenating $\vec{S}$ and $\vec{T}$. We use $vec(\vec{R})$ to denote the column vector resulting from stacking the columns of $\vec{R}$, while $\vec{S} \otimes \vec{T}$ denotes the Kronecker product of $\vec{S}$ and $\vec{T}$ for any matrices $\vec{S}$ and $\vec{T}$. 

Let $p(\vec{x})$ denote the probability density function (pdf) or probability mass function (pmf) of the random variable $\vec{x}$ and denote by $\mathcal{C}^q$ the set of positive definite symmetric (PDS) matrices of dimension $q\times q$. 
We use
$
    \vec{x} \sim N(\vec{\mu}, \vec{\Sigma})
$
to indicate that the random vector $\vec{x}\in \Re^k$ follows a Gaussian distribution with mean vector $\vec{\mu}\in \Re^k$ and  covariance matrix $\vec{\Sigma}\in \mathcal{C}^k$. We say an $n\times k$  matrix $\vec{R}$ follows a matrix normal distribution,
$
    \vec{R} \sim MN(\vec{M}, \vec{S}, \vec{T})
$
if and only if
$
    vec(\vec{R}) \sim N(vec(\vec{M}), \vec{T} \otimes \vec{S}), 
$
where $\vec{M}\in \Re^{n\times k}$ with  $\vec{S}\in \mathcal{C}^n$ and $\vec{T}\in \mathcal{C}^k$. 
We use $x \sim IG(a, b)$ to denote that the random variable $x > 0$ follows an inverse Gamma distribution with pdf
 \[
p(x \mid a,b) = \frac{b^{a}}{\Gamma(a)} x^{-(a+1)} \exp\left(-\frac{b}{x}\right)
\]
with shape parameter $a > 0$ and scale parameter $b > 0$. For a random matrix $\vec{X} \in \mathcal{C}^k$, we let $\vec{X} \sim IW(\nu, \vec{\Psi})$ denote that $\vec{X}$ follows an inverse Wishart distribution with pdf
\[
p(\vec{X} \mid \nu, \vec{\Psi}) = \frac{|\vec{\Psi}|^{\nu/2}}{2^{\nu k/2} \pi^{k(k-1)/4} \prod_{j=1}^k \Gamma\!\left(\frac{\nu + 1-j}{2} \right)} |\vec{X}|^{-(\nu + k + 1)/2} \exp\left( -\frac{1}{2} \mathrm{tr}\left( \vec{\Psi} \vec{X}^{-1} \right) \right)
\]
with degrees-of-freedom parameter $\nu > k-1$ and   scale matrix $\vec{\Psi} \in \mathcal{C}^k$.






\section{Fixed instruments} \label{sec:fixed_Z}

We also consider a fixed-instrument version of the model given by
\begin{align*}
    \vec{y} &= \alpha \vec{\iota} + \vec{X} \vec{\tau} + \vec{W} \vec{\beta} + \vec{\epsilon} \\
    \vec{X} &=  \vec{\iota} \vec{\Gamma} + \vec{[Z: W]} \vec{\Delta} + \vec{H},
\end{align*}
where $\vec Z$ is an $n \times p_1$ matrix of excluded instruments and $\vec W$ is an $n \times p_2$ matrix of covariates. The instruments are assumed to be valid a priori. All columns of $\vec{W}$, on the other hand, can still enter either model and can be used either as instruments if they are excluded from the outcome model or as covariates if they enter both models.

In this setting, we can also consider the two-component prior by \cite{zhang_two-component_2016} for the treatment model. This is primarily motivated by settings with very weak instruments, where it may be desirable to use different degrees of penalisation for instruments and covariates to ensure that weak instruments can still be included. We can only use the two-component prior with a single endogenous variable, as it does not yield a closed-form (conditional) posterior for $l>1$. Then, the prior on the parameter $\vec{\lambda}_M = (\gamma, \vec{\delta}_M)$ is
\begin{align*}
    \vec{\lambda}_M \mid M, \vec{\Sigma} \sim N\left(\vec{0}, \; \sigma_{xx} \vec{G}_M (\vec{V}_M^\intercal \vec{V}_M)^{-1} \vec{G}_M \right),
\end{align*}
where $\vec{G}_M$ is a diagonal matrix with entries $\sqrt{g_C}$ for the intercept and the covariates and $\sqrt{g_I}$ for the instruments. To encourage the inclusion of weak instruments, we specify $g_I = c g_C$, where $c \in (0, 1)$ is a constant, which does not depend on $n$. Our default choice is $c=1/2$ throughout the paper. This structure can be used for fixed or stochastic $g$'s.

The model prior has to be slightly adjusted as the treatment model has more potential variables in this setting. Our choice of the default prior mean model size reflects this by setting $m_L = p_2/2, m_M = (p_1+p_2)/2$.

The results in Sections \ref{sec:Posterior_Inference} and \ref{sec:consistency} of the main text still hold for this variant with $\vec{U} = [\vec{\iota : X : W}]$ and $\vec{V} = [\vec{\iota : Z : W}]$. The conditional posterior based on the two-component prior is derived in supplementary Section \ref{sec:posteriors}.

\section{Further details on the prior specification} 

\subsection{Prior specification for $\nu$}\label{sec:prior_nu}

\begin{figure}
    \centering
    \includegraphics[width=\linewidth]{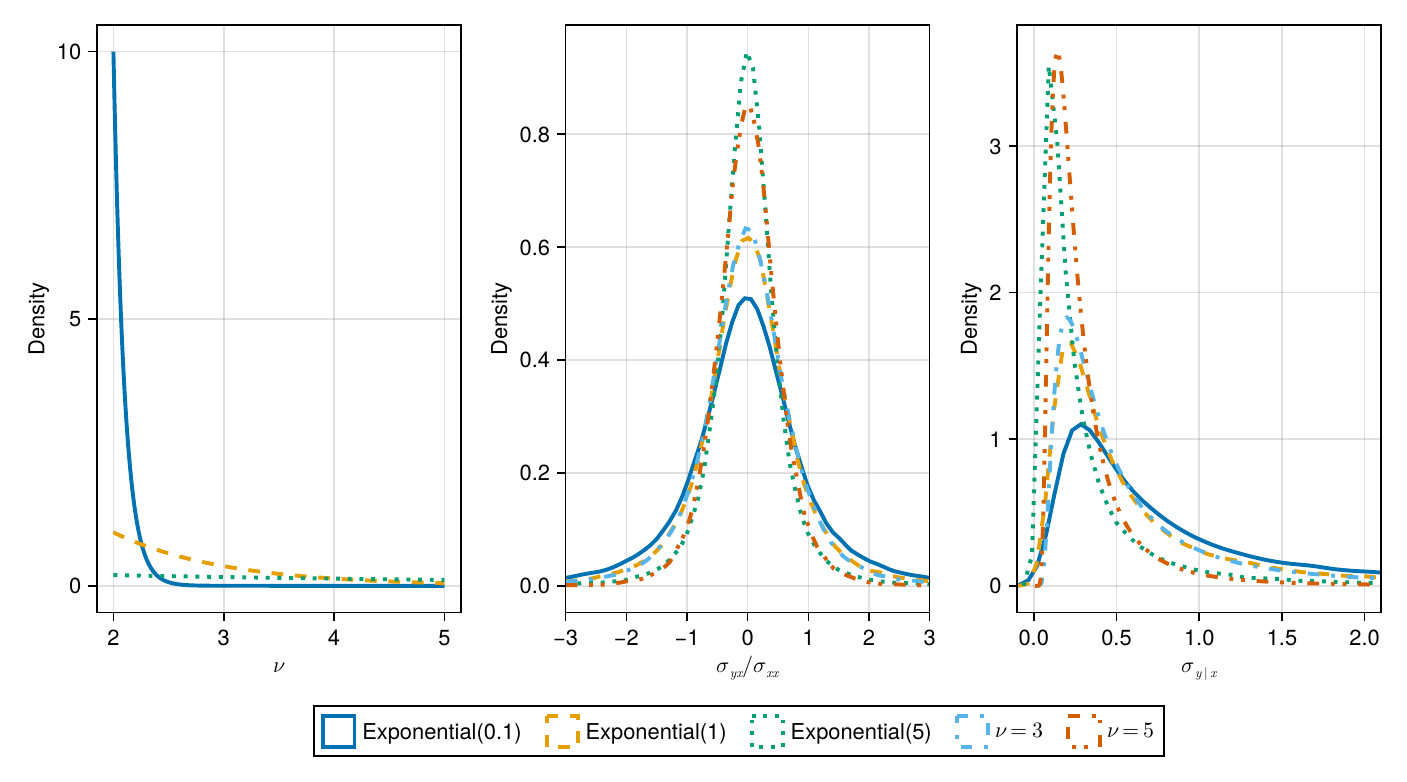}
    \caption{Exponential priors (scale parameterisation) on $\nu$ and the implied prior on the covariance ratio $\sigma_{yx} / \sigma_{xx}$ and the conditional variance $\sigma_{y \mid x}$ in the case of $l = 1$. The Exponential priors are shifted by $l+1 = 2$. 
    } 
\label{fig:implied_prior_sigma}
\end{figure}

For the case of the IW prior, Figure \ref{fig:implied_prior_sigma} presents different shifted Exponential priors on $\nu$ and their implied priors on the covariance ratio $\sigma_{yx} / \sigma_{xx}$ and the conditional outcome variance $\sigma_{y \mid x}$. The Exponential with scale $1$ strikes a good balance between being uninformative on $\nu$ itself while also keeping the implied prior on the covariance ratio relatively uninformative and is, therefore, our default choice. The priors on the covariance ratio and on $\sigma_{y|x}$ are very close to those implied by  fixing $\nu = 3$, which is done in \cite{karl_instrumental_2012}. For the Cholesky prior, we will choose the same prior on $\nu$. 

\subsection{Prior on the number of valid instruments}
\label{sec:priorValidInstr}

\begin{figure}
    \centering
    \includegraphics[width=0.8\linewidth]{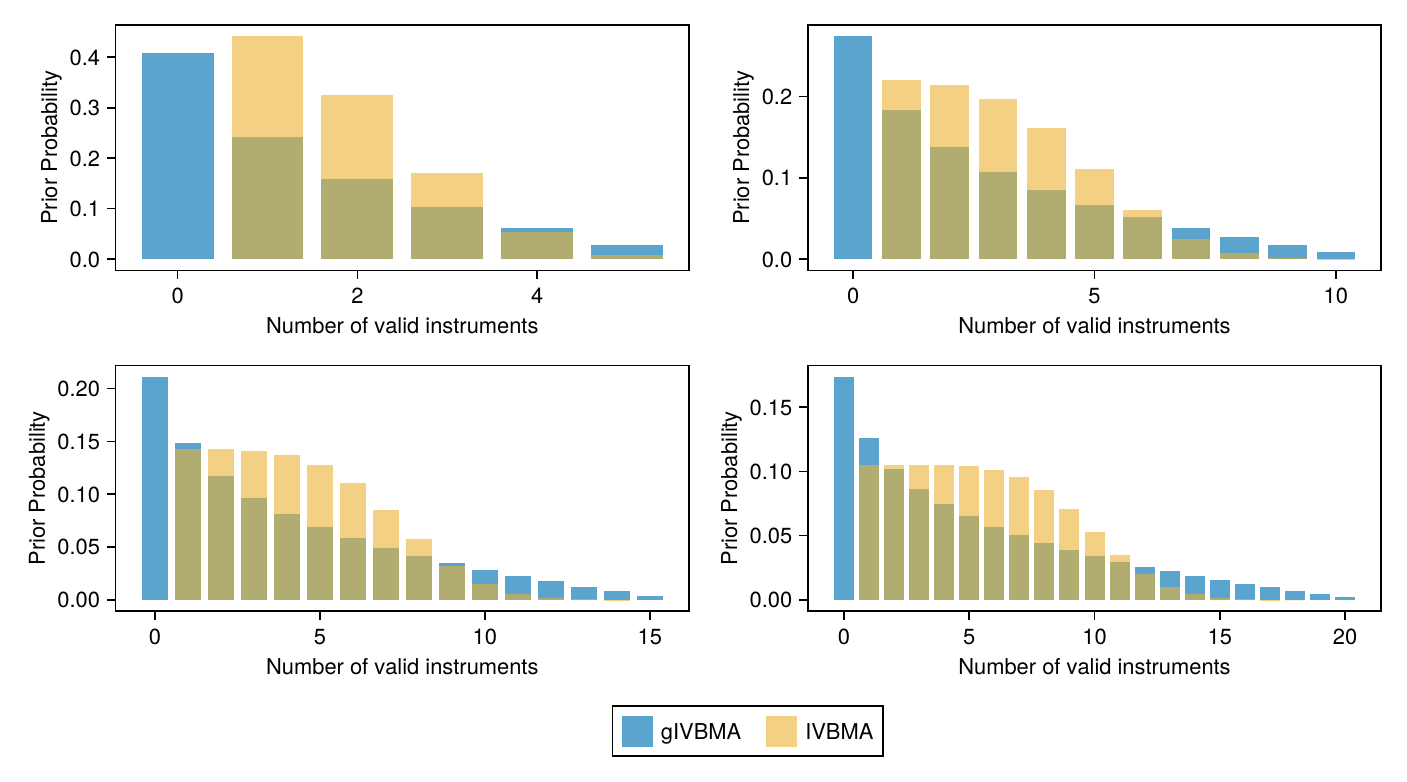}
    \caption{The implied prior on the number of valid and relevant instruments $N_Z$ for our proposed model  (gIVBMA) and the one used in \cite{karl_instrumental_2012} (IVBMA; assuming $l=1$) for $p = 5, 10, 15, 20$ (from top-left to bottom-right).}
    \label{fig:instrument_prior}
\end{figure}

An interesting question is what prior is induced on the number of valid and relevant instruments by the prior on the model space described in Subsection \ref{sec:priormodels} of the paper. Let $N_Z \in \{0, \ldots, p \}$ denote the number of valid and relevant instruments implied by a combination of outcome model and treatment model $(L_i, M_j)$, that is, all variables that are included in $M_j$, but not in $L_i$, denoted as the set $M_j \setminus L_i$ (in a slight abuse of notation). Then, the distribution of $N_Z$ is
\begin{align*}
    p(N_Z) &= \sum_{i, j: \: \# \lvert M_j \setminus L_i\rvert = N_Z} p(L_i) p(M_j) \\
    &= \frac{1}{p+1}\sum_{p_i=0}^{p}\sum_{k=0}^{p_i}
    1_{\{0\le N_Z\le p-p_i\}}
    \binom{p_i}{p_i -k} \binom{p-p_i}{N_Z}
    \frac{\Gamma(1+p_i+N_Z-k)\Gamma(1+p-p_i-N_Z+k)}{\Gamma(p+2)}.
\end{align*}
Figure~\ref{fig:instrument_prior} presents the prior distribution of $N_Z$ induced by our gIVBMA setting for various choices of $p$, and compares it to the prior implied by the IVBMA specification in \cite{karl_instrumental_2012} for $l=1$, which selects a uniform prior over all identified models\footnote{This identification condition does not seem to be enforced in the \texttt{ivbma} R package. Instead, it assumes a uniform prior over all models (rather than all identified models). Thus, IVBMA puts some mass on non-identified models, i.e., $N_Z = 0$, in our simulations and real-data applications.}. The main differences are that IVBMA puts zero mass on $N_Z=0$ and less prior mass on large values of $N_Z$. The IVBMA prior is more uniform on small and moderate $N_Z$ for larger $p$, whereas the gIVBMA prior induces a smooth geometric-shaped decline of the prior mass from $N_Z=0$ to $N_Z=p$ in all cases. 

\section{Derivations and Proofs} \label{sec:derivs_proofs}

\subsection{Derivation of the conditional outcome distribution} \label{sec:deriv_cond}

Here, we derive the conditional distribution of the outcome given all endogenous variables presented in Section \ref{sec:sampling} of the main text. Consider again the structural model
\begin{align*}
    \vec{y} &= \alpha \vec{\iota} + \vec{X} \vec{\tau} + \vec{Z} \vec{\beta} + \vec{\epsilon} \\
    \vec{X} &=  \vec{\iota} \vec{\Gamma} + \vec{Z} \vec{\Delta} + \vec{H},
\end{align*}
with
\begin{align*}
    \vec{[\epsilon : H]} \sim MN(\vec{0}, \vec{I_n}, \vec{\Sigma}),
\end{align*}
where the structural covariance matrix $\vec{\Sigma}$ can be partitioned into
\begin{align*}
    \vec{\Sigma} = \begin{pmatrix}
        \sigma_{yy} & \vec{\Sigma_{yx}} \\ \vec{\Sigma_{yx}}^\intercal & \vec{\Sigma_{xx}}
    \end{pmatrix}.
\end{align*}
First, note that the reduced form equation of $\vec{y}$ is given by
\begin{align*}
    \vec{y} = \alpha \vec{\iota} + \left(\vec{\iota \Gamma} + \vec{Z \Delta} \right) \vec{\tau} + \vec{W\beta }+ \vec{H \tau} + \vec{\epsilon}
\end{align*}
and therefore the joint distribution of $\vec{y}$ and $\vec{X}$ is given by
\begin{align*}
    \begin{pmatrix}
        \vec{y} \\ vec(\vec{X})
    \end{pmatrix} \sim N\left(  \begin{pmatrix}
        \alpha \vec{\iota} + \left(\vec{\iota \Gamma} + \vec{Z \Delta} \right) \vec{\tau + W \beta} \\ vec(\vec{\iota \Gamma + Z \Delta)}
    \end{pmatrix}, \vec{\Psi} \otimes \vec{I_n} \right),
\end{align*}
where we call
\begin{align*}
    \vec{\Psi} = \begin{pmatrix}
        1 & \vec{\tau}^\intercal \\
        0 & \vec{I_l}
    \end{pmatrix} \Sigma \begin{pmatrix}
        1 & \vec{\tau}^\intercal \\
        0 & \vec{I_l}
    \end{pmatrix}^\intercal = \begin{pmatrix}
        \sigma_{yy} + 2 \vec{\tau}^\intercal \vec{\Sigma_{yx}}^\intercal + \vec{\tau}^\intercal \vec{\Sigma_{xx}} \vec{\tau} & \vec{\Sigma_{yx}} + \vec{\tau}^\intercal \vec{\Sigma_{xx}} \\
        \vec{\Sigma_{yx}}^\intercal + \vec{\Sigma_{xx} \tau} & \vec{\Sigma_{xx}}
    \end{pmatrix}
\end{align*}
the reduced form covariance. Now, we can condition on $\vec{X}$ to get
\begin{align*}
    \vec{y} \mid \vec{X} \sim N(\vec{\mu_{y|x}}, \; \sigma_{y|x} \vec{I_n}),
\end{align*}
where, using well-known properties of the multivariate Normal distribution, we have
\begin{align*}
    \vec{\mu_{y|x}} &=  \alpha \vec{\iota} + \left(\vec{\iota \Gamma + Z \Delta} \right) \vec{\tau} + \vec{Z \beta} + \left((\vec{\Sigma_{yx}} \vec{\Sigma_{xx}^{-1}} + \vec{\tau}^\intercal) \otimes \vec{I_n} \right) vec(\vec{H}) \\
    &=  \alpha \vec{\iota} + \left(\vec{\iota \Gamma} + \vec{Z \Delta} \right) \vec{\tau} + \vec{Z \beta} + \vec{H} (\vec{\Sigma_{xx}^{-1} \Sigma_{yx}}^\intercal + \vec{\tau}) \\
    &= \alpha \vec{\iota} + \vec{X \tau} + \vec{Z \beta} + \vec{H \Sigma_{xx}^{-1} \Sigma_{yx}}^\intercal
\end{align*}
and
\begin{align*}
    \sigma_{y|x} = \sigma_{yy} - \vec{\Sigma_{yx} \Sigma_{xx}^{-1} \Sigma_{yx}}^\intercal.
\end{align*}

\subsection{Derivation of the conditional posteriors and marginal likelihoods} \label{sec:posteriors}

\subsubsection{The Gaussian case}

Here, we derive the conditional posterior distributions and marginal likelihoods. Let $\vec{\rho} = (\vec{\alpha}, \vec{\tau}^\intercal, \vec{\beta}^\intercal)$ be the outcome parameters and $\vec{U} = [\vec{\iota : X : Z}]$ be the outcome design matrix (we drop the model subscripts in this section for convenience). Similarly, define $\vec{\Lambda} = [\vec{\Gamma}^\intercal : \vec{\Delta}^\intercal]^\intercal$ and $\vec{V} = [\vec{\iota : Z}]$. Then, the likelihood is
\begin{align*}
    \vec{y} \mid \vec{X} &\sim N( \vec{U \rho} + \vec{H \Sigma_{xx}^{-1} \Sigma_{yx}}^\intercal, \sigma_{y|x} \vec{I_n}) \\
    \vec{X} &\sim MN(\vec{V \Lambda}, \vec{I_n}, \vec{\Sigma_{xx}})
\end{align*}
and the coefficient priors are given by
\begin{align*}
    \vec{\rho} \mid \vec{\Sigma} &\sim N(\vec{0}, g_L \sigma_{y|x} (\vec{U}^\intercal \vec{U})^{-1}) \\
    \vec{\Lambda} \mid \vec{\Sigma} &\sim MN(\vec{0}, g_M (\vec{V}^\intercal \vec{V})^{-1}, \vec{\Sigma_{xx}}).
\end{align*}
Define the corrected outcome $\vec{\Tilde{y}} = \vec{y} - \vec{H} \vec{\Sigma_{xx}^{-1}} \vec{\Sigma_{yx}}^\intercal$. Then, the conditional posterior for $\vec{\rho}$ is
\begin{align*}
    p(\vec{\rho} \mid \vec{\Lambda}, \vec{\Sigma}, \vec{y}, \vec{X}) &\propto p(\vec{y} \mid \vec{X}, \vec{\rho}, \vec{\Lambda, \Sigma}) p(\vec{X} \mid \vec{\Lambda, \Sigma}) p(\vec{\rho} \mid \vec{\Sigma}) \\
    &\propto p(\vec{y} \mid \vec{X, \rho, \Lambda, \Sigma}) p(\vec{\rho} \mid \vec{\Sigma}) \\
    &\propto \exp\left( - \frac{1}{2 \sigma_{y|x}} (\vec{\Tilde{y}} - \vec{U \rho})^\intercal (\vec{\Tilde{y}} - \vec{U \rho}) \right) \exp\left( - \frac{1}{2 \sigma_{y|x}} \vec{\rho}^\intercal \left(g_L^{-1} \vec{U}^\intercal\vec{U} \right) \vec{\rho} \right) \\
    &= \exp\left( - \frac{1}{2 \sigma_{y|x}} \left(\vec{\Tilde{y}}^\intercal \vec{\Tilde{y}} + \vec{\rho}^\intercal \left(\frac{g_L + 1}{g_L} \vec{U}^\intercal \vec{U} \right) \vec{\rho} - 2 \vec{\Tilde{y}}^\intercal \vec{U \rho} \right. \right. \\
    &+ \left. \left. \frac{g_L}{g_L + 1} \vec{\Tilde{y}}^\intercal \vec{P_U \Tilde{y}} - \frac{g_L}{g_L + 1} \vec{\Tilde{y}}^\intercal \vec{P_U \Tilde{y}} \right)  \right) \\
    &\propto N\left( \vec{\rho} \mid \frac{g_L}{g_L + 1} (\vec{U}^\intercal \vec{U})^{-1} \vec{U}^\intercal \vec{\Tilde{y}}, \frac{g_L}{g_L + 1} \sigma_{y|x} (\vec{U}^\intercal \vec{U})^{-1} \right) \\
    &\times \exp\left( - \frac{1}{2 \sigma_{y|x}} \vec{\Tilde{y}}^\intercal \left(\vec{I_n} - \frac{g_L}{g_L + 1} \vec{P_U} \right) \vec{\Tilde{y}} \right) \\
    &\propto N\left( \vec{\rho} \mid \frac{g_L}{g_L + 1} (\vec{U}^\intercal \vec{U})^{-1} \vec{U}^\intercal \vec{\Tilde{y}}, \frac{g_L}{g_L + 1} \sigma_{y|x} (\vec{U}^\intercal \vec{U})^{-1} \right).
\end{align*}
The (conditional) marginal likelihood is the normalising constant of this distribution (adjusted by the prior). So it will be the remaining exponential times the ratio of normalising constants of the prior and this posterior. Thus,
\begin{align*}
    p(\vec{y} \mid \vec{X}, \vec{\Lambda}, \vec{\Sigma}) &\propto \left( \frac{| g_L / (g_L + 1) \sigma_{y|x} (\vec{U}^\intercal \vec{U})^{-1} |}{| g_L \sigma_{y|x} (\vec{U}^\intercal \vec{U})^{-1} |} \right)^{-1/2} \exp\left( - \frac{1}{2 \sigma_{y|x}} \vec{\Tilde{y}}^\intercal \left(\vec{I_n} - \frac{g_L}{g_L + 1} \vec{P_U} \right) \vec{\Tilde{y}} \right) \\
    &\propto (g_L + 1)^{-d_U / 2} \exp\left( - \frac{1}{2 \sigma_{y|x}} \vec{\Tilde{y}}^\intercal \left(\vec{I_n} - \frac{g_L}{g_L + 1} \vec{P_U} \right) \vec{\Tilde{y}} \right),
\end{align*}
where $d_U$ is the number of columns included in $\vec{U}$. Note that we condition on $\vec{X}$ in the marginal likelihood. This is because when integrating out $\vec{\rho}$, it is sufficient to consider the conditional model as the marginal distribution of $\vec{X}$ does not depend on $\vec{\rho}$. However, this marginal likelihood is proportional to the one based on the full joint distribution, where the proportionality constant does not depend on the outcome model. Therefore, this is merely a matter of notation and does not affect the results.

For the treatment model, define
\begin{align*}
    \vec{B_\Sigma} &= \vec{I_l} + \frac{1}{\sigma_{y|x}} \vec{\Sigma_{yx}}^\intercal \vec{\Sigma_{yx} \Sigma_{xx}^{-1} }\\
    \vec{\Tilde{X}} &= \vec{X} - \frac{1}{\sigma_{y|x}} \vec{\epsilon \Sigma_{yx} }\left(\vec{B_\Sigma^{-1}}\right)^\intercal.
\end{align*}
Also note that $p(\vec{y} \mid \vec{X}, \vec{\rho, \Lambda, \Sigma})$ depends on $\vec{\Lambda}$ through the residual matrix $\vec{H}$, so we cannot drop this factor. Then, we have
\begin{align*}
    p(\vec{\Lambda} \mid \vec{\rho}, \vec{\Sigma}, y, X) &\propto p(\vec{y} \mid \vec{X}, \vec{\rho}, \vec{\Lambda}, \vec{\Sigma}) p(\vec{X} \mid \vec{\Lambda}, \vec{\Sigma}) p(\vec{\Lambda} \mid \vec{\Sigma}) \\
    &\propto \exp\left( -\frac{1}{2 \sigma_{y|x}} \left( \vec{\epsilon} - (\vec{X} - \vec{V \Lambda}) \vec{\Sigma_{xx}^{-1} \Sigma_{yx}}^\intercal \right)^\intercal \left( \vec{\epsilon} - (\vec{X} - \vec{V \Lambda}) \vec{\Sigma_{xx}^{-1} \Sigma_{yx}}^\intercal \right) \right) \\
    &\quad \times \exp\left( -\frac{1}{2} \text{tr}\left( \vec{\Sigma_{xx}^{-1}} (\vec{X} - \vec{V \Lambda})^\intercal (\vec{X} - \vec{V \Lambda}) \right) \right) \exp\left( -\frac{1}{2} \text{tr}\left( \vec{\Sigma_{xx}^{-1} \Lambda}^\intercal (g_M^{-1} \vec{V}^\intercal \vec{V}) \Lambda \right) \right) \\
    &\propto \exp\left( -\frac{1}{2} \text{tr} \bigg( \vec{\Sigma_{xx}^{-1}} (\vec{B_\Sigma} + g_M^{-1} \vec{I_l}) \bigg[ \vec{\Lambda}^\intercal \vec{V}^\intercal \vec{V} \vec{\Lambda} - 2 (\vec{I_l} + g_M^{-1} \vec{B_\Sigma^{-1}})^{-1} \vec{\Tilde{X}}^\intercal \vec{V \Lambda} \right. \\
    &\quad \left. + (\vec{I_l} + g_M^{-1} \vec{B_\Sigma^{-1}})^{-1} \vec{\Tilde{X}}^\intercal \vec{P_V \Tilde{X}} \left((\vec{I_l} + g_M^{-1} \vec{B_\Sigma^{-1}})^{-1} \right)^\intercal \right. \\
    &\quad \left. - (\vec{I_l} + g_M^{-1} \vec{B_\Sigma^{-1}})^{-1} \vec{\Tilde{X}}^\intercal \vec{P_V} \vec{\Tilde{X}} \left((\vec{I_l} + g_M^{-1} \vec{B_\Sigma^{-1}})^{-1} \right)^\intercal \bigg] \bigg) \right) \\
    &\quad \times \exp\left( -\frac{1}{2\sigma_{y|x}} (\vec{\epsilon}^\intercal \vec{\epsilon} - 2 \vec{\epsilon}^\intercal \vec{X} \vec{\Sigma_{xx}^{-1} \Sigma_{yx}}^\intercal) -\frac{1}{2} \text{tr}(\vec{\Sigma_{xx}^{-1} B_\Sigma} \vec{X}^\intercal \vec{X}) \right) \\
    &\propto MN \left( \vec{\Lambda} \mid (\vec{V}^\intercal \vec{V}) \vec{V}^\intercal \vec{\Tilde{X}} \left((\vec{I_l} + g_M^{-1} \vec{B_\Sigma^{-1}})^{-1} \right)^\intercal, (\vec{V}^\intercal \vec{V})^{-1}, (\vec{B_\Sigma} + g_M^{-1} \vec{I_l})^{-1} \vec{\Sigma_{xx}}  \right),
\end{align*}
Again, the (conditional) marginal likelihood is the normalising constant of this unnormalised distribution. If we omit all factors that do not depend on the design matrix and, therefore, the model (we only care about ratios between different models), we have
\begin{align*}
    p(\vec{y}, \vec{X} \mid \vec{\rho}, \vec{\Sigma}) \propto | g_M \vec{B_\Sigma} + \vec{I_l}|^{-d_V/2} \exp\left( \frac{1}{2} \text{tr} \left(\vec{A_\Sigma} \Tilde{\vec{X}}^\intercal \vec{P_V} \Tilde{\vec{X}} \right) \right),
\end{align*}
where $d_{V}$ is the total number of columns included in $\vec{V}$ and
\begin{align*}
    \vec{A_{\Sigma}} = \left( \left(\vec{I_l} + g_M^{-1} \vec{B_\Sigma^{-1}} \right)^{-1} \right)^\intercal \vec{\Sigma_{xx}^{-1}} \vec{B_\Sigma}.
\end{align*}
Note that in the treatment model, the marginal likelihood is now based on the joint distribution of $\vec{y}$ and $\vec{X}$. This is because the conditional outcome model depends on the treatment parameters $\vec{\Lambda}$. Therefore, when integrating out $\vec{\Lambda}$, we have to consider the joint distribution.

For the two-component prior (as defined in Section \ref{sec:fixed_Z}), the derivation is very similar. The squared term in $\vec{\Lambda}$ now becomes
\begin{align*}
    \vec{B_\Sigma} \vec{\Lambda}^\intercal \vec{V}^\intercal \vec{V} \vec{\Lambda} + \vec{\Lambda}^\intercal \vec{G}^{-1} \vec{V}^\intercal \vec{V} \vec{G}^{-1} \vec{\Lambda}
\end{align*}
In the one-dimensional case, $\vec{B_\Sigma} = b_\Sigma$ is a scalar and we can factorize this expression as
\begin{align*}
    \vec{\Lambda}^\intercal (b_\Sigma \vec{V}^\intercal \vec{V} + \vec{G}^{-1} \vec{V}^\intercal \vec{V} \vec{G}^{-1}) \vec{\Lambda}
\end{align*}
and then complete the square in $\vec{\Lambda}$. In this case, we have the following conditional posterior
\begin{align*}
    \vec{\Lambda} \mid \vec{\rho}, \vec{\Sigma}, \vec{y}, \vec{x} &\sim N \left( b_\Sigma \vec{A_G} \vec{V}^\intercal \vec{\Tilde{x}},\; \sigma_{xx} \vec{A_G} \right),
\end{align*}
where
\begin{align*}
    b_\Sigma &=  1 + \frac{1}{\sigma_{y|x}} \frac{\sigma_{yx}^2}{\sigma_{xx}} \\
    \vec{\Tilde{x}} &= \vec{x} - \frac{\sigma_{yx}}{b_\Sigma \sigma_{y|x}} \vec{\epsilon} \\
    \vec{A_G} &= \left( b_\Sigma \vec{V}^\intercal \vec{V} + \vec{G}^{-1} \vec{V}^\intercal \vec{V} \vec{G}^{-1} \right)^{-1}.
\end{align*}
The marginal likelihood is given by
\begin{align*}
    p(\vec{y}, \vec{x} \mid \vec{\rho}, \vec{\Sigma}) \propto \left(\frac{|\vec{A_G}|}{|\vec{G} (\vec{V}^\intercal \vec{V})^{-1} \vec{G} |}\right)^{1/2} \exp\left( \frac{b_\Sigma^2}{2 \sigma_{xx}} \vec{\Tilde{x}}^\intercal \vec{V} \vec{A_G} \vec{V}^\intercal \vec{\Tilde{x}} \right).
\end{align*}




\subsubsection{The non-Gaussian case}\label{sec:CBF_NG}

The conditional posterior distributions and marginal likelihoods change slightly in the non-Gaussian case. For the latent Gaussian outcome $\vec{q}$ define $\Tilde{\vec{q}} = \vec{q} - \left(\vec{Q} - \vec{V} \vec{\Lambda} \right) \vec{\Sigma_{xx}}^{-1} \vec{\Sigma_{yx}}^\intercal$. Now, the formulas from above remain valid after replacing $\Tilde{\vec{y}}$ with $\Tilde{\vec{q}}$. In particular, we have
\begin{align*}
    \vec{\rho} \mid \vec{q}, \vec{Q}, \vec{X}, \vec{\Lambda}, \vec{\Sigma} \sim N\left(\frac{g_L}{g_L + 1} (\vec{U}^\intercal \vec{U})^{-1} \vec{U}^\intercal \vec{\Tilde{q}}, \frac{g_L}{g_L + 1} \sigma_{y|x} (\vec{U}^\intercal \vec{U})^{-1} \right)
\end{align*}
and
\begin{align}\label{eq:ML_NG_q}
    p(\vec{q} \mid \vec{X}, \vec{Q}, \vec{\Lambda}, \vec{\Sigma}) &\propto (g_L + 1)^{-d_U / 2} \exp\left( - \frac{1}{2 \sigma_{y|x}} \vec{\Tilde{q}}^\intercal \left(\vec{I_n} - \frac{g_L}{g_L + 1} \vec{P_U} \right) \vec{\Tilde{q}} \right).
\end{align}
There is some subtlety in that the posterior quantities now depend on both the actual treatment $\vec{X}$ through the projection on $\vec{U} = [\vec{\iota}: \vec{X}: \vec{W}]$ and its latent Gaussian representation $\vec{Q}$ through the latent residual $\vec{H} = \vec{Q} - \vec{V} \vec{\Lambda}$.

For the treatment model, define $$\vec{\Tilde{Q}} = \vec{Q} - \frac{1}{\sigma_{y|x}}  \left( \vec{q} - \vec{U} \vec{\rho} \right)\vec{\Sigma_{yx}}\left(\vec{B_\Sigma^{-1}}\right)^\intercal$$ such that we have
\begin{align*}
    \vec{\Lambda} \mid \vec{q}, \vec{Q}, \vec{X}, \vec{\rho}, \vec{\Sigma} \sim MN \left( (\vec{V}^\intercal \vec{V}) \vec{V}^\intercal \vec{\Tilde{Q}} \left((\vec{I_l} + g_M^{-1} \vec{B_\Sigma^{-1}})^{-1} \right)^\intercal, (\vec{V}^\intercal \vec{V})^{-1}, (\vec{B_\Sigma} + g_M^{-1} \vec{I_l})^{-1} \vec{\Sigma_{xx}}  \right)
\end{align*}
and
\begin{align}\label{eq:ML_NG_qQ}
    p(\vec{q}, \vec{Q} \mid \vec{X}, \vec{\rho}, \vec{\Sigma}) \propto | g_M \vec{B_\Sigma} + \vec{I_l}|^{-d_V/2} \exp\left( \frac{1}{2} \text{tr} \left( \vec{A_\Sigma} \vec{\Tilde{Q}}^\intercal \vec{P_{V}} \vec{\Tilde{Q}} \right) \right),
\end{align}
where $d_{V}$ is the total number of columns included in $\vec{V}$. Note that this marginal likelihood still depends on $\vec{X}$ through the latent outcome residual $\vec{\epsilon} = \vec{q} - \vec{U} \vec{\rho}$, but not on the actual outcome $\vec{y}$. To identify the dependence on $\vec{X}$, it is helpful to factorise the joint treatment marginal likelihood as
\begin{align}    
    p(\vec{q}, \vec{Q} \mid \vec{X}, \vec{\rho}, \vec{\Sigma}) &\propto | g_M \vec{B_\Sigma} + \vec{I_l}|^{-d_V/2} \exp\left( \frac{1}{2} \text{tr}\left( \vec{A_\Sigma} \vec{Q}^\intercal \vec{P_V} \vec{Q} \right)\right) \notag \\
    &\cdot \exp \left(\frac{1}{2} \text{tr}\left( \frac{1}{\sigma_{y\mid x}} \vec{c_\Sigma} (\vec{q} - \vec{U \rho})^\intercal \vec{P_V} \left( (\vec{q} - \vec{U \rho}) \vec{\Sigma_{yx}} \left(\vec{B_\Sigma}^{-1}\right)^\intercal - 2 \vec{Q} \right)\right) \right) \notag \\
    &= | g_M \vec{B_\Sigma} + \vec{I_l}|^{-d_V/2} \exp\left( \frac{1}{2} \text{tr}\left( \vec{A_\Sigma} \vec{Q}^\intercal \vec{P_V} \vec{Q} \right)\right) \notag \\
    &\cdot \exp \left(-\frac{1}{2\sigma_{y\mid x}} \left(2 (\vec{q} - \vec{U \rho})^\intercal \vec{P_V} \vec{Q} \vec{c_\Sigma} - \vec{\Sigma_{yx}} \left(\vec{B_\Sigma}^{-1}\right)^\intercal \vec{c_\Sigma} (\vec{q} - \vec{U \rho})^\intercal \vec{P_V} (\vec{q} - \vec{U \rho}) \right) \right)\label{eq:ML_NG_fact},
\end{align}
where we have defined $\vec{c_\Sigma}=\vec{A_\Sigma} \vec{B_\Sigma}^{-1} \vec{\Sigma_{yx}}^\intercal$. 

The marginal distribution $p(\vec{Q} \mid \vec{\Sigma})$ is proportional to the first exponential, and the conditional distribution $p(\vec{q} \mid \vec{Q}, \vec{X}, \vec{\rho}, \vec{\Sigma})$ is proportional to the second exponential factor. The latter involves terms with both $\vec{q}, \vec{Q}$ and $\vec{U}$ and thus still depends on $\vec{X}$.

\subsection{Proof of Theorem \ref{Th:consistency}}\label{sec:consistencyProof}

Given the proper prior, we know that the posterior exists for all models under consideration, almost surely in the sampling distributions. One important condition that all models share is that the matrices $\vec{U}_L$ and $\vec{V}_M$ are of full column rank for any model we consider (an assumption made in Subsection 3.1). This can easily be imposed by the prior on the model space. First, we will examine model selection consistency for the Gaussian case, and then we will show that consistency extends to non-Gaussian settings. 

\subsubsection{Outcome model}\label{sec:consProof_outcome}
In order to prove model selection consistency of the procedure using CBFs for the outcome equation, we assume, without loss of generality, that model $L_i$ is the true model generating the data, and we consider CBF($L_i,L_j$), which measures the relative support for $L_i$ versus $L_j$. Thus, model selection consistency implies that CBF($L_i,L_j$) tends to $\infty$ for any $L_j\ne L_i$ as $n$ grows without bounds. We need to distinguish two cases:

\begin{itemize}
 \item{Model $L_j$ nests $L_i$.}
In this case, $L_j$ has all the important covariates but is overparameterised, i.e., some of its covariates are not in the true model. 

We use the expression for CBF($L_i,L_j$) in Subsection \ref{sec:CML}: 
\begin{equation}   
\label{eq:CBF_L}
    \text{CBF}(L_i, L_j) 
    = (g_L + 1)^{(d_{U_j} - d_{U_i})/2} \exp\left( -\frac{1}{2 \sigma_{y|x}} \frac{g_L}{g_L + 1} \vec{\Tilde{y}}^\intercal (\vec{P_{U_j}} - \vec{P_{U_i}}) \vec{\Tilde{y}} \right),
\end{equation}
where $\vec{U_i}, d_{U_i}$ and $\vec{U_j}, d_{U_j}$ are the design matrices and their number of columns in model $L_i$ and $L_j$ respectively. Clearly here we have that $d_{U_j}>d_{U_i}$. First, we show that the exponential factor does not tend to zero (or $\infty$) with $n$. From supplementary Section \ref{sec:deriv_cond} we know that the corrected outcome 
$\vec{\tilde{y}} = \vec{y} - \vec{H} \vec{\Sigma_{xx}^{-1}} \vec{\Sigma_{yx}}^\intercal$ given $\vec{X}$ is  distributed as 
\begin{equation}\vec{\tilde{y}} \mid \vec{X} \sim N( \vec{U_i \rho}_i, \sigma_{y|x} \vec{I_n})
\label{eq:ygivenx}\end{equation}
in the true model. 
This means that 
$$\frac{\vec{\tilde{y}}}{\sqrt{\sigma_{y|x}}} \mid \vec{X} \sim N( \vec{U_i \rho}_i, \vec{I_n})
$$
and we can now use standard results on quadratic forms of Normal random variables (see {\it e.g.}~Theorem 9.4 in \citealp[p.~175]{muller2006linear}) to establish that 
$\frac{1}{\sigma_{y|x}}
\vec{\Tilde{y}}^\intercal (\vec{P_{U_j} - P_{U_i}}) \vec{\Tilde{y}}$ is distributed as a central $\chi^2$ random variable with $d_{U_j}-d_{U_i}$ degrees of freedom, given that $\vec{P_{U_j}-P_{U_i}}$ is an idempotent matrix of rank $d_{U_j}-d_{U_i}$ and $(\vec{P_{U_j}-P_{U_i}})\vec{U_i}=0$. This ensures us that the exponential factor in (\ref{eq:CBF_L}) is almost surely a finite positive number and does not dictate the asymptotic behavior. 
Now consider the first factor in (\ref{eq:CBF_L}), where we distinguish between the following two prior options:

{\bf Fixed $g_L$:} Here a necessary and sufficient condition for CBF($L_i,L_j)\to\infty$ with $n$ is that $g_L$ increases without bound in $n$. This is the case in {\it e.g.} the BRIC prior we use, and would also hold for the unit information prior, and many more.    

{\bf Random $g_L$:} if we assume a hyperprior $p(g_L)$, the CBF above will asymptotically behave as 
$$I(L_i,L_j)=\int_{\Re_+} (1+g_L)^{c/2} p(g_L) dg_L,$$
where $c=d_{U_j}-d_{U_i}$. If we opt for, as done in this paper, the hyper-$g/n$ prior, then 
$$ I(L_i,L_j)=\frac{a-2}{2n}\int_{\Re_+} (1+g_L)^{c/2} (1+\frac{g_L}{n})^{-\frac{a}{2}} dg_L,$$
where $a>2$ for prior propriety. For any given $n$, the integral in $I(L_i,L_j)$ will be nonzero and will not explode for small $g_L$ but might do so for large values of $g_L$. If we focus on the right-hand tail, we obtain 
$$I(L_i,L_j)\approx n^{\frac{a}{2}-1} \frac{a-2}{2} \int g_L^{\frac{c-a}{2}} dg_L.$$
In case the degree of overparameterisation $c=d_{U_j}-d_{U_i}$ is smaller than $a-2$ ({\it{ e.g.}}~$d_{U_j}=d_{U_i}+1$ and $a>3$), the right hand tail will integrate and the integral wil be finite. But even then, the factor in front will increase without bound with $n$, leading the CBF to do the same thing. In case $d_{U_j}-d_{U_i}\ge a-2$ the integral will not be finite and thus CBF($L_i,L_j)\to \infty$ with $n$.\footnote{If we use a hyper-$g$ prior instead for $g_L$ a similar reasoning shows that model selection consistency does not hold for cases where $d_{U_j}-d_{U_i}< a-2$. This is qualitatively similar to the behaviour in the standard linear regression model, as in  \cite{liang_mixtures_2008}, but for different reasons, as a slightly different prior structure is used there.}

In summary, for the case where $L_j$ nests the true model $L_i$, we have that CBF$(L_i,L_j)\to \infty$ with $n$, thus showing model selection consistency for gIVBMA under the prior assumptions mentioned here. In fact, these are necessary and sufficient conditions. 
 \item{Model $L_j$ does not nest $L_i$.}
Here, the model $L_j$ lacks at least one important covariate. 
We slightly rewrite the expression for the CBF in (\ref{eq:CBF_L}) as
\begin{equation}   
\label{eq:CBF_L2}
    \text{CBF}(L_i, L_j) 
    = (g_L + 1)^{(d_{U_j} - d_{U_i})/2} \exp\left( -\frac{n}{2 \sigma_{y|x}} \frac{g_L}{g_L + 1} \Tilde{y}^\intercal \left(\frac{\vec{M_{U_i}}}{n} - \frac{\vec{M_{U_j}}}{n} \right) \Tilde{y} \right),
\end{equation}
using $\vec{M_U}=\vec{I_n}-\vec{P_U}$. We then apply the same reasoning as in Lemma A.1 of \cite{fernandez_benchmark_2001} to (\ref{eq:ygivenx}), and we use the fact  that for any model $L_j$ that does not nest $L_i$
\begin{equation}
\label{eq:ass1} 
\lim_{n\to\infty} \frac{\vec{\rho}_i^\intercal \vec{U_i}^\intercal \vec{M_{U_j}} \vec{U_i \rho}_i}{n}=b_j \in (0,\infty),
\end{equation}
which  means that $\vec{U_i}$ is not in the column space of $\vec{U_j}$, or we can not express $\vec{U_i}$ as 
$\vec{U_i}= \vec{U_j A}$
for some matrix $\vec{A}$ of dimension $d_{U_j}\times d_{U_i}$. In fact, (\ref{eq:ass1}) is satisfied if $U$ corresponding to the full model has full column rank, which was assumed in Subsection \ref{sec:priorCoeff}. Then we can state the following probability limit for any model $L_j$ not nesting the true $L_i$:
\begin{equation*}
p\lim_{n\to\infty} \frac{\vec{\Tilde y}^\intercal \vec{M_{U_j}} \vec{\Tilde y}}{n} = \sigma_{y|x} + b_j,
\end{equation*}
while for $L_i$ (or any model  nesting it), we obtain 
\begin{equation*}
p\lim_{n\to\infty} \frac{\vec{\Tilde y}^\intercal\vec{M_{U_i}} \vec{\Tilde y}}{n} = \sigma_{y|x}.
\end{equation*}
This immediately shows us that the CBF$\to\infty$
as the sample size grows, since the exponential term behaves as $\exp(c_jn)$ for some $c_j>0$. This is irrespective of $d_{U_j}$ and the value or prior for $g_L$. 

Thus, we also have model selection consistency for gIVBMA in the case that $L_j$ does not nest the true model $L_i$. 
\end{itemize}

Summarising, for any model $L_j$, we have shown that CBF($L_i,L_j$) in favour of the true model $L_i$ increases without bounds for the choice of the outcome model. Outcome model selection on the basis of gIVBMA is thus consistent, if and only if \begin{itemize}
\item $\lim_{n\to \infty} g_L=\infty$ for fixed $g_L$
\item we have that  $$\lim_{n\to\infty}\int_{\Re_+} (1+g_L)^{c/2} p(g_L) dg_L=\infty$$
for any $c\in\{1,\dots,p+k\}$ if we assume a hyperprior $p(g_L)$. The latter condition is always satisfied if it holds for $c=1$. 
\end{itemize}

\subsubsection{Treatment model}\label{sec:consProof_treatment}

Assume $M_i$ is the true treatment model generating the matrix of endogenous variables $X$. We show that $\text{CBF}(M_i, M_j)$ tends to infinity for all $M_j \neq M_i$. As above, we distinguish two cases:
\begin{itemize}
    \item Model $M_j$ nests $M_i$. We can rewrite the conditional Bayes factor for model $M_j$ versus $M_i$ as
\begin{align}\label{eq:CBF_M_app}
        \text{CBF}(M_i, M_j) = g_M^{l  (d_{V_j} - d_{V_i}) / 2} | \vec{B_\Sigma} + g_M^{-1} \vec{I_l}|^{(d_{V_j} - d_{V_i}) / 2} \exp \left( -\frac{1}{2} \text{tr} \left( \vec{A_\Sigma \Tilde{X}}^{\intercal} (\vec{P_{V_j}} - \vec{P_{V_i}}) \vec{\Tilde{X}} \right) \right),
    \end{align}
    where 
    \begin{align*}
    \vec{A_\Sigma} = \left( \left( \vec{I_l} + g_M^{-1} \vec{B_\Sigma}^{-1} \right)^{-1} \right)^\intercal \vec{\Sigma_{xx}}^{-1} \vec{B_\Sigma}.
    \end{align*}
    Since $M_j$ nests $M_i$ we have $d_{V_j} > d_{V_i}$. 
    
    We first show that the trace in the exponential term does not tend to infinity with $n$. Note that $\vec{\Tilde{X}}$ is a linear transformation of the residual matrix $[\vec{\epsilon : H}]$ and is therefore again a matrix Normal,
    \begin{align*}
        \vec{\Tilde{X}} \sim MN(\vec{V_i \Lambda_i}, \vec{I_n}, \vec{\Tilde{\Sigma}_{xx}}),
    \end{align*}
    where
    \begin{align*}
        \vec{\Tilde{\Sigma}_{xx}} = \begin{pmatrix}
            -\frac{1}{\sigma_{y|x} } \vec{\Sigma_{yx}} \left(\vec{B_\Sigma^{-1}}\right)^\intercal \\ \vec{I_l}
        \end{pmatrix}^\intercal \Sigma \begin{pmatrix}
            -\frac{1}{\sigma_{y|x} } \vec{\Sigma_{yx}} \left(\vec{B_\Sigma^{-1}}\right)^\intercal \\ \vec{I_l}
        \end{pmatrix}.
    \end{align*}
    As $\vec{P_{V_j} - P_{V_i}}$ is idempotent and $(\vec{P_{V_j} - P_{V_i}})\vec{V_i} = 0$, the matrix quadratic form $\vec{\Tilde{X}}^{\intercal} (\vec{P_{V_j} - P_{V_i}}) \vec{\Tilde{X}}$ follows a (central) Wishart distribution with $\text{tr}(\vec{P_{V_j} - P_{V_i}}) = d_{V_j} - d_{V_i}$ degrees of freedom \citep[see e.g. Corollary 10.8.2 and Theorem 10.9 in][p. 202]{muller2006linear}.

    Using linearity of the trace operator and Theorem 10.10 in \cite{muller2006linear}, we have
    \begin{align*}
        \mathbb{E} \left[ \text{tr} \left( \vec{A_\Sigma \Tilde{X}}^{\intercal}(\vec{P_{V_j} - P_{V_i}}) \vec{\Tilde{X} }\right)\right] &=   \text{tr} \left( \mathbb{E}\left[ \vec{A_\Sigma \Tilde{X}}^{\intercal}(\vec{P_{V_j} - P_{V_i}}) \vec{\Tilde{X}} \right]\right) \\
        &= (d_{V_j} - d_{V_i}) \text{tr} \left( \vec{A_\Sigma \Tilde{\Sigma}_{xx}} \right).
    \end{align*}
    For small $g_M$, $\vec{A_\Sigma}$ behaves like $g_M \vec{B_\Sigma}^\intercal \vec{\Sigma_{xx}}^{-1} \vec{B_\Sigma}$ and for large $g_M$ it tends to $\vec{\Sigma_{xx}}^{-1} \vec{B_\Sigma}$. The latter does not explode, and therefore, this expectation is finite for all values of $g_M$. Since the expectation is finite, the trace is finite almost surely, and the exponential term does not tend to zero.
    
    Therefore, similarly to the previous subsection, we need to establish that the first term tends to infinity with $n$. Note that the determinant $|\vec{B_\Sigma} + g_M^{-1} \vec{I_l}|$ behaves like $|\vec{B_\Sigma}|$ for large $g_M$ and therefore does not dictate the asymptotic behavior. Thus, it is sufficient to focus on the first term. For any fixed $g_M$, a necessary and sufficient condition is that $g_M\to\infty$ with $n$ (e.g., BRIC). If we use a hyper prior on $g_M$, we need to ensure that $$\lim_{n\to\infty}\int_{\Re_+} g_M^{c/2} p(g_M) dg_M=\infty$$
    for $c\in\{l,2l,\dots,pl\}$. Similarly to the situation for the outcome model, we can easily show that this is always satisfied for the hyper-$g/n$ prior.

    \item Model $M_j$ does not nest $M_i$. Again, we define the orthogonal projection $\vec{M_V} = \vec{I_n} - \vec{P_V}$ to rewrite the CBF as
    \begin{align}
        \text{CBF}(M_i, M_j) = g_M^{l  (d_{V_j} - d_{V_i}) / 2} | \vec{B_\Sigma} + g_M^{-1} \vec{I_l}|^{(d_{V_j} - d_{V_i}) / 2} \exp \left( -\frac{n}{2} \text{tr} \left( \vec{A_\Sigma \Tilde{X}}^{\intercal} \left( \frac{\vec{M_{V_i}}}{n} - \frac{\vec{M_{V_j}}}{n} \right) \vec{\Tilde{X}} \right) \right).
    \end{align}
    We then use a version of Lemma A.1. in \cite{fernandez_benchmark_2001} modified to the matrix setting:
    \begin{lemma} \label{lemma_a1_fls}
    If $M_i$ is nested within (or equal to) $M_j$,
    \begin{align*}
        p\lim_{n \to \infty} \frac{\vec{\Tilde{X}}^\intercal \vec{M_{V_j}} \vec{\Tilde{X}}}{n} = \vec{\Tilde{\Sigma}_{xx}}.
    \end{align*}
    If for any $M_j$ that does not nest $M_i$,
    \begin{align*}
        \lim_{n \to \infty} \frac{\vec{\Lambda_i}^\intercal \vec{V_i}^\intercal \vec{M_{V_j}} \vec{V_i} \vec{\Lambda_i}}{n} = \vec{D_j},
    \end{align*}
    where $\vec{D_j}$ is a positive definite $l \times l$ matrix, then
    \begin{align*}
        p\lim_{n \to \infty} \frac{\vec{\Tilde{X}}^\intercal \vec{M_{V_j}} \vec{\Tilde{X}}}{n} = \vec{\Tilde{\Sigma}_{xx}} + \vec{D_j}.
    \end{align*}
    \end{lemma}
    The assumption for the non-nested case in Lemma \ref{lemma_a1_fls} is satisfied if $\vec{V}$ corresponding to the full model has full column rank, which was assumed in Subsection \ref{sec:priorCoeff}. Applying Lemma \ref{lemma_a1_fls}, we have that the exponential term behaves (as $n \to \infty$) as
    \begin{align*}
        \exp \left( n \cdot  \text{tr} \left( \vec{A_\Sigma D_j} \right) \right).
    \end{align*}
    Note that $\vec{A_\Sigma}$ can be written as the quadratic form
\begin{align*}
    \vec{A_\Sigma} = \left( \left( \vec{I}_l + g_M^{-1} \vec{B_\Sigma}^{-1} \right)^{-1} \right)^\intercal \vec{\Sigma_{xx}}^{-1} \left(\vec{B_\Sigma}+g_M^{-1}\vec{I}_l\right) \left( \vec{I}_l + g_M^{-1} \vec{B_\Sigma}^{-1} \right)^{-1}.
    \end{align*}
  Since  
    \begin{align*}   \vec{\Sigma_{xx}^{-1}} \left( \vec{B_\Sigma} + g_M^{-1} \vec{I_l} \right) = \frac{g_M + 1}{g_M} \vec{\Sigma_{xx}^{-1}} + \frac{1}{\sigma_{y|x}} \vec{\Sigma_{xx}^{-1} \Sigma_{yx}}^\intercal \vec{\Sigma_{yx} \Sigma_{xx}^{-1}}
    \end{align*}
    is symmetric and positive-semidefinite, we can conclude that $\vec{A_\Sigma}$ is also positive-semidefinite, implying that
    \begin{align*}
        \text{tr} \left( \vec{A_\Sigma D_j} \right) \geq 0.
    \end{align*}
    The equality holds only if $\vec{A_\Sigma D_j}$ is the zero matrix, as the trace of the product of two positive-semidefinite matrices is a (squared) Frobenius norm. Therefore, the trace is positive almost surely, and the exponential tends to infinity as $n \to \infty$.
\end{itemize}
For any treatment model $M_j$, we have shown that $\text{CBF}(M_i, M_j)$ in favour of the true treatment model $M_i$ tends to infinity as the sample size $n$ tends to infinity. Thus, the treatment model selection is consistent under the following (necessary and sufficient) prior  assumptions on $g_M$: 
\begin{itemize}
\item $\lim_{n\to \infty} g_M=\infty$ for fixed $g_M$
\item under a hyperprior $p(g_M)$, we have 
$$\lim_{n\to \infty}\int_{\Re_+} g_M^{c/2} p(g_M) dg_M=\infty$$
for $c\in\{l,2l,\dots,(p+k)l\}$.
\end{itemize}

\textbf{The two-component prior: } The conditional Bayes factor under the two-component prior is given by
\begin{align*}
    \text{CBF}(M_i, M_j) &= \left(\frac{|\vec{A_G^{(i)}}|}{|\vec{A_G^{(j)}}|} \frac{|\vec{G}_j (\vec{V}_j^\intercal \vec{V}_j)^{-1} \vec{G}_j |}{|\vec{G}_i (\vec{V}_i^\intercal \vec{V}_i)^{-1} \vec{G}_i |}\right)^{1/2}  \exp\left( - \frac{b_\Sigma^2}{2 \sigma_{xx}} \vec{\Tilde{x}}^\intercal \left(\vec{V}_j \vec{A_G^{(j)}} \vec{V}_j^\intercal - \vec{V}_i \vec{A_G^{(i)}} \vec{V}_i^\intercal \right) \vec{\Tilde{x}} \right),
\end{align*}
where
\begin{align*}
    \vec{A_G^{(i)}} &= \left( b_\Sigma \vec{V}_i^\intercal \vec{V}_i + \vec{G_i}^{-1} \vec{V}_i^\intercal \vec{V}_i \vec{G_i}^{-1} \right)^{-1}.
\end{align*}
The exponential factor in this CBF behaves very similarly to the one in the single-$g$ case (only with $l=1$), as the quadratic form in the exponent can be bounded by that of the single-$g$ case \citep[][Appendix B]{zhang_two-component_2016}. 

Recall that the relation between $g_I$ for the instruments and $g_C$ for the exogenous covariates is given by
$$g_I=c g_C$$
for some constant $c<1$ which does not depend on $n$. Then, note that
\begin{align*}
    \left(\frac{| \vec{G}_j (\vec{V}_j^\intercal \vec{V}_j)^{-1} \vec{G}_j|}{|\vec{A_G}^{(j)}|} \right)^{1/2} \geq b_\Sigma^{d_{V_j}/2} |\vec{G}_j | + 1 \geq (b_\Sigma g_I)^{d_{V_j}/2},
\end{align*}
where the first inequality is based on Minkowski's determinant theorem \citep[see][Appendix B]{zhang_two-component_2016} and the second inequality holds since $g_C > g_I$. Similarly, we can get a lower bound on the inverse as follows:
\begin{align*}
    \left(\frac{|\vec{A_G}^{(j)}|}{| \vec{G}_j (\vec{V}_j^\intercal \vec{V}_j)^{-1} \vec{G}_j|} \right)^{1/2} = \left(\frac{| \vec{G}_j^{-1} (\vec{V}_j^\intercal \vec{V}_j) \vec{G}_j^{-1}|}{|b_\Sigma (\vec{V}_j^\intercal \vec{V}_j)+ \vec{G}_j^{-1} (\vec{V}_j^\intercal \vec{V}_j) \vec{G}_j^{-1}|} \right)^{1/2} \geq  \left(\frac{| \vec{G}_j^{-1} (\vec{V}_j^\intercal \vec{V}_j) \vec{G}_j^{-1}|}{|b_\Sigma (\vec{V}_j^\intercal \vec{V}_j)+ \vec{G}_{I,j}^{-1} (\vec{V}_j^\intercal \vec{V}_j) \vec{G}_{I,j}^{-1}|} \right)^{1/2},
\end{align*}
where $\vec{G}_{I,j}$ is a diagonal matrix with all entries equal to $\sqrt{g_I}$ and the inequality follows from the monotonicity of determinants with respect to positive semi-definite matrices and the fact that $\vec{G}_{I,j}^{-1} (\vec{V}_j^\intercal \vec{V}_j) \vec{G}_{I,j}^{-1} - \vec{G}_j^{-1} (\vec{V}_j^\intercal \vec{V}_j) \vec{G}_j^{-1}$ is positive semi-definite. We can then write
\begin{align*}
    \left(\frac{|\vec{A_G}^{(j)}|}{| \vec{G}_{j} (\vec{V}_j^\intercal \vec{V}_j)^{-1} \vec{G}_{j}|} \right)^{1/2} \geq \frac{|\vec{G}_j|^{-1}}{(b_\Sigma+g_I^{-1})^{d_{V_j}/2}}\geq \frac{g_C^{-d_{V_j}/2}}{(b_\Sigma+g_I^{-1})^{d_{V_j}/2}} =
    \left(\frac{c}{b_\Sigma g_I+1}\right)^{d_{V_j}/2}.
\end{align*}
Thus, we can bound the relevant ratio both ways: 
\begin{align*}
    (b_\Sigma g_I)^{d_{V_j}/2}\leq \left(\frac{| \vec{G}_j (\vec{V}_j^\intercal \vec{V}_j)^{-1} \vec{G}_j|}{|\vec{A_G}^{(j)}|} \right)^{1/2} \leq  \left(\frac{b_\Sigma g_I+1}{c}\right)^{d_{V_j}/2},
\end{align*}
so that it is clear the ratio behaves like $g_I^{d_{V_j}/2}$ and the same necessary and sufficient condition as for the single-$g$ case applies, with $g_M$ replaced by $g_I$ (and $l=1$).

\subsubsection{Extension to the non-Gaussian case}

A mild assumption we make is that the prior on any additional parameters in $r_y$ and $\vec{r}_x=(r_{x_1},\dots,r_{x_l})$ is proper and independent of $\vec{q},\vec{Q}, L_i, M_j$. 

The conditional Bayes factor for the outcome model will now be based on the marginal likelihood for 
$\vec{y}\mid \vec{X}, \vec{Q}, L_i, M_j, \vec{\Lambda}, \vec{\Sigma}$, which can be written as 
\begin{equation*}
p(\vec{y}\mid \vec{X}, \vec{Q}, L_i, M_j,  \vec{\Lambda}, \vec{\Sigma})=\int p(\vec{y}\mid \vec{q},r_y) p(\vec{q}\mid \vec{X}, \vec{Q}, L_i, M_j, \vec{\Lambda}, 
\vec{\Sigma})
p(r_y) d\vec{q}\, dr_y,
\end{equation*}
where the second factor on the right-hand side is (\ref{eq:ML_NG_q}). Now, assume that the true model is $L_i$ and write the conditional marginal likelihood of $L_i$ as 
\begin{equation*}
p(\vec{y}\mid \vec{X}, \vec{Q}, L_i, M_j,  \vec{\Lambda}, \vec{\Sigma})=\int p(\vec{y}\mid \vec{q},r_y) p(\vec{q}\mid \vec{X}, \vec{Q}, L_k, M_j, \vec{\Lambda}, 
\vec{\Sigma})
\,\mathrm{CBF}_q(L_i,L_k)
p(r_y) \,d\vec{q} \,dr_y,
\end{equation*}
where CBF$_q(\cdot,\cdot)$ is the expression in (\ref{eq:CBF_L}), but with  $\Tilde{\vec{y}}$ replaced by $\Tilde{\vec{q}}$. We can now replace the integral in $r_y$ by
\begin{equation}\label{eq:def_h(z)}
p(\vec{y} \mid \vec{q})= \int  p(\vec{y} \mid \vec{q}, r_y) p(r_y) dr_y , 
\end{equation}
which has to be finite a.s.~since the posterior exists. This gives us
\begin{equation*}
p(\vec{y}\mid \vec{X}, \vec{Q}, L_i, M_j,  \vec{\Lambda}, \vec{\Sigma})=\int p(\vec{y}\mid \vec{q}) p(\vec{q}\mid \vec{X}, \vec{Q}, L_k, M_j, \vec{\Lambda}, 
\vec{\Sigma})\,
\mathrm{CBF}_q(L_i,L_k)
\,d\vec{q}.
\end{equation*}
The expression in (\ref{eq:ML_NG_q}) is almost surely positive and finite, so that the ratio CBF$_q(L_i,L_k)$ is 
bounded from below by some positive number $B_{ik}$ almost surely in $\vec{q}$.
Thus, we can write 
\begin{equation*}
p(\vec{y}\mid \vec{X}, \vec{Q}, L_i, M_j,  \vec{\Lambda}, \vec{\Sigma})>B_{ik}\int p(\vec{y}\mid \vec{q}) p(\vec{q}\mid \vec{X}, \vec{Q}, L_k, M_j, \vec{\Lambda}, 
\vec{\Sigma})
d\vec{q}.
\end{equation*}
The integral on the right-hand side is the conditional marginal likelihood for $L_k$, so the conditional Bayes factor for the true $L_i$ versus the misspecified $L_k$
is larger than $B_{ik}$. Now $B_{ik}$ is the minimum conditional Bayes factor of $L_i$ versus $L_k$ for $\vec{q}$, and we know from the analysis in Subsection \ref{sec:consProof_outcome} that the underlying Gaussian model for $\vec{q}$ with the prior in Section 3 is model-selection consistent for the unit information prior and the hyper-$g/n$ prior. 
As a consequence, under these choices for $g$ (or any other choice that satisfies the criteria for  consistency in the Gaussian model), $\lim_{n\to \infty} \textrm{CBF}_q(L_i,L_k)=\infty$
almost surely in $\vec{q}$. Thus, it must be the case that 
$\lim_{n\to \infty} B_{ik}=\infty$, which immediately leads to model selection consistency for the outcome model. 

To examine the  consistent selection of the treatment model, we need to consider
\begin{equation}\label{eq:ML_treatment_NG}
   p(\vec{y}, \vec{X} \mid L_j, M_i, \vec{\rho}, \vec{\Sigma}) = \int p(\vec{y} \mid \vec{q}, r_y) p(\vec{X} \mid \vec{Q}, \vec{r}_x) p(\vec{q}, \vec{Q} \mid L_j, M_i, \vec{X}, \vec{\rho}, \vec{\Sigma}) p(r_y) p(\vec{r}_x)\, d\vec{q}\, d\vec{Q}\,
   dr_y \,d\vec{r}_x,
\end{equation}
where the third factor under the integral is given in (\ref{eq:ML_NG_qQ}), and we need to make sure that the expression we integrate characterizes a valid factorization of a probability distribution. Using (\ref{eq:ML_NG_fact}) in Section \ref{sec:CBF_NG} we can write this as 
\begin{align*}
    p(\vec{y}, \vec{X} \mid L_j, M_i, \vec{\rho}, \vec{\Sigma}) =& \int \left( \int p(\vec{y} \mid \vec{q}, r_y) p(\vec{q} \mid L_j, M_i, \vec{Q}, \vec{X}, \vec{\rho}, \vec{\Sigma}) p(r_y) d\vec{q}\, dr_y\right) \\ &p(\vec{X} \mid \vec{Q}, \vec{r}_x) p(\vec{Q}\mid M_i, \vec{\Sigma}) p(\vec{r}_x)\, d\vec{Q}\, d\vec{r}_x \,
\end{align*}
which ensures that (\ref{eq:ML_treatment_NG}) represents a valid probability statement. 

We can then rewrite (\ref{eq:ML_treatment_NG}) as 
\begin{align*}\label{eq:ML_treatment_NG2}
   p(\vec{y}, \vec{X} \mid L_j, M_i, \vec{\rho}, \vec{\Sigma}) = &\int p(\vec{y} \mid \vec{q}, r_y) p(\vec{X} \mid \vec{Q}, \vec{r}_x) p(\vec{q}, \vec{Q} \mid L_j, M_k, \vec{X}, \vec{\rho}, \vec{\Sigma})\\ 
&\textrm{CBF}_{qQ}(M_i,M_k) p(r_y) p(\vec{r}_x)\, d\vec{q}\, d\vec{Q}\,
   dr_y \,d\vec{r}_x,
\end{align*}
where CBF$_{qQ}(M_i,M_k)$ is the conditional Bayes factor in (\ref{eq:CBF_M_app}) with $\Tilde{\vec{X}}$ replaced by $\Tilde{\vec{Q}}$. 
Following the same reasoning as for the outcome model, we can then prove model selection consistency for the treatment model.



\section{Additional details on the simulation experiments} \label{sec:details_simulation}

\subsection{Performance measures}
\label{sec:performance}

This Section defines the performance measures used in our simulation experiments. Define the median absolute error (MAE) as the median $\ell_1$ difference between the point estimator and the true parameter,
\begin{align*}
    \text{MAE}(\vec{\tau}, \vec{\Hat{\tau}}) = \text{Median}_{i = 1,\ldots, M} \left\lVert \vec{\tau} - \vec{\Hat{\tau}}^{(i)} \right\rVert_1,
\end{align*}
where $\vec{\Hat{\tau}}^{(i)}$ is the posterior estimate obtained from the $i$-th simulated dataset and $M$ is the number of dataset replications. If $l = 1$ (i.e., $\tau$ is a scalar), the $\ell_1$ norm simplifies to the absolute value. Similarly, we estimate the median bias as
\begin{align*}
    \text{Bias}(\vec{\tau}, \vec{\Hat{\tau}}) = \left\lVert \text{Median}_{i = 1,\ldots, M} \left( \vec{\Hat{\tau}}^{(i)} \right) - \vec{\tau} \right\rVert_1 .
\end{align*}
For all Bayesian methods, the posterior mean is used as the point estimator for the MAE and median bias calculations. Finally, we define the coverage of each method as the proportion of credible intervals (or confidence intervals for the classical estimators) that include the true parameter value.

For the log-predictive score (LPS) calculation, let $\vec{\theta} = (\vec{\rho, \Lambda, \Sigma}, L, M)$ denote all parameters and model indices considered, and note that we only focus on the conditional distribution $\vec{y} \mid \vec{X}$ as this is of primary interest in most applications. Then, the LPS is given by (with some abuse of notation as $\vec{\theta}$ has continuous and discrete components)
\begin{align*}
    \text{LPS} = -\frac{1}{n^*}\sum_{i=1}^{n^*}\log \int p(y_i^* \mid \vec{X}_i^*, \vec{\theta}) p(\vec{\theta} \mid \vec{y, X}) d\vec{\theta},
\end{align*}
where the integral is the posterior predictive distribution for holdout observations $(y_i^*, \vec{X}_i^*), \; i = 1, \ldots, n^*$. We divide by the holdout sample size $n^*$ to make the results comparable across different sizes of the holdout dataset.  This integral is approximated by averaging over our posterior sample of $\{\vec{\theta}^{(s)}: s = 1, \ldots, S\}$ \citep[see e.g.][]{gelman_bayesian_2014}
\begin{align*}
    \text{LPS} \approx -\frac{1}{n^*}\sum_{i=1}^{n^*}\log \left( \frac{1}{S} \sum_{s=1}^S p(y_i^* \mid \vec{X}_i^*, \vec{\theta}^{(s)}) \right).
\end{align*}

For the classical estimators, we compute an analogue of the log predictive density by taking the log of the outcome density evaluated at the point estimates, i.e. $- (n^*)^{-1} \sum_{i=1}^{n^*} \log p({y_i^*} \mid \vec{X}_i^*, \vec{\Hat{\theta}})$. We generate an additional 20\% of holdout observations in each dataset to compute the LPS. We then compare the mean LPS values across all simulated datasets.

\subsection{Implementation of competing estimators}\label{sec:competing}

Here, we present some additional details on how the competing methods are implemented in our simulation study. Throughout this Section, define $\vec{U} = [\vec{\iota: X : W}]$ to be the outcome (or second stage) design matrix and $\vec{V} = [\vec{\iota: Z : W}]$ to be the treatment (or first stage) design matrix, where $\vec{W}$ is the matrix of all available exogenous control variables and $\vec{Z}$ is the matrix of all available instruments. Let $k_U$ and $k_V$ denote their number of columns, respectively. Also, let $\vec{P_V} = \vec{V} (\vec{V}^\intercal \vec{V})^{-1} \vec{V}^\intercal$ denote the first stage projection matrix. The aim is to estimate the outcome parameter $\vec{\rho}$.
\begin{itemize}

    \item \textbf{Two-Stage Least Squares (TSLS):} TSLS linearly regresses the endogenous variables on the instruments and uses the resulting fitted values in the outcome model. This is equivalent to using the linear projection $\vec{P_V X}$ instead of $\vec{X}$. Accordingly, the TSLS estimator is given by
    \begin{align*}
        \vec{\Hat{\rho}}_{TSLS} = (\vec{U}^\intercal \vec{P_V} \vec{U})^{-1} \vec{U}^\intercal \vec{P_V} \vec{y}
    \end{align*}
    and its variance is estimated as
    \begin{align*}
        \Hat{\text{var}}_{TSLS} =  \Hat{\sigma}^2 (\vec{U}^\intercal \vec{P_V} \vec{U})^{-1},
    \end{align*}
    where $\Hat{\sigma}^2 = \left \lVert \vec{y} - \vec{U} \vec{\Hat{\rho}}_{TSLS} \right \rVert_2^2 / (n - k_U)$. The oracle TSLS (O-TSLS) estimator refers to the one that uses the correct subset of columns in $\vec{U}$ and $\vec{V}$ in the formula above. For more details on TSLS estimation, we refer to any (graduate) Econometrics textbook \citep[e.g.][]{davidson_econometric_2004}. 

    \item \textbf{Jacknife instrumental variable estimation (JIVE):} Motivated by the bias of TSLS in overidentified settings, \cite{angrist_jackknife_1999} propose to use jackknife fitted values in the first-stage. The intuition behind their approach is that not using the $i$-th observation when predicting the first stage gets rid of the correlation between the predicted value and the residual for observation $i$. Let $\vec{U(i)}$ and $\vec{V(i)}$ be the design matrices with the $i$-th observation left out. Then, define $\vec{\Hat{U}}_{JIVE}$ as the matrix with row $i$ containing $\vec{V_i} (\vec{V(i)}^\intercal \vec{V(i)})^{-1} \vec{V(i)} \vec{U(i)}$ (the predicted value for the $i$-th observation). Then, the JIVE estimator is defined as\footnote{We use what they refer to as JIVE1. They also propose a slightly different estimator named JIVE2. However, the difference between JIVE1 and JIVE2 is minimal in their simulations, so we do not expect this to make a big difference.}
    \begin{align*}
        \vec{\Hat{\rho}}_{JIVE} = (\vec{\Hat{U}}_{JIVE}^\intercal  \vec{U})^{-1} \vec{\Hat{U}}_{JIVE}^\intercal \vec{y}.
    \end{align*}
    \cite{angrist_jackknife_1999} use TSLS standard errors based on a just-identified setting with instruments $\vec{\Hat{U}}_{JIVE}$ (which is asymptotically valid).

    \item \textbf{Regularised JIVE (RJIVE):} \cite{hansen_instrumental_2014} propose JIVE with regularisation (in particular, Ridge) in the first stage. The jackknife prediction in the first stage then becomes
    \begin{align*}
        \vec{V_i} (\vec{V(i)}^\intercal \vec{V(i)} + \lambda \vec{I_{n-1}})^{-1} \vec{V(i)} \vec{U(i)}.
    \end{align*}
    They propose to set $\lambda = c^2 \cdot p$, where the constant $c$ is set to the sample standard deviation of $\vec{X}$ (after partialling out control variables). We use this value of $\lambda$ in our simulation experiments. The second stage estimation remains unchanged from that of regular JIVE.

    \item \textbf{Post-Lasso:} \cite{belloni_sparse_2012} propose to use Lasso regularisation in the first stage. We have not implemented this manually, but use the \texttt{rlassoIV} function in the \texttt{hdm} R package \citep{hdm}. However, this function only provides estimates (and standard errors) for the endogenous variables but not for the exogenous covariates (which we need to compute the LPS). Also, when no instruments are selected, the software only returns a warning but no effect estimates. Thus, we do not provide LPS estimates and compute the other criteria only in those cases where at least one instrument is selected. 

    \item \textbf{Model-averaged TSLS (MATSLS):} \cite{kuersteiner_constructing_2010} propose to use model-averaged first-stage predictions of the endogenous variable. That is, they use a TSLS estimator with a weighted average first-stage projection, $\vec{P(w)} = \sum_{i=1}^M w_i \vec{P_{V_i}}$, where the sum goes over all possible models. The weights are estimated by minimising a mean-squared error criterion. We allow the weights to be unrestricted (i.e. they can be negative). Then, the TSLS estimator is computed as above using $\vec{P(w)}$ in place of $\vec{P_V}$. To implement this procedure, we have translated the MATLAB code found on Ryo Okui's website to Julia.

    \item \textbf{sisVIVE:} \cite{kang_instrumental_2016} propose the sisVIVE estimator for settings with potentially invalid instruments. Instruments can be included in the outcome model but are subject to an $\ell_1$ penalty. The estimation strategy is to minimise the moment condition subject to that $\ell_1$ penalty term. We use the \texttt{sisVIVE} R package \citep{sisvive} to implement the estimator. To the best of our knowledge, there is no simple way of computing valid standard errors for the sisVIVE estimator (the software does not provide standard errors, and \cite{kang_instrumental_2016} do not discuss how one would compute them). Therefore, our simulations do not provide any coverage results for sisVIVE.

    \item \textbf{BMA:} Naive BMA refers to using BMA on the outcome model only (and ignoring the treatment equation). We use a $g$-prior specification on the model-specific slope coefficients and uninformative priors on the intercept and the residual variance \citep[see e.g.][]{fernandez_benchmark_2001}. Note that this is different from the prior specification of gIVBMA, where the intercept is included in the $g$-prior. 

    \item \textbf{IVBMA:} To implement the IVBMA method by \cite{karl_instrumental_2012}, we use the most recent version (1.05, 2014-09-18) of the \texttt{ivbma} R package. This R package is not available on CRAN anymore (withdrawn on 2018-12-07) but can still be downloaded from the archive. As no intercept is included by default, we centre the outcome and treatments before using the IVBMA function. Their code does not accommodate the approximate extension to Poisson variables suggested in \cite{lee_incorporating_2022}, so when we use IVBMA in the presence of non-Gaussian variables, we use a Gaussian approximation for any non-Gaussian component.

    \item \textbf{BayesHS:} We fit a Bayesian IV model with all instruments included and put a Horseshoe prior \citep{carvalho2009handling} on both the outcome coefficients $\vec\beta$ and the treatment coefficients $\vec\Delta$. The horseshoe prior on $\vec\beta$ (analogously on $\vec\Delta$) corresponds to 
    $$
    \beta_i \sim N\left(0, \lambda_i^2 \mu^2 \right),\quad \mu \sim C^{+}(0, 1),\quad \lambda_i \sim C^{+}(0, 1),
    $$
    where $C^{+}$ denotes a half-Cauchy distribution. We refer to the $\lambda_i$ as the local and $\mu$ as the global shrinkage parameters and we choose them independently for the outcome and treatment model. We approximate the posterior distribution using variational inference in \texttt{Turing.jl} \citep{fjelde2025turing}. Our variational family is a Gaussian mean field and we perform $500$ optimisation steps of the distance over weighted gradients algorithm \citep{khaled2023dowg}, which is the default optimiser in \texttt{Turing.jl}. We have also tried sampling from the posterior using a No-U-Turn sampler. However, the convergence of the chains was poor in many cases, and the execution  was very slow.

\end{itemize}

Whenever we use purely data-driven instrument selection in our simulations, we also allow naive BMA 
to include all instruments in the outcome model. Letting gIVBMA use the instrument information in the outcome model while the naive method cannot would unfairly skew the comparison towards gIVBMA. This is particularly relevant when some of the instruments may be invalid.

\subsection{Invalid Instruments}

\begin{figure}
    \centering
    \includegraphics[width=0.8\linewidth]{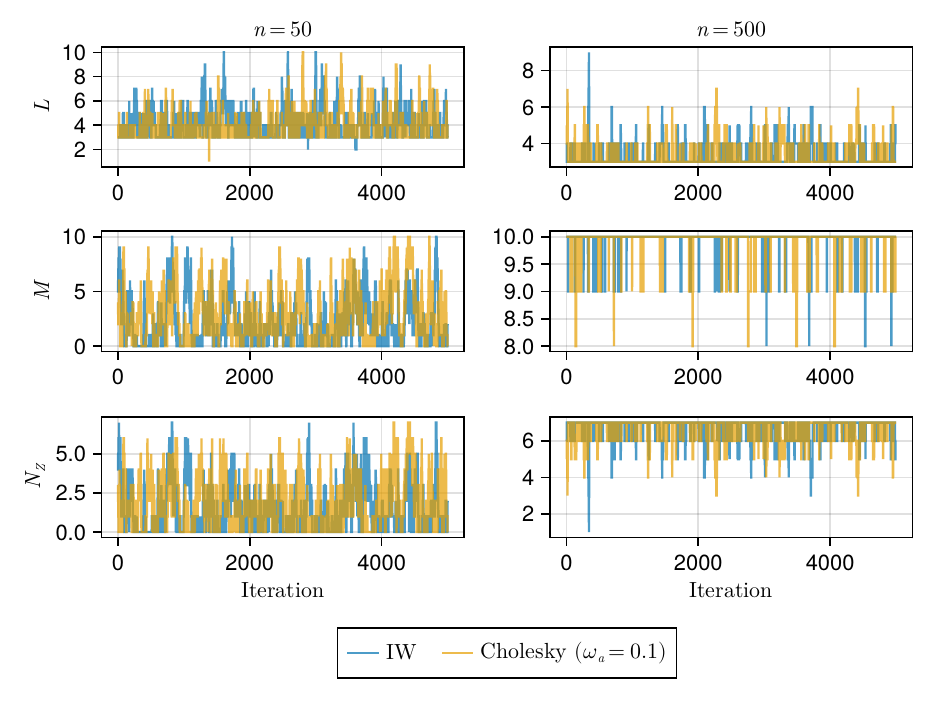}
    \caption{\textbf{Invalid Instruments:} A chain of the outcome model size ($L$), treatment model size ($M$), and the number of valid and relevant instruments $N_Z$ implied by outcome and treatment model at each iteration for both considered gIVBMA variants. These results are based on a single simulated dataset of size $n = 50$ (left) and $n = 500$ (right), respectively, with $s = 3$ out of $10$ instruments being invalid.}
    \label{fig:invalid_instruments_mixing}
\end{figure}

Figure~\ref{fig:invalid_instruments_mixing} displays traceplots of the included variables in both models and the number of implied instruments at each iteration of the chain for a single simulated dataset. We only consider the hyper-$g/n$ specifications with the full inverse Wishart prior and the Cholesky prior with $\omega_a = 0.1$, as these perform best in the simulation experiment (see results in Section~\ref{sec:sim-invalid} of the main text). For the smaller sample size $n=50$, the chains still visit non-identified models with high frequency (in line with the averages over 100 datasets in Table \ref{tab:Kang_Sim_Instruments} in the main text), but mix well. For $n=500$, the chains mix well and spend most of their time at $N_Z = 6$ or the true value, $N_Z = 7$. The difference in behaviour between the small and large sample is mainly driven by the treatment model. In the small sample, the treatment model includes too few instruments in most iterations. This is a consequence of the parsimony preference of Bayes factors and the small amount of data information.

\subsection{Multiple endogenous variables with correlated instruments} \label{sec:sim_mult}

We consider an example with two endogenous variables, one following a Gaussian distribution and one following a Beta distribution, and correlated instruments. We take inspiration from the simulation design in \cite{fernandez_benchmark_2001}: There are $15$ instruments where $Z_1, \ldots, Z_{10}$ are generated from independent standard Gaussians and $Z_{11}, \ldots, Z_{15}$ are generated according to
\begin{align*}
    \vec{[Z_{11}: \ldots : Z_{15}]} = \vec{[Z_{1}: \ldots : Z_{5}]} [0.3, 0.5, 0.7, 0.9, 1.1]^\intercal [1, 1, 1, 1, 1] + \vec{E},
\end{align*}
where $\vec{E}$ is an $n \times 5$ matrix of independent Gaussian errors. This data-generating process leads to moderate correlations between the groups $Z_1, \ldots, Z_{5}$ and $Z_{11}, \ldots, Z_{15}$ and relatively strong correlations within $Z_{11}, \ldots, Z_{15}$. Then, we generate the (latent Gaussian) endogenous variables according to
\begin{align*}
    \vec{Q} = \vec{\iota} [4, -1] + \vec{Z_1} [2, -2] + \vec{Z_5} [-1, 1] + \vec{Z_7} [1.5, 1] + \vec{Z_{11}} [1, 1] + \vec{Z_{13}} [0.5, -0.5] + \vec{H}
\end{align*}
and we set $\vec{X_1} = \vec{Q_1}$ and sample $\vec{X_2}$ from a Beta distribution with mean $\exp(Q_{i2}) / (1 + \exp(Q_{i2}))$ for observation $i$ and dispersion parameter $r = 1$. The outcome is then generated from 
$$
    \vec{y} = \vec{\iota} + \vec{X} [0.5, -0.5]^\intercal + \vec{\epsilon},
$$
where $\vec{[\epsilon : H]}$ are jointly Gaussian with mean zero and covariance matrix $\vec{\Sigma}$. 
The covariance is set to ${\Sigma}_{ij} = c^{|i-j|}, i, j = 1, 2, 3$ with $c = 2/3$, so the endogeneity is relatively strong. None of the instruments directly affects the outcome; therefore, they are all valid. Again, we vary the sample size $n \in \{50, 500\}$. 

Posterior results on MAE, bias, coverage and prediction are summarized in Table \ref{tab:SimResMultEndo}, which is discussed in the main text. 

Table \ref{tab:SimMultEnd_PIPs} presents the median posterior inclusion probabilities from the simulation with multiple endogenous variables. The gIVBMA variants always include the correct variables while only very sparingly including wrong instruments in the smaller sample and not at all in the larger sample. IVBMA correctly selects the included instruments, but also includes other unnecessary variables too often for the Gaussian endogenous variable. For the second (Beta)  endogenous variable, some of the true instruments are often excluded in the $n=50$ scenario, while the performance is good for $n=500$.

\begin{table}[H]
\centering
\begin{tabular}{lccccc}
\toprule
\multicolumn{6}{c}{$n=50$} \\
\midrule
 & \textbf{MAE} & \textbf{Bias} & \textbf{Cov. X1} & \textbf{Cov. X2} & \textbf{LPS} \\
\midrule
BMA (hyper-$g/n$) & 0.53 & 0.34 & 0.64 & 0.9 & 1.38 \\
gIVBMA (hyper-$g/n$, IW) & 0.27 & \textbf{0.05} & 0.97 & 0.97 & 1.19 \\
gIVBMA (hyper-$g/n$, $\omega_a = 0.1$) & \textbf{0.25} & \textbf{0.05} & 0.97 & 0.97 & \textbf{1.18} \\
IVBMA & 0.42 & 0.39 & 0.98 & 1.0 & 1.29 \\
TSLS & 0.3 & 0.16 & 0.96 & 0.94 & 1.46 \\
O-TSLS & 0.32 & 0.12 & 0.98 & 0.96 & 1.48 \\
MATSLS & 0.49 & 0.18 & \textbf{0.95} & \textbf{0.95} & 1.65 \\
\midrule
\multicolumn{6}{c}{$n=500$} \\
\midrule
 & \textbf{MAE} & \textbf{Bias} & \textbf{Cov. X1} & \textbf{Cov. X2} & \textbf{LPS} \\
\midrule
BMA (hyper-$g/n$) & 0.76 & 0.67 & 0.0 & \textbf{0.95} & \textbf{1.13} \\
gIVBMA (hyper-$g/n$, IW) & \textbf{0.08} & \textbf{0.02} & \textbf{0.92} & 0.92 & 1.14 \\
gIVBMA (hyper-$g/n$, $\omega_a = 0.1$) & 0.1 & \textbf{0.02} & 0.91 & 0.96 & \textbf{1.13} \\
IVBMA & 0.21 & 0.18 & 0.98 & 0.93 & 1.14 \\
TSLS & 0.11 & 0.06 & 0.9 & 0.96 & 1.42 \\
O-TSLS & 0.11 & 0.04 & 0.91 & 0.96 & 1.42 \\
MATSLS & 0.1 & 0.05 & 0.89 & \textbf{0.95} & 1.42 \\
\bottomrule
\end{tabular}
\caption{{\bf Multiple endogenous variables with correlated instruments:} Results with two endogenous variables (one Gaussian and one Beta) based on 100 simulated datasets. The best values in each column are printed in bold.}
\label{tab:SimResMultEndo}
\end{table}

\begin{table}[h]
\centering
\begin{tabular}{l*{8}{c}}
\toprule
& \multicolumn{2}{c}{gIVBMA (h-$g/n$, IW)} & \multicolumn{2}{c}{gIVBMA (h-$g/n$, $\omega_a = 0.1$)} & \multicolumn{2}{c}{IVBMA (X1)} & \multicolumn{2}{c}{IVBMA (X2)} \\
\cmidrule(lr){2-3} \cmidrule(lr){4-5} \cmidrule(lr){6-7} \cmidrule(lr){8-9}
Variable & n=50 & n=500 & n=50 & n=500 & n=50 & n=500 & n=50 & n=500 \\
\midrule
$\boldsymbol{Z_{1}}$ & $1.0$ & $1.0$ & $1.0$ & $1.0$ & $1.0$ & $1.0$ & $1.0$ & $1.0$ \\
$Z_{2}$ & $0.0$ & $0.0$ & $0.0$ & $0.0$ & $0.185$ & $0.055$ & $0.065$ & $0.015$ \\
$Z_{3}$ & $0.0$ & $0.0$ & $0.001$ & $0.0$ & $0.184$ & $0.058$ & $0.084$ & $0.015$ \\
$Z_{4}$ & $0.0$ & $0.0$ & $0.002$ & $0.0$ & $0.233$ & $0.066$ & $0.08$ & $0.016$ \\
$\boldsymbol{Z_{5}}$ & $1.0$ & $1.0$ & $1.0$ & $1.0$ & $1.0$ & $1.0$ & $0.325$ & $1.0$ \\
$Z_{6}$ & $0.0$ & $0.0$ & $0.005$ & $0.0$ & $0.184$ & $0.04$ & $0.062$ & $0.008$ \\
$\boldsymbol{Z_{7}}$ & $1.0$ & $1.0$ & $1.0$ & $1.0$ & $1.0$ & $1.0$ & $0.256$ & $1.0$ \\
$Z_{8}$ & $0.006$ & $0.0$ & $0.002$ & $0.0$ & $0.178$ & $0.051$ & $0.064$ & $0.014$ \\
$Z_{9}$ & $0.003$ & $0.0$ & $0.004$ & $0.0$ & $0.174$ & $0.05$ & $0.057$ & $0.011$ \\
$Z_{10}$ & $0.0$ & $0.0$ & $0.0$ & $0.0$ & $0.164$ & $0.04$ & $0.061$ & $0.016$ \\
$\boldsymbol{Z_{11}}$ & $1.0$ & $1.0$ & $1.0$ & $1.0$ & $1.0$ & $1.0$ & $0.507$ & $1.0$ \\
$Z_{12}$ & $0.0$ & $0.0$ & $0.004$ & $0.0$ & $0.144$ & $0.04$ & $0.073$ & $0.014$ \\
$\boldsymbol{Z_{13}}$ & $1.0$ & $1.0$ & $1.0$ & $1.0$ & $1.0$ & $1.0$ & $0.08$ & $1.0$ \\
$Z_{14}$ & $0.001$ & $0.0$ & $0.007$ & $0.0$ & $0.143$ & $0.044$ & $0.067$ & $0.009$ \\
$Z_{15}$ & $0.004$ & $0.0$ & $0.004$ & $0.0$ & $0.173$ & $0.045$ & $0.055$ & $0.008$ \\
\bottomrule
\end{tabular}
\caption{\textbf{Multiple endogenous variables with correlated instruments:} Median treatment posterior inclusion probabilities across 100 simulated datasets. The instruments included in the true model are printed in bold. Note that IVBMA uses separate treatment models for the two endogenous variables $X_1$ and $X_2$.}
\label{tab:SimMultEnd_PIPs}
\end{table}

\subsection{Many weak instruments} \label{sec:many_weak}

We now consider a setting with many individually weak instruments and $l=1$. Our simulation setup is similar to model C in \cite{kuersteiner_constructing_2010}, where half of the instruments are relevant but increasingly weaker, and the other half is irrelevant. The main difference from their design is that we also consider exogenous covariates. We consider $p_1=20$ instruments and $p_2=10$ covariates generated from independent standard normal distributions and two different sample sizes $n \in \{50, 500\}$. Then, we multiply all instruments and covariates with even indices by $100$ to have more variation in the variables' scale. We expect IVBMA to struggle as its prior variance does not account for the scale, while the $g$-prior used in gIVBMA is scale-invariant. The instruments' coefficients are generated by $$\delta_i = c(p_1) \left( 1 - \frac{i}{p_1/2 + 1}\right)^4, \quad i = 1,\dots, p_1/2,$$and all other coefficients are set to zero. The constant $c(p_1)$ is chosen so that the first-stage $R^2$ of the instruments 
is approximately $0.01$ or $0.1$. 
The coefficients of all instruments with even indices are then divided by $100$ to adjust for the scaling. The covariates' coefficients are set to $1/10$ or $1/1000$ for the first $p_2/2$ odd and even covariates, respectively, and zero else, and the treatment effect is set to $\tau = 1/10$. The covariance matrix $\vec{\Sigma}$ is chosen to have $\sigma_{yy}=\sigma_{xx}=1$ and $\sigma_{yx}=1/2$.

We compare gIVBMA's performance to naive BMA (not accounting for endogeneity), IVBMA, Bayesian IV with a horseshoe prior, TSLS on the full model, oracle TSLS (O-TSLS) where we know the correct model specifications, which is of course not attainable in empirical work, TSLS with a jackknife-type fitted value in the first stage \citep[JIVE, ][]{angrist_jackknife_1999}, TSLS with a jackknife and ridge-regularised first-stage \citep[RJIVE, ][]{hansen_instrumental_2014}, the post-LASSO estimator by \cite{belloni_sparse_2012}, and the model-averaged TSLS estimator (MATSLS) with unrestricted weights by \cite{kuersteiner_constructing_2010}. For gIVBMA, we consider three different $g$-prior specifications: hyper-$g/n$ with the full inverse Wishart prior and the Cholesky-based prior with $\omega_a = 0.1$, and the two-component prior (2C) with a hyper-$g/n$ prior on the covariate component $g_C$ and $c=1/2$ and the full inverse Wishart covariance prior. In this setting, the instruments are assumed to be valid a priori and cannot be included in the outcome model, as described in Section \ref{sec:fixed_Z}.

\begin{table}[H]
\centering
\begin{tabular}{l*{8}{r}}
\toprule
 & \multicolumn{8}{c}{$n = 50$} \\
 & \multicolumn{4}{c}{$R^2_f = 0.01$} & \multicolumn{4}{c}{$R^2_f = 0.1$} \\
\cmidrule(lr){2-5}\cmidrule(lr){6-9}
 & \textbf{MAE} & \textbf{Bias} & \textbf{Cov.} & \textbf{LPS} & \textbf{MAE} & \textbf{Bias} & \textbf{Cov.} & \textbf{LPS} \\
\midrule
BMA (hyper-$g/n$) & 0.46 & 0.46 & 0.08 & \textbf{1.38} & 0.45 & 0.45 & 0.08 & 1.48 \\
gIVBMA (IW) & 0.17 & 0.08 & \textbf{0.93} & 1.39 & 0.16 & 0.06 & 1.0 & 1.48 \\
gIVBMA ($\omega_a = 0.1$) & 0.36 & 0.36 & 0.85 & \textbf{1.38} & 0.37 & 0.37 & 0.89 & 1.48 \\
gIVBMA (IW, 2C) & 0.19 & 0.07 & 0.92 & \textbf{1.38} & 0.2 & 0.07 & 1.0 & 1.48 \\
BayesHS & \textbf{0.03} & \textbf{0.03} & \textbf{0.93} & 2.2 & \textbf{0.03} & 0.03 & 1.0 & 2.23 \\
TSLS & 0.47 & 0.47 & 0.29 & 1.47 & 0.44 & 0.44 & 0.36 & 1.56 \\
O-TSLS & 0.51 & 0.51 & 0.48 & 1.43 & 0.37 & 0.37 & 0.7 & 1.5 \\
JIVE & 0.76 & 0.47 & 0.86 & 1.89 & 0.83 & 0.52 & \textbf{0.91} & 2.01 \\
RJIVE & 0.58 & 0.5 & 0.81 & 1.84 & 0.71 & 0.54 & 0.89 & 2.04 \\
MATSLS & 0.79 & 0.36 & 0.9 & 2.03 & 0.47 & \textbf{0.02} & 0.88 & 1.98 \\
IVBMA & 0.13 & 0.11 & 0.99 & 1.41 & 0.11 & 0.05 & 0.99 & \textbf{1.46} \\
Post-LASSO & 0.53 & 0.53 & 0.8 & - & 0.34 & 0.18 & 1.0 & - \\
\midrule
 & \multicolumn{8}{c}{$n = 500$} \\
 & \multicolumn{4}{c}{$R^2_f = 0.01$} & \multicolumn{4}{c}{$R^2_f = 0.1$} \\
\cmidrule(lr){2-5}\cmidrule(lr){6-9}
 & \textbf{MAE} & \textbf{Bias} & \textbf{Cov.} & \textbf{LPS} & \textbf{MAE} & \textbf{Bias} & \textbf{Cov.} & \textbf{LPS} \\
\midrule
BMA (hyper-$g/n$) & 0.49 & 0.49 & 0.0 & \textbf{1.28} & 0.45 & 0.45 & 0.0 & 1.3 \\
gIVBMA (IW) & 0.31 & 0.25 & 0.9 & \textbf{1.28} & 0.12 & 0.04 & 0.9 & \textbf{1.28} \\
gIVBMA ($\omega_a = 0.1$) & 0.41 & 0.41 & 0.78 & \textbf{1.28} & 0.09 & 0.07 & 0.92 & \textbf{1.28} \\
gIVBMA (IW, 2C) & 0.34 & 0.32 & 0.84 & \textbf{1.28} & 0.12 & 0.02 & 0.93 & \textbf{1.28} \\
BayesHS & \textbf{0.02} & 0.02 & 1.0 & 2.14 & \textbf{0.02} & 0.02 & 1.0 & 2.14 \\
TSLS & 0.38 & 0.38 & 0.44 & 1.32 & 0.12 & 0.1 & 0.83 & 1.39 \\
O-TSLS & 0.34 & 0.33 & 0.65 & 1.35 & 0.09 & 0.04 & 0.9 & 1.41 \\
JIVE & 0.9 & 0.56 & 0.73 & 1.89 & 0.22 & 0.19 & 0.97 & 1.62 \\
RJIVE & 0.92 & 0.56 & 0.73 & 1.9 & 0.22 & 0.2 & 0.97 & 1.62 \\
MATSLS & 0.41 & 0.1 & \textbf{0.92} & 1.64 & 0.1 & \textbf{0.0} & \textbf{0.95} & 1.45 \\
IVBMA & 0.12 & \textbf{0.01} & 0.8 & 1.29 & 0.1 & 0.08 & 0.71 & 1.29 \\
Post-LASSO & 0.48 & 0.26 & 1.0 & - & 0.1 & \textbf{0.0} & 0.98 & - \\
\bottomrule
\end{tabular}
\caption{\textbf{Many Weak Instruments:} MAE, median bias, coverage, and mean LPS with a Gaussian endogenous variable based on 100 simulated datasets. The best values in each column are printed in bold. Post-Lasso only returns estimates for $\tau$, but not for the other coefficients, so we cannot compute the LPS. When no instrument is selected, no effect estimates are provided, therefore we do not consider those cases. The number of times no instruments were selected  in the first stage is (from top-left to bottom-right): 95, 85, 76, 3.}
\label{tab:KO2010_Sim}
\end{table}

Table \ref{tab:KO2010_Sim} shows the results. The full inverse Wishart with the regular hyper-$g/n$ prior is the best gIVBMA variant with good MAE and coverage in all scenarios. The performance of the two-component (2C) prior is very similar, while the Cholesky prior with $\omega_a=0.1$ is the worst variant of gIVBMA in three out of four scenarios. The Bayesian horseshoe does best in terms of MAE and bias, but overcovers in three out of four scenarios and predicts badly. IVBMA tends to do slightly better than gIVBMA in terms of MAE, but has bad coverage for $n=500$. TSLS and O-TSLS only come close in the larger sample size with stronger instruments. Naive BMA is not too far from TSLS in terms of MAE and bias, but does considerably worse with $n=500$ and stronger instruments. In line with the discussion at the beginning of Section \ref{sec:simulation}, the predictive performance of BMA is comparable to gIVBMA, but coverage is far too low. The jackknife-based methods tend to have higher MAE and bias but better coverage than regular TSLS. MATSLS performs very well overall in the $n=500$ case but also has good coverage for $n=50$. Post-Lasso performs well with the stronger instruments. However, only in the $n=500$ scenario with stronger instruments does the Lasso tend to select instruments, whereas in the other cases, it often does not select any. 

To investigate the performance of our method if the treatment is a count variable, we consider the same setting, but the treatment $x_i$ is now simulated from a Poisson with rate parameter $\exp(q_i)$, where the Gaussian $q_i$ is generated as before.

\begin{table}[H]
\centering
\begin{tabular}{l*{8}{r}}
\toprule
 & \multicolumn{8}{c}{$n = 50$} \\
 & \multicolumn{4}{c}{$R^2_f = 0.01$} & \multicolumn{4}{c}{$R^2_f = 0.1$} \\
\cmidrule(lr){2-5}\cmidrule(lr){6-9}
 & \textbf{MAE} & \textbf{Bias} & \textbf{Cov.} & \textbf{LPS} & \textbf{MAE} & \textbf{Bias} & \textbf{Cov.} & \textbf{LPS} \\
\midrule
BMA (hyper-$g/n$) & 0.064 & 0.064 & 0.32 & \textbf{1.42} & 0.049 & 0.049 & 0.39 & \textbf{1.47} \\
gIVBMA (IW) & 0.023 & 0.007 & 0.93 & 1.51 & 0.028 & \textbf{0.001} & 0.96 & 1.53 \\
gIVBMA ($\omega_a = 0.1$) & 0.03 & 0.026 & 0.91 & 1.45 & 0.025 & 0.018 & 0.93 & 1.48 \\
gIVBMA (IW, 2C) & 0.026 & \textbf{0.003} & \textbf{0.94} & 1.5 & 0.028 & 0.002 & 0.94 & 1.52 \\
BayesHS & \textbf{0.021} & 0.02 & 0.97 & 2.2 & \textbf{0.018} & 0.017 & 0.99 & 2.17 \\
TSLS & 0.063 & 0.063 & 0.58 & 1.55 & 0.05 & 0.05 & 0.62 & 1.54 \\
O-TSLS & 0.063 & 0.061 & 0.75 & 1.55 & 0.045 & 0.044 & 0.79 & 1.54 \\
JIVE & 0.091 & 0.065 & 0.91 & 1.9 & 0.104 & 0.053 & 0.98 & 2.12 \\
RJIVE & 0.127 & 0.049 & 0.98 & 2.54 & 0.112 & 0.031 & 1.0 & 2.82 \\
MATSLS & 0.136 & 0.054 & 0.96 & 2.11 & 0.075 & 0.019 & \textbf{0.95} & 2.06 \\
IVBMA & 0.099 & 0.076 & 0.55 & 1.52 & 0.099 & 0.091 & 0.51 & 1.52 \\
Post-LASSO & 0.066 & 0.066 & 1.0 & - & 0.107 & 0.08 & 1.0 & - \\
\midrule
 & \multicolumn{8}{c}{$n = 500$} \\
 & \multicolumn{4}{c}{$R^2_f = 0.01$} & \multicolumn{4}{c}{$R^2_f = 0.1$} \\
\cmidrule(lr){2-5}\cmidrule(lr){6-9}
 & \textbf{MAE} & \textbf{Bias} & \textbf{Cov.} & \textbf{LPS} & \textbf{MAE} & \textbf{Bias} & \textbf{Cov.} & \textbf{LPS} \\
\midrule
BMA (hyper-$g/n$) & 0.056 & 0.056 & 0.0 & \textbf{1.35} & 0.047 & 0.047 & 0.0 & \textbf{1.36} \\
gIVBMA (IW) & \textbf{0.007} & \textbf{0.001} & \textbf{0.94} & 1.42 & \textbf{0.006} & \textbf{0.0} & 0.96 & 1.42 \\
gIVBMA ($\omega_a = 0.1$) & \textbf{0.007} & 0.003 & 0.93 & 1.41 & \textbf{0.006} & 0.003 & 0.94 & 1.42 \\
gIVBMA (IW, 2C) & \textbf{0.007} & \textbf{0.001} & \textbf{0.94} & 1.42 & \textbf{0.006} & \textbf{0.0} & 0.96 & 1.42 \\
BayesHS & 0.035 & 0.035 & 1.0 & 2.28 & 0.031 & 0.031 & 1.0 & 2.28 \\
TSLS & 0.052 & 0.052 & 0.61 & 1.38 & 0.02 & 0.019 & 0.84 & 1.4 \\
O-TSLS & 0.048 & 0.048 & 0.86 & 1.38 & 0.018 & 0.012 & 0.9 & 1.41 \\
JIVE & 0.079 & 0.062 & 0.88 & 1.71 & 0.071 & 0.018 & 1.0 & 1.78 \\
RJIVE & 0.066 & 0.053 & 0.89 & 1.7 & 0.097 & 0.018 & 0.98 & 1.86 \\
MATSLS & 0.089 & 0.013 & 0.98 & 1.85 & 0.021 & 0.003 & \textbf{0.95} & 1.46 \\
IVBMA & 0.1 & 0.066 & 0.33 & 1.36 & 0.082 & 0.041 & 0.46 & 1.38 \\
Post-LASSO & 0.358 & 0.121 & 1.0 & - & 0.019 & 0.013 & 1.0 & - \\
\bottomrule
\end{tabular}
\caption{\textbf{Many Weak Instruments:} MAE, median bias, coverage, and mean LPS with a Poisson endogenous variable based on 100 simulated datasets. The best values in each column are printed in bold. Post-Lasso only returns estimates for $\tau$, but not for the other coefficients, so we cannot compute the LPS. When no instrument is selected, no effect estimates are provided, therefore we do not consider those cases. The number of times no instruments were selected  in the first stage is (from top-left to bottom-right): 97, 98, 95, 44.}
\label{tab:PLN_Sim}
\end{table}

Table \ref{tab:PLN_Sim} presents the results, which are qualitatively somewhat similar to the ones above. Notably, the differences are smaller than in the Gaussian setting. This is perhaps a consequence of the additional variance induced by the non-linear transformation, diluting the degree of endogeneity. All gIVBMA variants display strong performance. Its MAE is only beaten by BayesHS in the small $n$ scenario, but its coverage and predictions are much better. IVBMA's performance is now much worse, in particular its coverage is very low. Post-Lasso has even more instances of not selecting a single instrument, so its performance is based on very few iterations.

\begin{figure}
    \centering
    \includegraphics[width=0.8\linewidth]{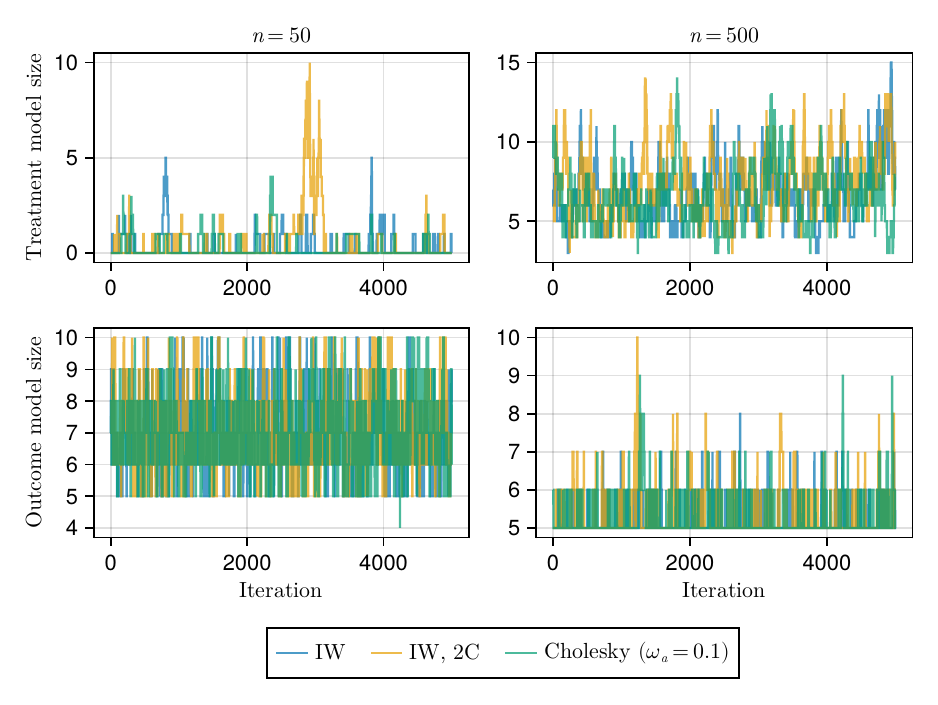}
    \caption{\textbf{Many Weak Instruments:} A chain of treatment (top) and outcome (bottom) model sizes of length $5,000$ (after $1,000$ burn in iterations) for the three considered gIVBMA variants. These results are based on a single simulated dataset for $n = 50$ (left) and $n = 500$ (right), respectively, with instrument strength of $R_f^2 \approx 0.1$.}
    \label{fig:many_weak_mixing}
\end{figure}

For a single realisation of the data-generating process with a Gaussian endogenous variable, Figure~\ref{fig:many_weak_mixing} shows traceplots of the treatment and outcome model sizes. In the treatment model, the $n=50$ chains spend quite some time at the intercept only model, where no instruments are included. However, for $n=500$, the chains for the treatment models mix nicely with most of their mass between $5$ and $10$ included variables. The outcome models' chains mix well between $5-10$ included covariates (the true model includes $5$) for $n=50$, whereas for the larger sample they concentrate on $5$ or $6$ included covariates.

\section{Additional details on the empirical examples} \label{sec:details_examples}

\subsection{Returns to schooling} \label{sec:details_schooling}

Here, we present additional details for the returns to schooling example based on the \cite{card1995collegeproximity} dataset. Table~\ref{tab:card_def} includes definitions of all the variables used. 

\begin{figure}
    \centering
    \includegraphics[width=0.9\linewidth]{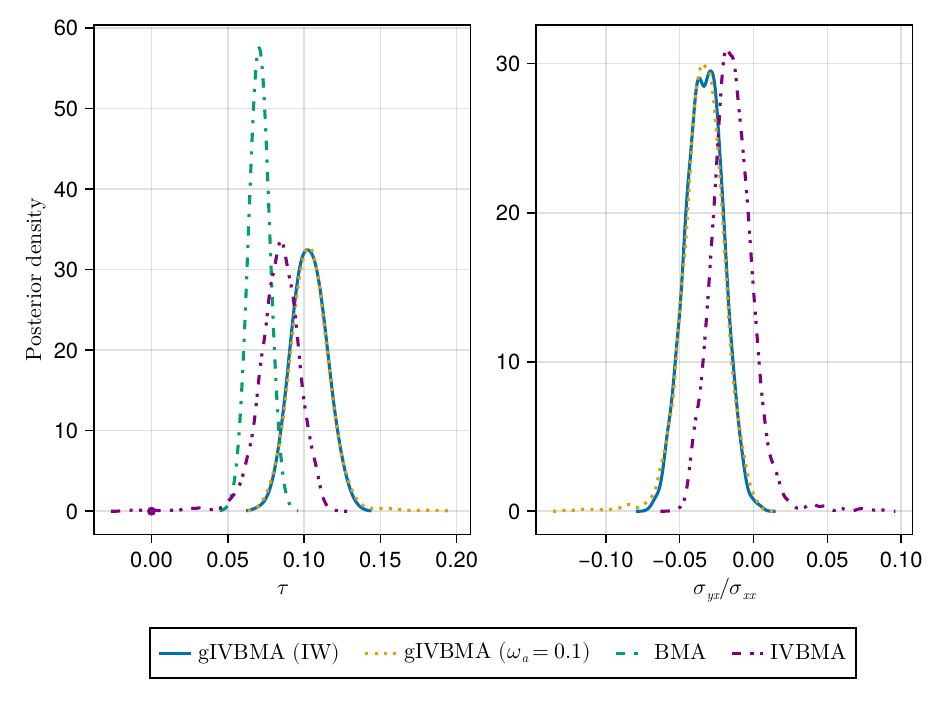}
    \caption{{\bf {Returns to schooling}}: Posterior distributions of the treatment effect of education (Rao-Blackwellized for gIVBMA; Kernel density estimate for IVBMA) and the covariance ratio $\sigma_{yx} / \sigma_{xx}$ based on the \cite{card1995collegeproximity} dataset only using the observations that have complete entries for parental education ($n=2,215$). The algorithm was run for $5,000$ iterations, the first $500$ of which were discarded as burn-in.}
    \label{fig:card_posterior_small}
\end{figure}

\begin{figure}
    \centering
    \includegraphics[width=0.8\linewidth]{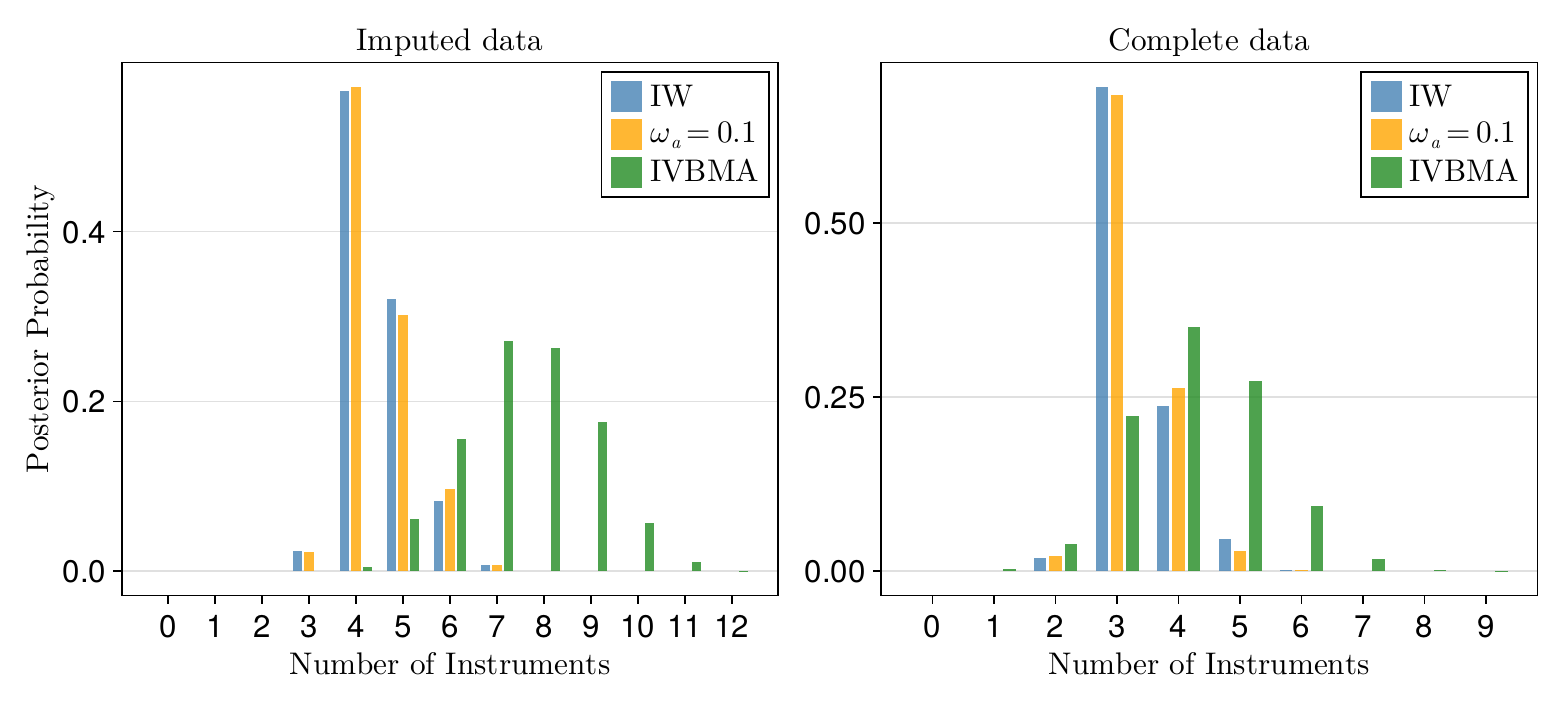}
    \caption{\textbf{Returns to schooling:} The implied posterior distribution of the number of valid and relevant instruments for gIVBMA with the inverse Wishart prior (IW), the Cholesky ($\omega_a = 0.1$) prior, and IVBMA.}
    \label{fig:card_instruments}
\end{figure}

Figure \ref{fig:card_posterior_small} presents the posterior results on the dataset only including observations with complete parental education. Tables \ref{tab:card_pip_1} and \ref{tab:card_pip_2} present the posterior inclusion probabilities obtained with the full \cite{card1995collegeproximity} dataset and the smaller complete dataset, respectively, by gIVBMA, IVBMA, and naive BMA (only the outcome model). Figure~\ref{fig:card_instruments} presents the implied posterior distributions of the number of valid and relevant instruments.

\begin{table}[H]
\centering
\begin{tabular}{lp{0.7\textwidth}}
\toprule
Variable & Definition \\
\midrule
lwage & Logarithm of the hourly wage (in cents) in 1976 \\
educ & Years of schooling in 1976 \\
exper & Experience as measured by age - educ - 6 \\
expersq & Experience squared \\
nearc2 & =1 if near 2-year college in 1966 \\
nearc4 & =1 if near 4-year college in 1966 \\
fatheduc & Father's schooling in years\\
motheduc & Mother's schooling in years\\
fathmiss & =1 if fatheduc is missing \\
mothmiss & =1 if motheduc is missing \\
momdad14 & =1 if live with mom and dad at 14 \\
sinmom14 & =1 if with single mom at 14 \\
step14 & =1 if with step parent at 14 \\
black & =1 if black \\
south & =1 if in south in 1976 \\
smsa & =1 if in a standard metropolitan statistical area (SMSA) in 1976 \\
married & =1 if married in 1976 \\
reg66i & =1 for region i in 1966 \\
\bottomrule
\end{tabular}
\caption{\textbf{Returns to schooling:} Definitions of all variables used from the \cite{card1995collegeproximity} dataset.}
\label{tab:card_def}
\end{table}

\begin{table}[H]
\centering
\begin{tabular}{lccccccc}
\toprule
& \multicolumn{2}{c}{gIVBMA (IW)} & \multicolumn{2}{c}{gIVBMA ($\omega_a = 0.1 $)} & \multicolumn{2}{c}{IVBMA} & \multicolumn{1}{c}{BMA} \\
& L & M & L & M & L & M & L \\
\midrule
exper & 1.0 & 1.0 & 1.0 & 1.0 & 0.006 & 1.0 & 1.0 \\
expersq & 1.0 & 0.0 & 1.0 & 0.014 & 0.0 & 0.003 & 1.0 \\
nearc2 & 0.024 & 0.009 & 0.058 & 0.0 & 0.218 & 0.032 & 0.704 \\
nearc4 & 0.002 & 0.971 & 0.0 & 0.984 & 0.048 & 1.0 & 0.092 \\
momdad14 & 0.005 & 1.0 & 0.005 & 1.0 & 0.008 & 1.0 & 0.176 \\
sinmom14 & 0.008 & 0.009 & 0.002 & 0.023 & 0.048 & 0.218 & 0.079 \\
step14 & 0.0 & 0.003 & 0.014 & 0.01 & 0.333 & 0.208 & 0.164 \\
black & 1.0 & 1.0 & 1.0 & 1.0 & 1.0 & 1.0 & 1.0 \\
south & 1.0 & 0.041 & 1.0 & 0.073 & 1.0 & 0.192 & 1.0 \\
smsa & 1.0 & 0.927 & 1.0 & 0.9 & 1.0 & 0.991 & 1.0 \\
married & 1.0 & 0.982 & 1.0 & 0.958 & 1.0 & 0.989 & 1.0 \\
reg662 & 0.0 & 0.009 & 0.002 & 0.0 & 0.074 & 0.143 & 0.113 \\
reg663 & 0.101 & 0.0 & 0.093 & 0.007 & 0.086 & 0.104 & 0.782 \\
reg664 & 0.048 & 0.0 & 0.013 & 0.023 & 0.581 & 0.477 & 0.205 \\
reg665 & 0.002 & 0.014 & 0.019 & 0.003 & 0.032 & 0.164 & 0.174 \\
reg666 & 0.0 & 0.03 & 0.0 & 0.0 & 0.033 & 0.17 & 0.148 \\
reg667 & 0.0 & 0.002 & 0.0 & 0.004 & 0.023 & 0.173 & 0.108 \\
reg668 & 0.771 & 0.087 & 0.773 & 0.064 & 0.893 & 0.687 & 0.947 \\
reg669 & 0.0 & 0.31 & 0.0 & 0.295 & 0.056 & 0.853 & 0.22 \\
fatheduc & 0.0 & 1.0 & 0.007 & 1.0 & 0.0 & 1.0 & 0.14 \\
motheduc & 0.0 & 1.0 & 0.007 & 1.0 & 1.0 & 1.0 & 0.467 \\
fathmiss & 0.0 & 0.095 & 0.0 & 0.15 & 0.026 & 0.829 & 0.083 \\
mothmiss & 0.009 & 0.032 & 0.0 & 0.016 & 0.012 & 0.545 & 0.11 \\
\bottomrule
\end{tabular}
\caption{\textbf{Returns to schooling:} Posterior inclusion probabilities for the \cite{card1995collegeproximity} dataset using the full (imputed) dataset ($n=3,003$).}
\label{tab:card_pip_1}
\end{table}

\begin{table}[H]
\centering
\begin{tabular}{lccccccc}
\toprule
& \multicolumn{2}{c}{gIVBMA (IW)} & \multicolumn{2}{c}{gIVBMA ($\omega_a = 0.1 $)} & \multicolumn{2}{c}{IVBMA} & \multicolumn{1}{c}{BMA} \\
& L & M & L & M & L & M & L \\
\midrule
exper & 1.0 & 1.0 & 1.0 & 1.0 & 1.0 & 1.0 & 1.0 \\
expersq & 1.0 & 0.001 & 1.0 & 0.001 & 0.389 & 0.006 & 1.0 \\
nearc2 & 0.128 & 0.032 & 0.111 & 0.0 & 0.324 & 0.106 & 0.754 \\
nearc4 & 0.0 & 0.12 & 0.001 & 0.096 & 0.022 & 0.42 & 0.09 \\
momdad14 & 0.0 & 0.007 & 0.0 & 0.0 & 1.0 & 0.144 & 0.154 \\
sinmom14 & 0.004 & 0.003 & 0.0 & 0.007 & 0.799 & 0.258 & 0.166 \\
step14 & 0.01 & 1.0 & 0.0 & 1.0 & 1.0 & 1.0 & 0.284 \\
black & 1.0 & 0.2 & 1.0 & 0.21 & 1.0 & 0.682 & 1.0 \\
south & 1.0 & 0.018 & 1.0 & 0.052 & 1.0 & 0.094 & 1.0 \\
smsa & 1.0 & 0.784 & 1.0 & 0.805 & 1.0 & 0.972 & 1.0 \\
married & 1.0 & 0.421 & 1.0 & 0.406 & 1.0 & 0.728 & 1.0 \\
reg662 & 0.0 & 0.026 & 0.0 & 0.0 & 0.035 & 0.123 & 0.12 \\
reg663 & 0.066 & 0.008 & 0.046 & 0.004 & 0.126 & 0.084 & 0.605 \\
reg664 & 0.011 & 0.001 & 0.009 & 0.016 & 0.276 & 0.273 & 0.314 \\
reg665 & 0.0 & 0.0 & 0.0 & 0.0 & 0.026 & 0.118 & 0.132 \\
reg666 & 0.026 & 0.012 & 0.0 & 0.0 & 0.028 & 0.25 & 0.115 \\
reg667 & 0.0 & 0.0 & 0.0 & 0.014 & 0.037 & 0.144 & 0.148 \\
reg668 & 0.364 & 0.016 & 0.342 & 0.03 & 0.9 & 0.449 & 0.846 \\
reg669 & 0.0 & 0.137 & 0.0 & 0.175 & 0.026 & 0.813 & 0.097 \\
fatheduc & 0.015 & 1.0 & 0.018 & 1.0 & 0.01 & 1.0 & 0.095 \\
motheduc & 0.012 & 1.0 & 0.004 & 1.0 & 0.024 & 1.0 & 0.826 \\
\bottomrule
\end{tabular}
\caption{\textbf{Returns to schooling:} Posterior inclusion probabilities for the \cite{card1995collegeproximity} dataset only using observations that have complete parental education values ($n = 2,215$).}
\label{tab:card_pip_2}
\end{table}

We evaluate the predictive performance on this dataset in a 5-fold cross-validation exercise comparing gIVBMA under the hyper-$g/n$ and BRIC prior with IVBMA, naive BMA, and TSLS. We do not believe any of the other classical methods in our simulation experiments are particularly suited to this setting, as we neither have many weak instruments nor evidence for the intended instruments being invalid. As shown in Table \ref{tab:schooling_5_fold_LPS}, the performance of both gIVBMA variants, IVBMA and naive BMA, is relatively similar on all three subsets of the data. TSLS performs the worst in this predictive comparison. However, this assessment is not entirely fair, both due to the reasons discussed in Section 6  and because TSLS is the only method lacking flexibility in covariate and instrument selection.

\begin{table}[H]
\centering
\begin{tabular}{lcc}
\toprule
Method & Imputed ($n = 3,003$) & Complete ($n=2,215$) \\
\midrule
gIVBMA (IW) & 0.425 & \textbf{0.434} \\
gIVBMA ($\omega_a = 0.1 $) & \textbf{0.424} & \textbf{0.434} \\
BMA & 0.426 & 0.435 \\
IVBMA & 0.439 & 0.442 \\
TSLS & 0.584 & 0.584 \\
\bottomrule
\end{tabular}
\caption{\textbf{Returns to schooling:} The mean LPS calculated over each fold of the imputed data ($n = 3,003$) and the complete parental education data ($n = 2,215$) in a 5-fold cross-validation procedure.}
\label{tab:schooling_5_fold_LPS}
\end{table}

\end{document}